\documentclass[12pt]{article}
\usepackage[T1]{fontenc}
\usepackage[utf8]{inputenc}
\usepackage{comment}
\usepackage{multicol}
\usepackage{cuted}
\usepackage{float}
\usepackage{tikz}
\usetikzlibrary{knots}
\usepackage{braids}
\usetikzlibrary{decorations.markings,hobby,knots,celtic,arrows,shapes,snakes,cd,decorations.pathreplacing,shapes.geometric,calc,shadings,decorations.pathmorphing,arrows,scopes}
\usepackage{pgfplots}
\usepackage{tikz-quantumgates}
\usepackage{rotating}
\usepackage{cancel}
\usepackage{bm}
\usepackage{relsize}
\expandafter\let\csname equation*\endcsname\relax
\expandafter\let\csname endequation*\endcsname\relax
\usepackage{amsmath,esint}
\usepackage{amsthm}
\usepackage{braket}
\usepackage{amssymb}
\usepackage{amstext}
\usepackage{enumerate}
\usepackage{mathtools}
\usepackage[mathscr]{euscript}
\usepackage{hyperref}
\usepackage{physics}
\usepackage{simplewick}
\usepackage{graphicx}
\usepackage{xcolor,colortbl}
\usepackage{caption}
\usepackage{subcaption}
\graphicspath{{images/}}
\usepackage{color}
\usepackage[a4paper, width=170mm, top=25mm, bottom=25mm, bindingoffset=6mm]{geometry}
\usepackage{fancyhdr}
\fancypagestyle{plain}{}
\newcommand{\eq}[1]{\begin{align}#1\end{align}}
\newcommand{\fig}[1]{\begin{figure}#1\end{figure}}
\newcommand{\tik}[1]{\begin{tikzpicture}#1\end{tikzpicture}}

\usepackage{tocloft}
\usepackage{titlesec}
\providecommand{\keywords}[1]{\textbf{Keywords:} #1}
\usepackage{authblk}
\usepackage{hyperref}
\newcommand*{\email}[1]{%
	\normalsize\href{mailto:#1}{#1}
}
\title{\textbf{Ternary Logic Design in Topological Quantum Computing}}
\author{$^{1,2}$Muhammad Ilyas, $^3$Shawn Cui, $^4$Marek Perkowski}
\affil{$^1$Portland State University, Department of Physics, 1719 SW 10th Ave, Portland, Oregon, US\\
$^2$National Institute of Lasers and Optronics College, Pakistan Institute of Engineering and Applied Sciences, Nilore, Islamabad 45650, PK\\	 \email{ilyas.edu@gmail.com} \\
	$^3$Purdue University, Department of Mathematics, and Department of Physics and Astronomy, 150 N University Street, West Lafayette, Indiana 47907, US \\ \email{cui177@purdue.edu} \\
$^4$Portland State University, Department of ECE, 1900 SW Fourth Ave, Portland, Oregon, US\\ \email{mperkows@ee.pdx.edu}}
\begin{document}
\let\cleardoublepage\clearpage
\maketitle
\thispagestyle{empty}

\begin{abstract}
A quantum computer can perform exponentially faster than its classical counterpart. It works on the principle of superposition. But due to the decoherence effect, the superposition of a quantum state gets destroyed by the interaction with the environment. It is a real challenge to completely isolate a quantum system to make it free of decoherence. This problem can be circumvented by the use of topological quantum phases of matter. These phases have quasiparticles excitations called anyons. The anyons are charge-flux composites and show exotic fractional statistics. When the order of exchange matters, then the anyons are called non-Abelian anyons. Majorana fermions in topological superconductors and quasiparticles in some quantum Hall states are non-Abelian anyons. Such topological phases of matter have a  ground state degeneracy. The fusion of two or more non-Abelian anyons can result in a superposition of several anyons. The topological quantum gates are implemented by braiding and fusion of the non-Abelian anyons. The fault-tolerance is achieved through the topological degrees of freedom of anyons. Such degrees of freedom are non-local, hence inaccessible to the local perturbations.
In this paper, the Hilbert space for a topological qubit is discussed. The Ising and Fibonacci anyonic models for binary gates are briefly given.
Ternary logic gates are more compact than their binary counterparts and naturally arise in a type of anyonic model called the metaplectic anyons. The mathematical model, for the fusion and braiding matrices of metaplectic anyons, is the quantum deformation of the recoupling theory.
We proposed that the existing quantum ternary arithmetic gates can be realized by braiding and topological charge measurement of the metaplectic anyons.
\end{abstract}

\section{Introduction}\label{Intro}
Two of the greatest revolutions of the twentieth century were the discovery of quantum mechanics and the invention of computers. At the end of the twentieth century, these two fields merged and a new field of quantum information was born. The quantum information science ends Moore's law, according to which the computing power would double every eighteen months. This law governed silicon chip-based computers, for which the density of chips can be increased. Such computers obey the laws of classical mechanics. But we cannot reduce the physical size of chips infinitely. At the atomic level, particles behave according to the laws of quantum mechanics rather than the laws of classical mechanics.

In 1982, Richard Feynman pointed out that there is a fundamental limit with the ability of classical computers to efficiently simulate a quantum system \cite{feynman1982simulating}. He showed that some problems can be solved exponentially faster on a quantum computer using exponentially large-sized Hilbert space than they could be solved on a classical computer. David Deutsch showed that classical computers cannot efficiently simulate a quantum computer \cite{deutsch1985quantum}. Hence, a quantum computer is important for two reasons; it can perform faster, and it can answer questions about nature.

The building blocks of a classical computer are \textit{bits}. These bits are based on classical logic that has values of either $0$ or $1$. Operations on these bits are performed by a series of gates. These gates change their values and answer the operations. The classical circuits are composed in space from gates connected by wires. But quantum computers are based on quantum logic, which has values in the superposition of $0$ and $1$. The quantum gates manipulate the quantum superposition and give outputs with some probability. A ternary quantum gate is a three-valued logic design, based on the superposition of $0$,$1$, and $2$.

Many methods of encryption on a classical computer are based on difficulty in finding the prime factors of a large number. Peter Shor \cite{shor1994algorithms} invented an algorithm to find the prime factors of a number on a quantum computer with an exponential speed up. This algorithm created widespread interest in quantum computers. Many other quantum algorithms have already been proposed. Grover's search algorithm for an unstructured search \cite{grover1996fast, grover1997quantum} offers a quadratic speed up compared with a classical counterpart. These algorithms are implemented on a particular model of quantum computation.

Building a quantum computer is a great challenge due to its susceptibility to errors. The quantum superposition is destroyed due to its interaction with the environment. This process is called \textit{decoherence}. Moreover, we cannot measure the state and look for errors. In doing so we would kill the superposition. Errors can also be in the phase of a state. There are quantum error correction codes \cite{calderbank1996good, steane1996error,calderbank1997quantum}, but a quantum system needs to be completely isolated from the environment. In 1997, Alexei Kitaev proposed a model for the fault-tolerant quantum computation \cite{kitaev2003fault}. Information is encoded in some non-local degrees of freedom of particles, hence inaccessible to local perturbations \cite{gottesman1998theory,nayak2008non,freedman2003topological}. This is done using the systems which are \textit{topological} in nature. 
The topology is a study of spaces that are continuously deformable to each other. Such spaces are called manifolds. A manifold is a space that is Euclidean flat space locally when a small patch is taken, but it has some non-Euclidean structure globally. To compare two spaces, some properties of the spaces are computed. These properties remain invariant under the continuous deformation of one space to the other. Such properties are called \textit{topological invariants}. We will discuss topology in \ref{Knot}.

The topological nature of particles can be studied through their exchange statistics. Let $\psi(\bm{r}_i,\bm{r}_j)$ be the wave function of two particles at positions $\bm{r}_i$ and $\bm{r}_j$. In three dimensions, when two particles exchange their places, the wave function gets multiplied with a phase factor. That is, 
\eq{\psi(\bm{r}_i,\bm{r}_j) = e^{i\theta}\psi(\bm{r}_j,\bm{r}_i) \label{Exchange},}
where the values $\theta = 0, 2\pi$ correspond to the exchange of bosons and $\theta = \pi$ corresponds to the exchange of fermions. The phase acquired by the wave function is $+1$ for bosons and $-1$ for fermions. Bosons are integral spin particles and obey Bose-Einstein statistics, whereas fermions are half-integer spin particles and obey Fermi-Dirac statistics. A double exchange of these particles is equivalent to no exchange. If the particles are distinguishable, then their statistics is described by the \textit{permutation group} $\mathcal{S}_N$. A group is a mathematical structure to study symmetries. The permutation group is used to study the exchange symmetry.

Jon Magne Leinaas in 1977  \cite{leinaas1977theory} suggested that in a two-dimensional space, another statistic may occur, called \textit{fractional statistics}. For this new kind of statistics, $\theta$ has an arbitrary value between $0$ and $\pi$. Bosons and fermions do not obey fractional statistics even in two-dimensional space. Wilczek \cite{wilczek1982magnetic, wilczek1982quantum} proposed a model for the realization of the fractional statistics. He also named these particles as \textit{anyons} (neither boson nor fermion but any on) \cite{wilczek1982quantum}.

An anyon is not an elementary particle, but a collective phenomenon or a local disturbance in two-dimensional topological materials in a high magnetic field and at a very low temperature. A large number of elementary particles behave in a coordinated way to make quasiparticles. These particles can exist only inside a material, not in free space. Magnetic fluxes are attached to quasiparticles and make charge-flux composites. These quasiparticles obey fractional statistics. The Chern-Simons gauge theory is used as an effective field theory to describe these materials. Quasiparticles have a \textit{topological charge} which is a topological quantum number and is a generalization of the conventional charge. It is a topological invariant and changes on topological phase transition. We will discuss the topological invariants in \ref{Knot}.

The anyons are quasiparticles in quantum Hall states \cite{girvin1987quantum,moore1991nonabelions,read1999beyond} and as Majorana fermions in topological superconductors \cite{lahtinen2017short}.
Anyons are detected in laboratory \cite{camino20073, willett2009measurement,stern2006proposed,rosenow2016current,nakamura2020direct,willett2013magnetic}, and more recently \cite{rosenow2016current,nakamura2020direct}. The measurement of an anyon is done by interference as described in \cite{stern2008anyons,nayak2008non}. The discussion of topological materials, and their effective field theory, is beyond the scope of this paper, see \cite{nayak2008non,ilyas2021topological} and the references therein.

\begin{figure*}[h!]
	\centering
	\begin{tikzpicture}
		\draw[fill=green!20,thick] (-0.3,1.1) -- (4.4,1.1) -- (3.3,-0.2) -- (-1.5,-0.2)-- cycle;
		\draw [ultra thick,opacity=0.8,red] (0.5,0.5) ellipse (0.8 and 0.2);
		\draw [ultra thick,opacity=0.8,blue] (1.2,0.5) ellipse (1.5 and 0.4);
		\filldraw[blue] (0.5,0.5) circle (3pt);
		\filldraw[red] (2,0.5) circle (3pt);
		\node at (1.5,0.3) {$C_1$};
		\node at (2.9,0.3) {$C_2$};
	\end{tikzpicture}	
	\begin{tikzpicture}
		\draw[fill=green!20,thick] (-0.3,1.1) -- (4.4,1.1) -- (3.3,-0.2) -- (-1.5,-0.2)-- cycle;
		\draw [ultra thick,opacity=0.8,red] (0.5,0.5) ellipse (0.8 and 0.2);
		\draw [ultra thick,opacity=0.8,blue,rotate=45] (1.5,0.55) ellipse (1.4 and 0.3);
		\filldraw[blue] (0.5,0.5) circle (3pt);
		\filldraw[red] (2,0.5) circle (3pt);
		\node at (1.5,0.3) {$C_1$};
		\node at (2,2) {$C_2$};
	\end{tikzpicture}
	\caption[In general, two closed curves $C_1$ and $C_2$ are topologically distinct in two dimensions]{Two closed paths $C_1$ and $C_2$ are topologically distinct in two dimensions, but they can be deformed to each other in three dimensions.}
	\label{TwoPaths}
\end{figure*}
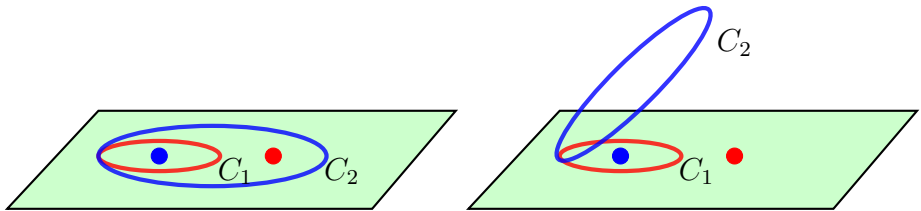

The fundamental difference between 2D and 3D is the difference in the topology of spacetime. The motion of particles makes \textit{knots and links} in spacetime. Two paths are topologically equivalent if one can be deformed to the other. In two dimensions, in general, we cannot transform one path to the other without cutting, as shown in Fig. \ref{TwoPaths}. All smoothly deformable trajectories are in the same equivalence class. Fermions and bosons do not obey the fractional statistics even in 2D, but the change in the wave function of the system in two-dimensional topological materials, when two quasiparticles are exchanged, is independent of the distance and speed of exchange. In contrast, the evolution may depend on some global characteristic of the path. Therefore, the statistics of anyons are topological. Instead of the permutation group, the double exchange of anyons is not equal to the identity. The exchange statistics of anyons is described by a \textit{braid group} in $(2+1)$-dimensional spacetime. The braid group is defined in \ref{Knot}. When the order of composition of two elements of a group does not matter then the group is \textit{Abelian}, otherwise, it is \textit{non-Abelian}.

One of the properties of the topological phases of matter is the existence of \textit{ground state degeneracy}. The degenerate ground states have a large energy gap to the excited states. The degeneracy depends on the topology of the two-dimensional system and the types of anyons present. The ground state is unique for trivial topology.
For Abelian anyons, the braid operators commute and the ground state is unique, but for non-Abelian anyons, the braiding corresponds to the evolution of the system in the degenerate ground state. The change of the system from one ground state to the other is studied using the Berry phase \cite{berry1984quantal} as described in \ref{GeoPhase}.
Let $g$ be degenerate states $\psi_a$ with $a=1,2,...,g$ of particles at positions $x_1,x_2,...,x_n$. Exchanging particles 1 and 2 may not just change the phase but may rotate it into a different state $\psi_b$. Braiding of 1 and 2 and that of 2 and 3 are given as
\eq{\psi_a \rightarrow M_{ab}\psi_b, \qquad \psi_a \rightarrow N_{ab} \psi_b \label{Exchange2},} 
where $M_{ab}$ and $N_{ab}$ are $g\cross g$ dimensional unitary matrices. For Abelian anyons, the $\theta$ in Eq. \ref{Exchange} is arbitrary and clockwise and anticlockwise exchanges commute. Which means that even the clockwise and anticlockwise exchanges may not be the same, but if we exchange particles clockwise then anticlockwise, it will be the same as if we perform anticlockwise exchange first then clockwise. In contrast, $M_{ab}$ and $N_{ab}$ in Eq. \ref{Exchange2} do not commute in general, that is $M_{ab}N_{ab}- N_{ab}M_{ab} \ne 0$ and particles obey non-Abelian statistics. 

Since the unitary evolution only depends on the topology of the path, wiggles of the path would not affect the outcome. No local perturbation can split the degeneracy, hence the system is decoherence-free. The topological nature of anyons is the source of the fault tolerance in a quantum computer.
A topological quantum computer is based on three steps; the creation of anyon-antianyon pairs from the vacuum, braiding, and fusion \cite{pachos2012introduction,nayak2008non}. Anyons can be combined by bringing them close to each other. This is called \textit{fusion}. The fusion is an inverse of the creation of the particles. The fusion of an anyon with its antiparticle gives the total topological charge zero, but the fusion of an anyon with another different type of anyon or antianyon may give the third particle or a superposition of a collection of several particles. The resultant types of particles depend on the \textit{fusion rules}. The topological charge of an anyon is assigned with respect to its fusion with other particles to get a vacuum.
There might not be a unique way to combine anyons. Different ways of the fusion of multiple anyons to get an outcome are called the \textit{fusion channels}. These fusion channels provide the basis states of the Hilbert space for quantum gates. The dimension of the Hilbert space is equal to the degeneracy of the ground state. The transformation between different fusion channels is given by \textit{$F$-matrices}.
The internal degrees of freedom of the anyons are changed by braiding and can put the system in another ground state. The phases acquired by anyons during the braiding are computed through \textit{R-matrices}.
A quantum superposition of states can be created by a suitable combination of $F$ and $R$ matrices.

From the path integral point of view, the anyon's trajectories make knots whose invariants are the probability amplitudes from an initial to a final configuration of the system of anyons. The orientations of knots correspond to the direction of particle trajectories, and the twist in a ribbon knot corresponds to the \textit{topological spin}. The topological spin is the phase due to the rotation of a topological charge around its magnetic flux attached to it. 
To specify the braiding statistics, we need the data such as; particle species, fusion rules, F-matrices, R-matrices, and topological spin.
The mathematical model for such data is the \textit{category theory} and the quantum deformation of the recoupling theory of angular momenta. 

The ternary logic gates and circuits, or the ones that consist of a combination of binary and ternary, are more compact than their binary counterparts \cite{haghparast2017towards}. There are some non-Abelian anyons for which the ternary structures naturally arise. Such anyons are called metaplectic anyons. In this paper, we proposed improved ternary arithmetic circuit designs that can be implemented with the metaplectic anyons.

This paper is organized as follows. The concept of the Hilbert space and topological qubit in topological quantum computation is described in Section \ref{TQC}.
Topological ternary logic design is based on the $SU(2)_k$ anyon model which is the quantum deformation of the recoupling theory of angular momentum. This model will be discussed in Section \ref{QDeform}.
Section \ref{Meta} is on one-qutrit and two-qutrit topological gates.
Our proposed topological ternary arithmetic circuit designs are presented in Section \ref{Arith}. We concluded this paper in Section \ref{Conclusion}. Appendices are added as the background on topological quantum computation. In \ref{QC}, the basic of quantum binary and ternary logic is explained. The topology and knot theory, and geometric phases are discussed in \ref{Knot} and \ref{GeoPhase}. The recoupling theory of angular momentum is given in \ref{Rec}.

\section{Topological Quantum Computation}\label{TQC}
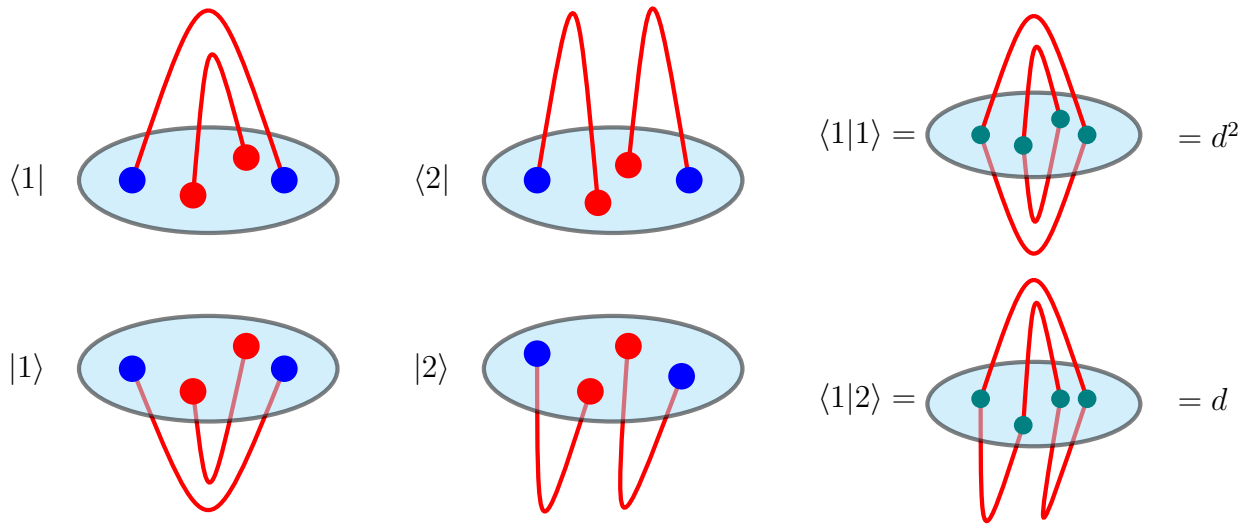
\begin{figure*}[h!]
	\centering
	\begin{tikzpicture}
		\draw [ultra thick,fill=cyan!30,opacity=0.5] (0,0) ellipse (1.7 and 0.7);
		\path[draw,ultra thick,red,postaction={decorate,decoration={markings,mark=between positions 0.15 and 1 step 0.25 with {\arrow[blue]{};}}}] (-1,0) .. controls(0,3) ..(1,0);
		\path[draw,ultra thick,red,postaction={decorate,decoration={markings, mark=between positions 0.15 and 1 step 0.25 with {\arrow[blue]{};}}}](-0.2,-0.2) .. controls(0,2.2) ..(0.5,0.3);
		\fill[blue] (-1,0) circle(5pt);
		\fill[blue] (1,0) circle(5pt);
		\fill[red] (-0.2,-0.2) circle(5pt);
		\fill[red] (0.5,0.3) circle(5pt);
		\path[draw,ultra thick,red,postaction={decorate,decoration={markings, mark=between positions 0.15 and 1 step 0.25 with {\arrow[blue]{};}}}] (-1,-2.5) .. controls(0,-5) ..(1,-2.5);
		\path[draw,ultra thick,red,postaction={decorate,decoration={markings, mark=between positions 0.15 and 1 step 0.25 with {\arrow[blue]{};}}}] (-0.2,-2.8) .. controls(0,-4.5) ..(0.5,-2.2);
		\draw [ultra thick,fill=cyan!30,opacity=0.5] (0,-2.5) ellipse (1.7 and 0.7);
		\fill[blue] (-1,-2.5) circle(5pt);
		\fill[blue] (1,-2.5) circle(5pt);
		\fill[red] (-0.2,-2.8) circle(5pt);
		\fill[red] (0.5,-2.2) circle(5pt);
		\node [left] at (-2,0) {$\bra{1}$};
		\node [left] at (-2,-2.5) {$\ket{1}$};
	\end{tikzpicture}\qquad
	\begin{tikzpicture}
		\draw [ultra thick,fill=cyan!30,opacity=0.5] (0,0) ellipse (1.7 and 0.7);
		\path[draw,ultra thick,red,postaction={decorate,decoration={markings, mark=between positions 0.15 and 1 step 0.25 with {\arrow[blue]{};}}}] (-1,0) .. controls(-0.5,3) ..(-0.2,-0.3);
		
		\path[draw,ultra thick,red,postaction={decorate,decoration={markings, mark=between positions 0.15 and 1 step 0.25 with {\arrow[blue]{};}}}] (0.2,0.2) .. controls(0.5,3) ..(1,0);
		\fill[blue] (-1,0) circle(5pt);
		\fill[red] (-0.2,-0.3) circle(5pt);	
		\fill[red] (0.2,0.2) circle(5pt);
		\fill[blue] (1,0) circle(5pt);	
		\path[draw,ultra thick,red,postaction={decorate,decoration={markings, mark=between positions 0.15 and 1 step 0.25 with {\arrow[blue]{};}}}] (-1,-2.3) .. controls(-1,-5) ..(-0.3,-2.8);	
		\path[draw,ultra thick,red,postaction={decorate,decoration={markings, mark=between positions 0.15 and 1 step 0.25 with {\arrow[blue]{};}}}] (0.2,-2.1) .. controls(0,-5) ..(0.9,-2.6);
		\draw [ultra thick,fill=cyan!30,opacity=0.5] (0,-2.5) ellipse (1.7 and 0.7);
		\fill[blue] (-1,-2.3) circle(5pt);
		\fill[red] (-0.3,-2.8) circle(5pt);	
		\fill[red] (0.2,-2.2) circle(5pt);
		\fill[blue] (0.9,-2.6) circle(5pt);
		\node [left] at (-2,0) {$\bra{2}$};
		\node [left] at (-2,-2.5) {$\ket{2}$};
	\end{tikzpicture}\qquad
	\begin{tikzpicture}[scale=0.7]
		\path[draw,ultra thick,red,postaction={decorate,decoration={markings,
				mark=between positions 0.15 and 1 step 0.25 with {\arrow[blue]{};}}}] (-1,0) .. controls(0,-3) ..(1,0);
		\path[draw,ultra thick,red,postaction={decorate,decoration={markings,
				mark=between positions 0.15 and 1 step 0.25 with {\arrow[blue]{};}}}] (-0.2,-0.2) .. controls(0,-2.2) ..(0.5,0.3);
		\draw [ultra thick,fill=cyan!30,opacity=0.5] (0,0) ellipse (2 and 0.8);
		\path[draw,ultra thick,red,postaction={decorate,decoration={markings,
				mark=between positions 0.15 and 1 step 0.25 with {\arrow[blue]{};}}}] (-1,0) .. controls(0,3) ..(1,0);
		\path[draw,ultra thick,red,postaction={decorate,decoration={markings,
				mark=between positions 0.15 and 1 step 0.25 with {\arrow[blue]{};}}}] (-0.2,-0.2) .. controls(0,2.2) ..(0.5,0.3);	
		\fill[teal] (-1,0) circle(5pt);
		\fill[teal] (1,0) circle(5pt);
		\fill[teal] (0.5,0.3) circle(5pt);
		\fill[teal] (-0.2,-0.2) circle(5pt);
		\node [left] at (-2,0) {$\bra{1}\ket{1}=$};
		\node [right] at (2.5,0) {$=d^2$};
		\path[draw,ultra thick,red,postaction={decorate,decoration={markings,
				mark=between positions 0.15 and 1 step 0.25 with {\arrow[blue]{};}}}] (-1,-5) .. controls(-1,-8) ..(-0.2,-5.5);
		\path[draw,ultra thick,red,postaction={decorate,decoration={markings,
				mark=between positions 0.15 and 1 step 0.25 with {\arrow[blue]{};}}}] (0.5,-5) .. controls(0,-8) ..(1,-5);
		\draw [ultra thick,fill=cyan!30,opacity = 0.5] (0,-5.1) ellipse (2 and 0.8);
		\path[draw,ultra thick,red,postaction={decorate,decoration={markings,
				mark=between positions 0.15 and 1 step 0.25 with {\arrow[blue]{};}}}] (-1,-5) .. controls(0,-2) ..(1,-5);
		\path[draw,ultra thick,red,postaction={decorate,decoration={markings,
				mark=between positions 0.15 and 1 step 0.25 with {\arrow[blue]{};}}}] (-0.2,-5.5) .. controls(0,-2.5) ..(0.5,-5);
		\fill[teal] (-1,-5) circle(5pt);
		\fill[teal] (-0.2,-5.5) circle(5pt);
		\fill[teal] (0.5,-5) circle(5pt);		
		\fill[teal] (1,-5) circle(5pt);
		\node [left] at (-2,-5) {$\bra{1}\ket{2}=$};
		\node [right] at (2.5,-5) {$=d$};
	\end{tikzpicture}
	\caption{States of spacetime histories \cite{simon2010quantum}.}
	\label{Histories}
\end{figure*}

Topological quantum computing is a fault-tolerant quantum computing, proposed by Alexei Kitaev \cite{kitaev2003fault}, manifested by manipulating quantum information using anyons. A quantum computation model involves three main steps; initialization, unitary evolution, and measurement \cite{divincenzo2000physical}.
In quantum theory, the time evolution of a state is represented by the unitary time evolution operator $U(t)$. When the initial state $\ket{\psi_i}$ evolves unitarily to the final state, it is written as $\ket{\psi_f}=U(t)\ket{\psi_i}$.
The initial state is an input and the final state is an output of a quantum gate, and the readout is a measurement in a certain basis to get a classical result \cite{Nielsen2002quantum}. See \ref{QC} for further details.
The gate operation $U(t)$ is equivalent to the rotation of states in the Hilbert space. Analogous to conventional quantum computing, topological quantum computing has three steps; the creation of pairs of anyons from the vacuum, their braiding, and their fusion. The result of fusion corresponds to the measurement, and the braiding corresponds to the unitary transformation $\Psi_f = (Braid) \Psi_i$. The braids cause rotation within degenerate N-particles space. The change of state by braiding can be explained by geometric phases. The braid group and the geometric phase are discussed in \ref{Knot} and \ref{GeoPhase}.
To see how topological quantum computing is a fault-tolerant quantum computing, let two spacetime histories $\ket{1}$ and $\ket{2}$ in Fig. \ref{Histories} have time reversed states $\bra{1}$ and $\bra{2}$. By the Kauffman bracket we have $\bra{1}\ket{1} = \bra{2}\ket{2} = d^2$, $\bra{2}\ket{1} = d$. The number $d$ is assigned to a loop as we discussed in context of the Kauffman bracket in \ref{Knot}. So $\ket{1}$ and $\ket{2}$ are distinct states as long as $\abs{d} \ne 1$ \cite{simon2010quantum}. The states $\ket{1}$ and $\ket{2}$ locally look the same, but the outcomes of their fusion with $\ket{1}$ or $\ket{2}$ are different. Therefore, disturbing one of the particles would not affect the outcome if the topology of a spacetime trajectory is not changed. The state $\ket{1}$ can also be interpreted as the creation of two pairs of anyons and the state $\bra{1}$ as a fusion of the two pairs of anyons.

For a basic introduction on topological quantum computation, see \cite{field2018introduction,pachos2012introduction}. The topological gates with Ising anyons, using the Kauffman version of the recoupling theory \cite{kauffman1994temperley}, are proposed by \cite{fan2010braid}. For the implementation of gates with Ising anyons in the quantum Hall phase, see \cite{georgiev2008towards,ahlbrecht2009implementation} and for gates with Fibonacci anyons, see \cite{hormozi2007topological}.

\subsection{Hilbert Space}
According to the axioms of topological quantum field theory, a vector space $V(\Sigma)$ is associated with a $d$-dimensional oriented manifold $\Sigma$, which depends only on the topology of $\Sigma$ \cite{atiyah1990geometry}.
The vector space of two disjoint vector spaces of $\Sigma_1$ and $\Sigma_2$, is the tensor product of the spaces of each $\Sigma$.
Reversing the orientation of the surface $\Sigma$ gives dual vector space $V^*$.
Here $\partial M= \Sigma$ can be thought of as a time slice of the system, and $V(\partial M)$ is some possible Hilbert space of the ground state. The interior of the rest of the manifold, other than a boundary, is the spacetime history of the system.

\begin{figure}[h!]
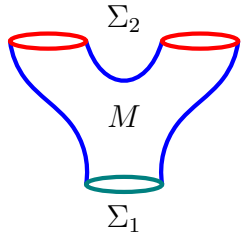

	\centering
	\tik{
		[use Hobby shortcut,scale=1]
		\begin{knot}[
			consider self intersections=true,
			ignore endpoint intersections=false,
			flip crossing=2,
			only when rendering/.style={}]
			\strand [ultra thick,knot=blue](0.5,-0.4).. (0.7,0.4).. (1.3,1).. (1.5,1.5);
			\strand [ultra thick,knot=blue](-0.5,-0.4).. (-0.7,0.4)..(-1.3,1).. (-1.5,1.5);
		\end{knot}	
		\draw[ultra thick,blue](-0.5,1.5) .. controls(-0.3,0.8) and (0.3,0.8)..(0.5,1.5);
		\draw [ultra thick,teal] (0,-0.4) ellipse (0.5 and 0.1);
		\draw [ultra thick,red] (-1,1.5) ellipse (0.5 and 0.1);
		\draw [ultra thick,red] (1,1.5) ellipse (0.5 and 0.1);
		\node at (0,0.5) {$M$};
		\node at (0,1.8) {$\Sigma_2$};
		\node at (0,-0.85) {$\Sigma_1$};}
	\caption{A manifold $M$ that is a spacetime history of a time slices $\Sigma_1$ and $\Sigma_2$.}
	\label{Cob}
\end{figure}

Let us have two punctures on a \textit{Riemann sphere} which is a complex manifold. The gluing axiom of topological quantum field theory \cite{atiyah1990geometry} states that we can glue the two punctures when they have opposing orientations. If we imagine one puncture as a particle, then the other hole must be considered as an antiparticle. It is explained in \ref{GeoPhase} that the ground state degeneracy for an $m$ number of particles on a torus is $m^g$, where $g$ is a genus. A genus is a handle in a manifold. From Fig. \ref{PunturesTori}, we get a genus-one torus $T^2$ when we glue the two opposing particles together. Therefore, the dimension of Hilbert space on a torus is equal to the number of particles types.
This idea can be generalized to $n$ punctures and higher genus torus, written as $n$-torus or $T^n$. Any Riemann surface can be formed by the composition of three punctured Riemann spheres, which is also called \textit{pants} \cite{nayak2008non,fuchs2002tft} shown in Fig. \ref{Cob}. If two of them are fused, a two-punctured sphere will result.
Since the opposite orientations of punctures in TQFT are the opposite charges in anyonic models, two punctures on a sphere with labels $a$ and $\bar{a}$ should have the same topological charge to fuse into the vacuum. The fusion of two particles requires that for $k$ charges, there are $k+1$ different possible allowed boundary conditions. These charges can be identified as $j=0,1/2,...,k/2$ and the corresponding anyonic model is $SU(2)_k$ model with $k+1$ quasiparticles \cite{elitzur1989remarks,nayak2008non}. 

\fig{[h!]
	\centering
	\tik{[scale=0.9]
		\draw[bend right,ultra thick,red] (-1,0) to (1,0);
		\draw [ultra thick,blue] (0,0.1) ellipse (1 and 1);
		\draw [ultra thick,red] (-0.5,0.5) ellipse (0.2 and 0.2);
		\draw [ultra thick,red] (0.5,0.5) ellipse (0.2 and 0.2);
		\node [below] at (-0.5,0.3) {$a$};
		\node [below] at (0.5,0.3) {$\bar{a}$};
		\draw[ultra thick,->] (1.2,0) -- (2,0);
	}
	\tik{[use Hobby shortcut,scale=0.9]
		\draw[bend right,ultra thick,red] (-1,0) to (1,0);
		\draw [ultra thick,blue] (0,0) ellipse (1 and 1);
		\begin{knot}[
			consider self intersections=true,
			ignore endpoint intersections=false,
			flip crossing=2,
			only when rendering/.style={}]
			\strand [ultra thick,knot=blue](-0.8,0.2) ..(-0.7,0.5).. (-0.7,1.3).. (0,1.7).. (0.7,1.3).. (0.7,0.5).. (0.8,0.2);
			\strand [ultra thick,knot=blue](-0.2,0.2).. (-0.3,0.5).. (-0.3,1.2).. (0,1.4).. (0.3,1.2).. (0.3,0.5).. (0.1,0.2);
		\end{knot}		
		\draw[ultra thick,red] (0,1.37) to (0,1.73);
	}
	\caption{Punctures on tori equivalent to the types of particles present}
	\label{PunturesTori}}

\begin{figure}[h!]
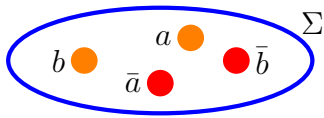

	\centering
	\tik{\draw [ultra thick,blue] (2,3) ellipse (2 and 0.7);
		\fill[orange] (1,3) circle(5pt);
		\fill[red] (2,2.7) circle(5pt);
		\fill[orange] (2.4,3.3) circle(5pt);
		\fill[red] (3,3) circle(5pt);
		\node [] at (4,3.5) {$\Sigma$};
		\node [left=3] at (1,3) {$b$};
		\node [left=3] at (2,2.7) {$\bar{a}$};
		\node [left=3] at (2.4,3.3) {$a$};
		\node [right=3] at (3,3) {$\bar{b}$};
	}
	\caption{Time slice of a manifold with particles.}
	\label{TimeSlice}
\end{figure}

A sphere with one hole is topologically equivalent to a two-dimensional manifold. Let we have a two-dimensional manifold as a disc. When there are no particles on this manifold, then the spacetime history will make a cylinder $M$. The time direction is taken upward in this article.
Here the boundary $\partial M= \Sigma$ can be thought of as a time slice of the system, and $V(\partial M)$ some possible Hilbert space of the ground state. In Fig. \ref{TimeSlice}, we have particles on a boundary of a manifold $M$. When we move these particles around each other, the trajectories or worldlines of the particles will make braids in $(2+1)$-dimensional spacetime and the spacetime history will have a nontrivial topology. This evolution is corresponding to the change of the system from one ground state to the other. Unless a particle is its antiparticle, the worldlines are assumed to be directed.

The types of anyons are categorized by quantum numbers attached to them called the \textit{topological charges}. When the charges are created from the vacuum, their total charge must be zero. Therefore, the value of the topological charge is assigned with respect to its fusion with other anyons. The topological charge zero is assigned to the vacuum. The vacuum is also called a \textit{trivial charge}.
For example, in Ising anyon model, there are three charges $\left\{\textbf{1},\sigma,\psi \right\}$. In some anyonic models, the vacuum is represented by $0$ or $I$. The $\textbf{1}$ represents a vacuum, or a trivial particle, whereas the $\sigma$ and $\psi$ are nontrivial particles. Since the anyons are created from the vacuum, to conserve the total charge, they must be fused to vacuum. The number of ways these anyons are fused to vacuum is called the \textit{fusion channels or fusion trees}. The number of the fusion channels is equal to the ground state degeneracy of the system. The fusion space is a shared property of a collection of non-Abelian anyons regardless of where they are located. Therefore, local perturbations do not affect the degeneracy of the system. In quantum topology, anyons on a two-dimensional manifold can be identified as the punctures or holes on a sphere.

The system with zero anyon is called the vacuum. It is a trivial particle represented by $0$, $\textbf{1}$ or $I$. Since the anyons are created from the vacuum, to conserve the charge, they must be fused to vacuum. The number of ways these anyons are fused to vacuum is called the \textit{fusion channels or fusion trees}. Number of the fusion channels are equal to the ground state degeneracy of the system. The fusion space is a shared property of a collection of the non-Abelian anyons regardless of where they are located. Therefore, the local perturbations do not affect the degeneracy of the system. 

Let several anyons be created from the vacuum, and let us consider a subtree consisting of two anyons $a$ and $b$. A fusion tree diagram is equivalent to the anyons' creation tree if the time direction is reversed.
When these anyons are Abelian, they fused to only one outcome and the dimension of Hilbert space for two anyons is one. But when the particles $a$ and $b$ are non-Abelian, there is more than one fusion outcome, that is $a\cross b = \sum_c N_{ab}^c$.
Their Hilbert space is denoted as $V_{ab}^{c}$. The dimension of Hilbert space is given as $N_{ab}^c = dim(V_{ab}^c)$. The numbers $N_{ab}^c$ are also called the \textit{fusion rules}. These fusion rules appear in conformal field theory and category theory in the form as $\phi_a\cross \phi_b = \sum_c N_{ab}^c\phi_c$. The fusion rules put some restrictions on what types of anyons a particular anyonic model can have. As the fusion of two non-Abelian anyons can result in several anyons, in general, $N_{ab}^c$ could have more values than one. Most of the anyonic models are built by considering only the two fusion outcomes; vacuum or another anyon, that is $N_{ab}^c=0,1$. When the fusion of $a$ and $b$ gives the topological charge $c$ then $N_{ab}^c=1$, but when $a$ and $b$ cannot be fused to $c$ then $N_{ab}^c=0$.

The dimension of Hilbert space increases with the number of anyons, analogous to the addition of spins $1/2 \otimes 1/2 = 0 \oplus 1 $. This analogy is not exact, because the anyons are not elementary particles, but they have internal degrees of freedom. The dimension of Hilbert space of $N$ particles of type $a$ is roughly $\sim d_a^N$, where $d$ is the dimension of Hilbert space of one anyon and is called the \textit{quantum dimension}. It is a measure of how much the Hilbert space is increased by adding this anyon. Therefore, it is an asymptotic degeneracy per particle. It needs not to be an integer. The quantum dimension of vacuum is one. In terms of knot theory, the $d$ is a number assigned to a loop. For the process $a\cross b = \sum_c N_{ab}^c c$, the quantum dimension can be defined to satisfy $d_ad_b = \sum_c N_{ab}^c d_c$.

\begin{figure}[h!]
	\centering
	\begin{subfigure}{0.4\textwidth}
		\centering
		\begin{tikzpicture}[scale=0.5]
			\draw [blue, ultra thick] (2,0) -- (2,1); 
			\draw [blue, ultra thick] (2,1) -- (4,3);
			\draw [blue, ultra thick] (2,1) -- (0,3);
			\draw [blue, ultra thick] (1,2) --(2,3);
			\node [above] at (0,3) {$a$};
			\node [above] at (2,3) {$b$};
			\node [above] at (4,3) {$c$};
			\node [below] at (2,0) {$d$};
			\node [below left] at (1.5,1.5) {$i$};
			\node [] at (6,1) {$= \sum_{j}(F_{acb}^d)^i_j $};
			\draw [blue, ultra thick] (10,0) -- (10,1); 
			\draw [blue, ultra thick] (10,1) -- (12,3);
			\draw [blue, ultra thick] (10,1) -- (8,3);
			\draw [blue, ultra thick] (11,2) --(10,3);		
			\node [above] at (8,3) {$a$};
			\node [above] at (10,3) {$b$};
			\node [above] at (12,3) {$c$};
			\node [below] at (10,0) {$d$};
			\node [below right] at (10.5,1.5) {$j$};
		\end{tikzpicture}
		\caption{}
	\end{subfigure}
	\begin{subfigure}{0.4\textwidth}
		\centering
		\begin{tikzpicture}[scale=0.5]
			\draw [blue, ultra thick] (1,0) -- (1,1);
			\draw [blue, ultra thick] (1,1) -- (0,2);
			\draw [blue, ultra thick] (1,1) -- (2,2);
			\draw [blue, ultra thick] (0,2) -- (2,4);
			\draw [blue, ultra thick] (2,2) -- (1.1,2.9);
			\draw [blue, ultra thick] (0.9,3.1) -- (0,4);
			\node [below] at (1,0) {$i$};
			\node [above] at (0,4) {$a$};
			\node [above] at (2,4) {$b$};
			\node [] at (4,1) {$= R_{ab}^{i}$};
			\draw [blue, ultra thick] (6,0) -- (6,1);
			\draw [blue, ultra thick] (6,1) -- (7,4);
			\draw [blue, ultra thick] (6,1) -- (5,4);
			\node [below] at (6,0) {$i$};
			\node [above] at (5,4) {$a$};
			\node [above] at (7,4) {$b$};
		\end{tikzpicture}
		\caption{}
	\end{subfigure}	
	\caption{(a) F and (b) R-moves.}
	\label{FRMatrices}
\end{figure}
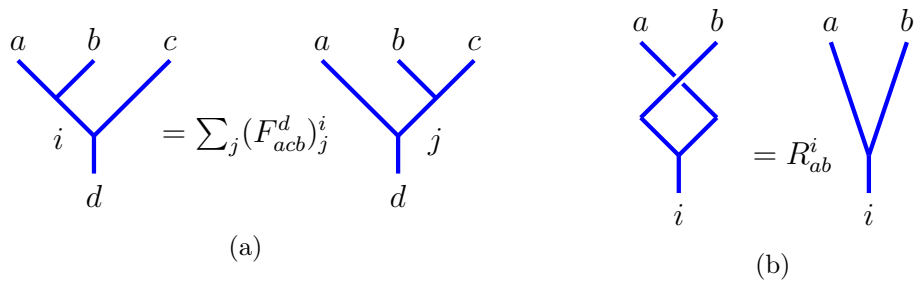
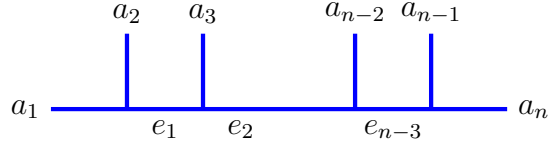
\begin{figure}[h!]
	\centering
	\begin{tikzpicture}
		\draw[ultra thick,blue] (0,0)--(6,0);
		\draw[ultra thick,blue] (1,0)--(1,1);
		\draw[ultra thick,blue] (2,0)--(2,1);
		\draw[ultra thick,blue] (4,0)--(4,1);
		\draw[ultra thick,blue] (5,0)--(5,1);
		\node [left] at (0,0) {$ a_1 $};
		\node [above] at (1,1) {$ a_2 $};
		\node [below] at (1.5,0) {$ e_1 $};
		\node [above] at (2,1) {$ a_3 $};
		\node [below] at (2.5,0) {$ e_2 $};
		\node [above] at (4,1) {$ a_{n-2} $};
		\node [below] at (4.5,0) {$ e_{n-3} $};
		\node [above] at (5,1) {$ a_{n-1} $};
		\node [right] at (6,0) {$ a_n $};
	\end{tikzpicture}
	\caption{Fusion space of anyons \cite{pachos2012introduction}}
	\label{FusionSpace}
\end{figure}

Let there be a situation when the fusion outcome of three non-Abelian anyons $a,b,c$ is $d$. (It can be a subtree of another fusion tree in which $d$ can be fused with another anyon and the outcome is vacuum or some other anyons).
There are more different fusion channels than one. For example, $a$ can be fused to $b$ first, then their outcome $i$ is fused with $c$ to get $d$. Or $b$ can be fused to $c$ first, then their fusion outcome $j$ can be fused with $a$ to get $d$. The fusion channels $i$ and $j$ make two sets of basis. The transformation between these bases $i$ and $j$ is given by $F$-symbols or $F$-moves. As for the non-Abelian anyons, $i$ and $j$ occur in more ways than one, the $F$-moves between different $i$'s and $j$'s will be a matrix called the \textit{$F$-matrix} shown in Fig. \ref{FRMatrices} (a). The vector space for these anyons can be written as $V_{abc}^d=\bigoplus_i V_{ab}^i \otimes V_{ic}^d=\bigoplus_j V_{bc}^j \otimes V_{ja}^d$. The fusion diagrams which can be continuously deformed into each other are equivalent and represent the same state.
For a diagram of $n$ anyons shown in Fig. \ref{FusionSpace}, the dimension of Hilbert space can be written in terms of the fusion rules as $N_{a_1a_2}^{e_1}N_{e_1a_3}^{e_2}...N_{e_{n-3}a_{n-1}}^{a_n}$.

Now let us braid two anyons $a$ and $b$ by exchanging their places. Abelian anyons get a complex phase that depends on the types of anyons and on whether the exchange is clockwise or counterclockwise. But it does not depend on the order of exchange. Therefore, for a Abelian anyon we have the $R$-move given as $R_{ab} = e^{i\theta_{ab}}$. But for non-Abelian anyons, the $R$-move is a matrix $R_{ab}^{i} = e^{i \theta_{ab}^{i}}$ and it depends on the order of the exchange \cite{lahtinen2017short}. Topologically equivalent braids have the same outcomes. The braid matrix is shown in Fig. \ref{FRMatrices} (b).

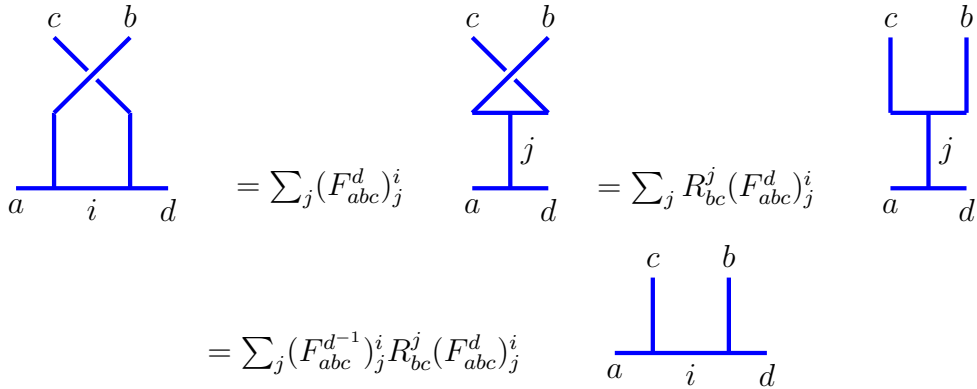
\begin{figure*}[h!]
	\centering
	\begin{tikzpicture}[scale=1]
		\draw[ultra thick,blue] (0.5,0)--(2.5,0);
		\draw[ultra thick,blue] (1,0)--(1,1);
		\draw[ultra thick,blue] (2,0)--(2,1);
		\node [below] at (0.5,0) {$ a $};
		\node [above] at (1,2) {$ c $};
		\node [below] at (1.5,0) {$ i $};
		\node [above] at (2,2) {$ b $};
		\node [below] at (2.5,0) {$ d $};
		\draw[ultra thick,knot=blue] (2,1)--(1,2);
		\draw[ultra thick,knot=blue] (1,1)--(2,2);
		\node [] at (4.5,0) {$= \sum_{j}(F_{abc}^d)^i_j $};
		\draw[ultra thick,blue] (6.5,0)--(7.5,0);
		\draw[ultra thick,blue] (7,0)--(7,1);	
		\draw[ultra thick,knot=blue] (7.5,1)--(6.5,2);
		\draw[ultra thick,knot=blue] (6.5,1)--(7.5,2);
		\draw[ultra thick,blue] (6.5,1)--(7.5,1);
		\node [below] at (6.5,0) {$ a $};
		\node [above] at (6.5,2) {$ c $};
		\node [right] at (7,0.5) {$ j $};
		\node [above] at (7.5,2) {$ b $};
		\node [below] at (7.5,0) {$ d $};
		\node [right] at (8,0) {$= \sum_{j}R_{bc}^j(F_{abc}^d)^i_j $};
		\draw[ultra thick,blue] (12,0)--(13,0);
		\draw[ultra thick,blue] (12.5,0)--(12.5,1);	
		\draw[ultra thick,knot=blue] (12,1)--(12,2);
		\draw[ultra thick,knot=blue] (13,1)--(13,2);
		\draw[ultra thick,blue] (12,1)--(13,1);
		\node [below] at (12,0) {$ a $};
		\node [above] at (12,2) {$ c $};
		\node [right] at (12.5,0.5) {$ j $};
		\node [above] at (13,2) {$ b $};
		\node [below] at (13,0) {$ d $};
	\end{tikzpicture}\\
	\begin{tikzpicture}		
		\node [right] at (13.5,0) {$= \sum_{j}(F_{abc}^{d^{-1}})^i_jR_{bc}^j(F_{abc}^d)^i_j $};
		\draw[ultra thick,blue] (19,0)--(21,0);
		\draw[ultra thick,blue] (19.5,0)--(19.5,1);
		\draw[ultra thick,blue] (20.5,0)--(20.5,1);
		\node [below] at (19,0) {$ a $};
		\node [above] at (19.5,1) {$ c $};
		\node [below] at (20,0) {$ i $};
		\node [above] at (20.5,1) {$ b $};
		\node [below] at (21,0) {$ d $};
	\end{tikzpicture}
	\caption{Braiding for the superposition of the two fusion channels.}
	\label{BMatrix}
\end{figure*}

The fusion channel of a pair of anyons cannot be changed by $R$-move only. In other words, the system would not evolve from one ground state to the other by exchanging the two anyons of the same pair. To change their fusion channel, we need to braid $b$ and $c$. For this braiding, we have to transform $i$ bases to the $j$ bases by an $F$-move. Therefore, having only two anyons is not enough for making a topological qubit.
To see the effect of the exchange of $b$ and $c$ in the basis $i$, first an $F$-matrix is applied to change the basis from $i$ to $j$, then an $R$ matrix is applied, and then an $F^{-1}$-matrix is applied to change the basis back to $i$. This process is shown in Fig. \ref{BMatrix} and can be written as
\eq{B_{ab} = {F_{acb}^d}^{-1} R_{ab} F_{acb}^d.}
The matrix $B$ creates the superposition of the fusion channels, with a distinct phase factor for different fusion channels.

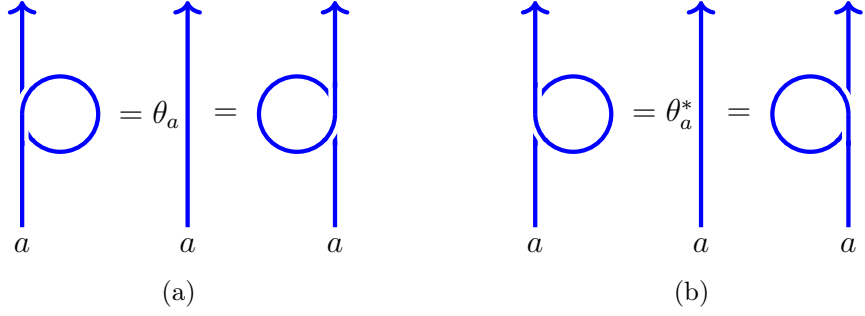
\begin{figure}[h!]
	\centering
	\begin{subfigure}{0.3\textwidth}
		\centering	
		\begin{tikzpicture}[use Hobby shortcut,scale=1]
			\begin{knot}[consider self intersections=true,
				ignore endpoint intersections=false,
				flip crossing=2,
				only when rendering/.style={}]		
				\strand [ultra thick,blue] (0,1) ..(0.5,0.5).. (1,1) .. (0.5,1.5) .. (0,1);
				\strand [ultra thick,blue] (0,-0.5) .. (0,1);
				\strand [ultra thick,blue,->] (0,1) .. (0,2.5);		
				\node [below] at (0,-0.5) {$a$};
			\end{knot}
			\path (0,0);
		\end{tikzpicture}	
		\begin{tikzpicture}
			\node at (0,1) {$=\theta_a$};
			\node at (1,1) {$=$};
			\draw[ultra thick,blue,->] (0.5,-0.5) -- (0.5,2.5);
			\node [below] at (0.5,-0.5) {$a$};
		\end{tikzpicture}
		\begin{tikzpicture}[use Hobby shortcut,scale=1]
			\begin{knot}[consider self intersections=true,
				ignore endpoint intersections=false,
				flip crossing=2,
				only when rendering/.style={}]
				\strand [ultra thick,blue,->] (0,1) .. (0,2.5);
				\strand [ultra thick,blue] (0,-0.5) .. (0,1);
				\strand [ultra thick,blue] (0,1) ..(-0.5,0.5).. (-1,1) .. (-0.5,1.5) .. (0,1);
				\node [below] at (0,-0.5) {$a$};
			\end{knot}
			\path (0,0);
		\end{tikzpicture}
		\caption{}
	\end{subfigure}\qquad \qquad
	\begin{subfigure}{0.3\textwidth}
		\centering
		\begin{tikzpicture}[use Hobby shortcut,scale=1]
			\begin{knot}[consider self intersections=true,
				ignore endpoint intersections=false,
				flip crossing=2,
				only when rendering/.style={}]
				\strand [ultra thick,blue] (0,-0.5) .. (0,1);
				\strand [ultra thick,blue,->] (0,1) .. (0,2.5);
				
				\strand [ultra thick,blue] (0,1) ..(0.5,0.5).. (1,1) .. (0.5,1.5) .. (0,1);
				\node [below] at (0,-0.5) {$a$};
			\end{knot}
			\path (0,0);
		\end{tikzpicture}
		\begin{tikzpicture}
			\node at (0,1) {$=\theta^*_a$};
			\node at (1,1) {$=$};
			\draw[ultra thick,blue,->] (0.5,-0.5) -- (0.5,2.5);
			\node [below] at (0.5,-0.5) {$a$};
		\end{tikzpicture}
		\begin{tikzpicture}[use Hobby shortcut,scale=1]
			\begin{knot}[consider self intersections=true,
				ignore endpoint intersections=false,
				flip crossing=2,
				only when rendering/.style={}]
				
				\strand [ultra thick,blue] (0,1) ..(-0.5,0.5).. (-1,1) .. (-0.5,1.5) .. (0,1);
				\strand [ultra thick,blue] (0,-0.5) .. (0,1);
				\strand [ultra thick,blue,->] (0,1) .. (0,2.5);		
				\node [below] at (0,-0.5) {$a$};
			\end{knot}
			\path (0,0);
		\end{tikzpicture}
		\caption{}
	\end{subfigure}
	\caption{Removing a twist is equivalent to adding a phase (a) $\theta_a = e^{2\pi i h_a}$ and (b) $\theta^*_a = e^{-2\pi ih_a}$.}
	\label{TwistPhase}
\end{figure}

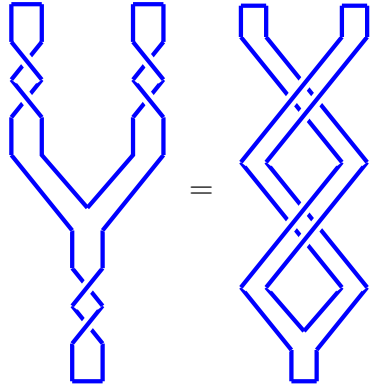
\begin{figure}[h!]
	\centering
	\begin{tikzpicture}	
		\draw[ultra thick,knot=blue] (0.2,0)--(0.2,0.5);
		\draw[ultra thick,knot=blue](-0.2,0)--(-0.2,0.5);
		\draw[ultra thick,knot=blue](0.2,0.5)--(-0.2,1);
		\draw[ultra thick,knot=blue](-0.2,0.5)--(0.2,1);
		\draw[ultra thick,knot=blue](0.2,1)--(-0.2,1.5);
		\draw[ultra thick,knot=blue](-0.2,1)--(0.2,1.5);
		\draw[ultra thick,knot=blue](-0.2,1.5)--(-0.2,2);
		\draw[ultra thick,knot=blue](0.2,1.5)--(0.2,2);
		\draw[ultra thick,knot=blue](-0.2,2)--(-1,3);
		\draw[ultra thick,knot=blue](0.2,2)--(1,3);
		
		\draw[ultra thick,knot=blue](-1,3)--(-1,3.5);
		\draw[ultra thick,knot=blue](1,3)--(1,3.5);
		
		\draw[ultra thick,blue](0,2.3)--(0.6,3)--(0.6,3.5);
		\draw[ultra thick,blue](0,2.3)--(-0.6,3)--(-0.6,3.5);
		\draw[ultra thick,knot=blue](0.6,3.5)--(1,4);
		\draw[ultra thick,knot=blue](1,3.5)--(0.6,4);
		\draw[ultra thick,knot=blue](-1,3.5)--(-0.6,4);
		\draw[ultra thick,knot=blue](-0.6,3.5)--(-1,4);
		\draw[ultra thick,knot=blue](0.6,4)--(1,4.5);
		\draw[ultra thick,knot=blue](1,4)--(0.6,4.5);
		\draw[ultra thick,knot=blue](-1,4)--(-0.6,4.5);
		\draw[ultra thick,knot=blue](-0.6,4)--(-1,4.5);
		\draw[ultra thick,blue](-1,4.5)--(-1,5);
		\draw[ultra thick,blue](-0.6,4.5)--(-0.6,5);
		\draw[ultra thick,blue](1,4.5)--(1,5);
		\draw[ultra thick,blue](0.6,4.5)--(0.6,5);
		\draw[ultra thick,blue](-0.2,0)--(0.2,0);
		\draw[ultra thick,blue](-1,5)--(-0.6,5);
		\draw[ultra thick,blue](0.6,5)--(1,5);
		\node at (1.5,2.5) {$=$};
	\end{tikzpicture}
	\begin{tikzpicture}[scale=0.83]	
		\draw[ultra thick,knot=blue](-0.2,0)--(-0.2,0.5);
		\draw[ultra thick,knot=blue](0.2,0)--(0.2,0.5);
		\draw[ultra thick,knot=blue](-0.2,0.5)--(-1,1.5);
		\draw[ultra thick,knot=blue](0.2,0.5)--(1,1.5);
		\draw[ultra thick,blue](0,0.8)--(0.6,1.5);
		\draw[ultra thick,blue](0,0.8)--(-0.6,1.5);
		
		\draw[ultra thick,knot=blue](0.6,1.5)--(-1,3.5);	
		\draw[ultra thick,knot=blue](1,1.5)--(-0.6,3.5);	
		\draw[ultra thick,knot=blue](-0.6,1.5)--(1,3.5);
		\draw[ultra thick,knot=blue](-1,1.5)--(0.6,3.5);
		
		\draw[ultra thick,knot=blue](0.6,3.5)--(-1,5.5);	
		\draw[ultra thick,knot=blue](1,3.5)--(-0.6,5.5);
		
		\draw[ultra thick,knot=blue](-1,3.5)--(0.6,5.5);
		\draw[ultra thick,knot=blue](-0.6,3.5)--(1,5.5);
		
		\draw[ultra thick,knot=blue](-0.6,5.5)--(-0.6,6);
		\draw[ultra thick,knot=blue](-1,5.5)--(-1,6);
		\draw[ultra thick,knot=blue](0.6,5.5)--(0.6,6);
		\draw[ultra thick,knot=blue](1,5.5)--(1,6);
		\draw[ultra thick,blue](-0.2,0)--(0.2,0);
		\draw[ultra thick,blue](-1,6)--(-0.6,6);
		\draw[ultra thick,blue](0.6,6)--(1,6);
	\end{tikzpicture}
	\caption{Braid-twist or spin-statistics correspondence.}
	\label{BraidTwist}
\end{figure}

The \textit{twist factor} or spin factor of an anyon is a phase corresponding to the rotation of a charge around its own magnetic flux. It can be thought of as a twist in a framed ribbon, as discussed in \ref{Knot}. This factor is written as $\theta_a = e^{2 \pi i h_a}$ when an anyon is rotated by $2\pi$ as shown in Fig. \ref{TwistPhase}, where $h_a$ is the \textit{topological spin} of an anyon $a$. It is an integer for bosons and gives spin factor identity. It is a half-integer for fermions that would give the spin factor $-1$. Its value is between 0 and 1 for an anyon. For vacuum, $h_0 = 0$. Through the ribbon equation, the spin factor can also be used to derive the entries of an $R$-matrix for a particular anyonic model. For example, when two anyons $a$ and $b$ are fused to $c$, the spin factor is given by the ribbon equation pictorially shown in Fig. \ref{BraidTwist} \cite{pachos2012introduction,eliens2014diagrammatics}
\eq{[R_{ab}^c]^2=\frac{\theta_c}{\theta_a \theta_b}= \frac{e^{2 \pi i h_c}}{e^{2 \pi i h_a} e^{2 \pi i h_b}} = e^{2 \pi i(h_c - h_a -h_b)}.}
This is interpreted as the full twist of fusion product $c$ combined with the full twist of the charges $a$ and $b$ in the opposite direction, and is equal to the double exchange of the $a$ and $b$. 

\subsection{Topological Qubit}
Now we will summarize what we discussed in the last section and build a qubit from this discussion. A pair of non-Abelian anyons cannot be used directly as a qubit, because the two states belong to different topological charge sectors $\ket{ab;i_1}$ and $\ket{ab;i_2}$, and cannot be superposed by braiding.
Let three anyons $a$, $b$ and $c$ be fused to $d$. The first two are fused to $i$, then their outcome is fused with the third gives $d$.
We can write the two different fusion channels as the two states of our qubit as
\eq{\ket{i} = \ket{a,b \rightarrow i} \ket{i,c \rightarrow d},}
where the tensor product symbol is omitted. Alternatively, when the last two anyons $b$ and $c$ are fused to $j$, the $j$ is fused with $c$ to make $d$. $i$ and $j$ are two sets of bases. This is shown in the Fig. \ref{FRMatrices}. A qubit can also be formed as
\eq{\ket{j} = \ket{b,c \rightarrow j}\ket{j,a \rightarrow d}.}
The change of basis is performed by using the $F$-matrices as
\eq{\ket{i} = \sum_j (F_{abc}^d)_j^i \ket{j}.}
The $(F^d_{abc})_j^i$ are the matrix elements of $F_{abc}^d$ summed over $j$. $F$ and $R$ matrices are obtained for a particular anyon model from the solution of the pentagon and hexagon equations \cite{pachos2012introduction}. Ising and Fibonacci anyons are the most popular systems to make the topological quantum computing logic gates. These anyons are found as quasiparticles in non-Abelian fractional quantum Hall effect and topological superconductors.

A possible error in topological quantum computation is due to the braiding or fusion with some unattended anyon in the system. This error can be avoided by carefully accounting for the charges participating in the encoding. Another type of error could be due to the energy of the system, such that the gap between the ground state and the excited state gets filled. This error can be minimized by keeping the system at a very low temperature. The measurement of outcome is either interference or the projective measurement as discussed in Ref. \cite{nayak2008non,bonderson2007non}.

\subsubsection{Example 1: Fibonacci Anyon}

The simplest non-Abelian anyon model consists of only two particles $\textbf{1}$ and $\tau$ \cite{trebst2008short,wang2010topological,delaney2016local}. The fusion rules for this anyon model are
\eq{\tau \cross \tau = \textbf{1} + \tau}
The basis states can be written as 
\eq{\ket{0} = \ket{\tau,\tau \rightarrow \textbf{1}}, \qquad \ket{1} = \ket{\tau,\tau \rightarrow \tau}}
The dimension is the different number of ways the fusion of all anyons can result in topological charge $\textbf{1}$ or $\tau$. In the fusion outcome of two $\tau$ we get $\textbf{1}$ with probability $p_0 = 1/\phi^2$ and $\tau$ with probability $p_1 = \phi/\phi^2 = 1/\phi$ \cite{delaney2016local}.
The dimension grows as the Fibonacci series, in which the next number is a sum of the last two numbers. The quantum dimension $d_\tau = \phi = (1+ \sqrt{5})/2$ is the golden mean.
\eq{\tau \cross \tau \cross \tau &= \textbf{1} + 2\tau \nonumber \\
	\tau \cross \tau\cross \tau\cross \tau &= 2 \cdot 1 + 3 \cdot \tau \nonumber \\
	\tau \cross \tau\cross \tau\cross \tau\cross \tau &= 3 \cdot 1 + 5 \cdot \tau \nonumber}
As we discussed above, no amount of braiding can change one qubit state to the other. Therefore, we need more than two $\tau$ particles for the qubit. Also, topological quantum computation has no tensor product structure. That means, if three $\tau$ anyons are used to make a qubit then the six anyons have only the five dimensional fusion space. Therefore, only a subspace is used to encode a qubit. Three Fibonacci anyons are required for the qubit and the fusion of four particles results in the vacuum \cite{lahtinen2017short}. $F$ and $R$ matrices for this model are obtained by consistency conditions, as in Ref. \cite{pachos2012introduction} are given as  
\eq{[F_{\tau\tau 1}^\tau]^\tau_\tau = \begin{pmatrix}
		\frac{1}{\phi} & \frac{1}{\sqrt{\phi}} \\
		\frac{1}{\sqrt{\phi}} & -\frac{1}{\phi}
	\end{pmatrix} \label{FFib}}
\eq{R_{\tau\tau} = \begin{pmatrix}
		R^1_{\tau\tau} & 0 \\
		0 & R^\tau_{\tau \tau}
	\end{pmatrix} = \begin{pmatrix}
		e^{4\pi i/5} & 0 \\
		0 & -e^{3\pi i/5}
	\end{pmatrix} \label{RFib}}
Quantum computing with Fibonacci anyons is done as follows.
The fusion of two $\tau$ particles gives either $\bm{1}$ or $\tau$. These two orthogonal states are represented as $\ket{(\bullet, \bullet)_\mathbf{1}}$ and $\ket{(\bullet, \bullet)_\tau}$. The addition of the third $\tau$ particle to the state $\ket{(\bullet, \bullet)_\mathbf{1}}$ gives $\tau$ and is denoted as $\ket{((\bullet, \bullet)_\mathbf{1}, \bullet)_\tau} \equiv \ket{0}$. But when the third particle is added to the state $\ket{(\bullet, \bullet)_\mathbf{1}}$, we get either $\bm{1}$ or $\tau$. These states are represented as $\ket{((\bullet,\bullet)_\tau,\bullet)_\tau}\equiv\ket{1}$ and
$\ket{((\bullet,\bullet)_\tau,\bullet)_{\bm{1}}}\equiv \ket{N}$, here $\ket{N}$ stands for the non-computational state. The amplitude in this state is considered as the \textit{leakage error} \cite{nayak2008non,hormozi2007topological}. The states $\ket{((\bullet, \bullet)_\mathbf{1}, \bullet)_\tau} \equiv \ket{0}$ and $\ket{((\bullet,\bullet)_\tau,\bullet)_\tau}\equiv\ket{1}$ are the basis states for a qubit and which are interchanged by an $F$ matrix \eqref{FFib}. The braiding of these particles is represented by an $R$ matrix \eqref{RFib}. These basis states and the fusion of Fibonacci anyons is shown in Fig.\ref{FusionSpaceFib}. The set of gates required to build any kind of circuit is called the universal quantum gate set. The Fibonacci anyonic model is the universal for quantum computing. These kinds of anyons are proposed to be found in the Read-Rezayi state $\nu = 12/5$ which is a very fragile state, so other anyon models are also under consideration \cite{nayak2008non}.

\begin{figure}[h!]
	\centering
	\begin{subfigure}{0.3\textwidth}
		\centering
		\begin{tikzpicture}
			\draw[ultra thick,blue] (0,0)--(3,0);
			\draw[ultra thick,blue] (1,0)--(1,1);
			\draw[ultra thick,blue] (2,0)--(2,1);
			\node [left] at (0,0) {$ \tau $};
			\node [above] at (1,1) {$ \tau $};
			\node [below] at (1.5,0) {$ i $};
			\node [above] at (2,1) {$ \tau $};
			\node [right] at (3,0) {$ \tau $};
		\end{tikzpicture}
		\caption{}
	\end{subfigure}
	\begin{subfigure}{0.4\textwidth}
		\centering
		\begin{tikzpicture}
			\draw[ultra thick,blue] (0,0)--(6,0);
			\draw[ultra thick,blue] (1,0)--(1,1);
			\draw[ultra thick,blue] (2,0)--(2,1);
			\draw[ultra thick,blue] (4,0)--(4,1);
			\draw[ultra thick,blue] (5,0)--(5,1);
			\node [left] at (0,0) {$ \tau $};
			\node [above] at (1,1) {$\tau $};
			\node [below] at (1.5,0) {$ e_1 $};
			\node [above] at (2,1) {$ \tau $};
			\node [below] at (2.5,0) {$ e_2 $};
			\node [above] at (4,1) {$ \tau $};
			\node [below] at (4.5,0) {$ e_{n-3} $};
			\node [above] at (5,1) {$ \tau $};
			\node [right] at (6,0) {$ \tau $};
		\end{tikzpicture}
		\caption{}
	\end{subfigure}	
	\caption[Fusion space of Fibonacci anyons.]{(a) Two orthogonal qubit states, $i$ would be either $1$ or $\tau$. (b) fusion space for $n$ Fibonacci anyons.}
	\label{FusionSpaceFib}
\end{figure}
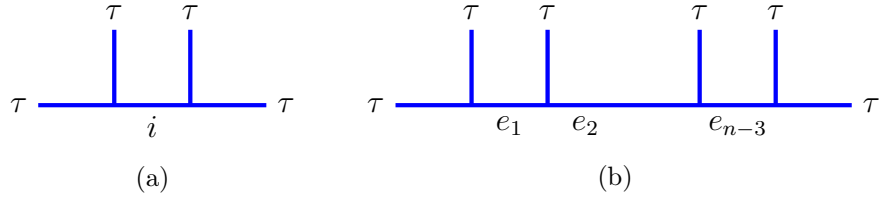

\begin{figure}[h!]
	\centering
	\begin{tikzpicture}
		\node at (-1.5,0) {$\ket{0} = \ket{((\bullet,\bullet)_1,\bullet)_\tau} =$};
		\draw [ultra thick,blue] (1.9,0) ellipse (1.5 and 0.5);
		\draw [ultra thick,blue] (1.55,0) ellipse (1 and 0.3);
		\draw [ultra thick,blue] (1.55,0) ellipse (1 and 0.3);
		\node[red] at (1,0) {$\bullet$};
		\node[red] at (2,0) {$\bullet$};
		\node[red] at (3,0) {$\bullet$};
		\node[below] at (2.65,0) {$\bm{1}$};
		\node[below] at (3.55,0) {$\tau$};
		\node at (4,0) {$=$};
		\draw [blue, ultra thick] (5,-1) -- (5,-0.5); 
		\draw [blue, ultra thick] (5,-0.5) -- (6,0.5);
		\draw [blue, ultra thick] (5,-0.5) -- (4,0.5);
		\draw [blue, ultra thick] (4.5,0) --(5,0.5);
		\node [above] at (6,0.5) {$\tau$};
		\node [above] at (4,0.5) {$\tau$};
		\node [above] at (5,0.5) {$\tau$};
		\node [below] at (5,-1) {$\tau$};
		\node [below] at (4.5,0) {$\bm{1}$};
	\end{tikzpicture}\\
	\begin{tikzpicture}
		\node at (-1.5,0) {$\ket{1} = \ket{((\bullet,\bullet)_\tau,\bullet)_\tau} =$};
		\draw [ultra thick,blue] (1.9,0) ellipse (1.5 and 0.5);
		\draw [ultra thick,blue] (1.55,0) ellipse (1 and 0.3);
		\draw [ultra thick,blue] (1.55,0) ellipse (1 and 0.3);
		\node[red] at (1,0) {$\bullet$};
		\node[red] at (2,0) {$\bullet$};
		\node[red] at (3,0) {$\bullet$};
		\node[below] at (2.65,0) {$\tau$};
		\node[below] at (3.55,0) {$\tau$};
		\node at (4,0) {$=$};
		\draw [blue, ultra thick] (5,-1) -- (5,-0.5); 
		\draw [blue, ultra thick] (5,-0.5) -- (6,0.5);
		\draw [blue, ultra thick] (5,-0.5) -- (4,0.5);
		\draw [blue, ultra thick] (4.5,0) --(5,0.5);
		\node [above] at (6,0.5) {$\tau$};
		\node [above] at (4,0.5) {$\tau$};
		\node [above] at (5,0.5) {$\tau$};
		\node [below] at (5,-1) {$\tau$};
		\node [below] at (4.4,0) {$\tau$};
	\end{tikzpicture}\\
	\begin{tikzpicture}
		\node at (-1.5,0) {$\ket{N} = \ket{((\bullet,\bullet)_\tau,\bullet)_{\bm{1}}}=$};
		\draw [ultra thick,blue] (1.9,0) ellipse (1.5 and 0.5);
		\draw [ultra thick,blue] (1.55,0) ellipse (1 and 0.3);
		\node[red] at (1,0) {$\bullet$};
		\node[red] at (2,0) {$\bullet$};
		\node[red] at (3,0) {$\bullet$};
		\node[below] at (2.65,0) {$\tau$};
		\node[below] at (3.55,0) {$\bm{1}$};
		\node at (4,0) {$=$};
		\draw [blue, ultra thick] (5,-1) -- (5,-0.5); 
		\draw [blue, ultra thick] (5,-0.5) -- (6,0.5);
		\draw [blue, ultra thick] (5,-0.5) -- (4,0.5);
		\draw [blue, ultra thick] (4.5,0) --(5,0.5);
		\node [above] at (6,0.5) {$\tau$};
		\node [above] at (4,0.5) {$\tau$};
		\node [above] at (5,0.5) {$\tau$};
		\node [below] at (5,-1) {$\bm{1}$};
		\node [below] at (4.4,0) {$\tau$};
	\end{tikzpicture}
	\caption[The orthogonal states of three Fibonacci particles.]{Orthogonal states of three Fibonacci particles \cite{nayak2008non}}
	\label{FibQubit}
\end{figure}
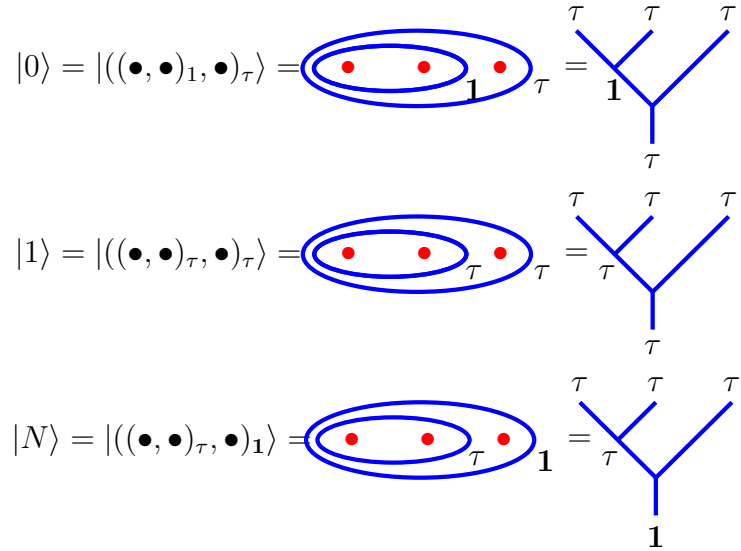

\subsubsection{Example 2: Ising Anyon}
This model has three anyons $\textbf{1}$, $\sigma$ and $\psi$.
The fusion rules for these anyons are;
\eq{\sigma \cross \sigma = \textbf{1} + \psi, \qquad \psi \cross \psi = \textbf{1}, \qquad \psi \cross \sigma = \sigma}
Two basis states can be written as
\eq{\ket{0}=\ket{\sigma, \sigma \rightarrow \textbf{1}}, \qquad \ket{1}=\ket{\sigma, \sigma \rightarrow \psi}}
Since two fusions belong to different topological charge sectors, at least three anyons are needed that can fuse to $\sigma$ in two different ways. For every added $\sigma$ the dimension of fusion space doubles, hence for $2N$ anyons the dimension is $2^{N-1}$. This model is non-universal, so non-topological schemes are also devised in addition to topological computation. The $F$ and $R$ matrices for this model \cite{lahtinen2017short} are,
\eq{F_{\sigma \sigma \sigma}^\sigma = \frac{1}{\sqrt{2}}\begin{pmatrix}
		1 & 1 \\
		1 & -1
	\end{pmatrix},\qquad
	R_{\sigma \sigma} = e^{-i \pi/8} \begin{pmatrix}
		1 & 0 \\
		0 & i
	\end{pmatrix},}
where $R_{\sigma \sigma}^1 = e^{-i\pi/8}$ and $R_{\sigma \sigma}^\psi = e^{i3\pi/8}$. The topological charges are labeled as $0,1/2, 1$ corresponds to $\textbf{1}, \sigma, \psi$. The total charge of two particles with charges $1/2$ is either $1$ or $0$. The total charge is $0$ when both particles have charges $1$, but the total charge is $1/2$ when one particle has charge $1$ and other has $1/2$. The quantum dimensions, $d_1=d_\psi =1$ and $d_\sigma=\sqrt{2}$, for these anyons are computed in Ref. \cite{pachos2012introduction} using the fusion rules.   

Let us consider four particles of charge $1/2$ with a total charge of $0$. The first two are fused either to $0$ or $1$. In case it is zero, then the total of the third and fourth must be zero. If the total of first and second is 1 then the total of third and fourth must be equal to 1. In this way, we have two states of four $1/2$ quasiparticles. There are $2^{n-1}$ states for $2n$ particles \cite{nayak19962n,nayak2008non}. 
When particles of the same pair, say $i$, are braided, only the phase is changed, but when a particle of pair $i$ is braided with the particle of other pair $j$, a NOT gate is applied \cite{nayak2008non}. Taking both particles of $i$ around both particles in $j$ then the basis state is multiplied by $+1$ if $j$ has a charge $0$, but it gets multiplied by $-1$ if it has a charge $1$. Six $\sigma$ anyons are required for two-qubit encoding. See Ref. \cite{nayak2008non,lahtinen2017short} for the implementation of CNOT and phase gates. The Ising anyon model is implemented by using the quantum Hall state $\nu = 5/2$ and Majorana particle in topological superconductors.

\section{Q-Analog of Recoupling Theory}\label{QDeform}
The fusion of quasiparticles is similar to the recoupling theory of addition of angular momentum in quantum mechanics \cite{rose1995elementary,varshalovich1988quantum,biedenharn1981angular}. Ternary logic gates are designed using the q-deformed version of the recoupling theory. Therefore, for the intuition, we will discuss the quantum deformation and the recoupling of angular momenta and then discuss the quantum deformation of recoupling theory \cite{biedenharn1995quantum,kirillov1988representations,kirillov1991clebsch}. The quantum deformed quantities are also called q-analogs. 

\subsection{Quantum Deformation}
In classical mechanics, states on phase space make a manifold represented by say $(q,p)$. Physical quantities are observables that are the functions of $(q,p)$. The Abelian algebra formed for these observables and associated geometry is commutative.
In quantum mechanics, due to the Heisenberg uncertainty principle, there is no arbitrary precision of the quantities. Algebra is non-commutative and classical mechanics is the limiting case when Planck constant $h \rightarrow 0$. Therefore, quantum mechanics is a kind of deformation of classical mechanics. 
In quantum mechanics, the commuting classical observables are replaced with the noncommuting Hermitian operators. Hence we can say that we deform classical algebra and the deformation parameter is $h$. 
The noncommutativity of the variables $X$ and $Y$ in the deformed space is written as 
\eq{XY = qYX,}
where $q$ is a complex number in general. It is called the \textit{deformation parameter}. Let $q$ be a number different from $1$, and $h$ be a number different from $0$. If we take $x= qx_0$ or $x=x_0+h$ we can have the classical values when $h\to 0$ or $q\to 1$. These two are related as $q= e^h$ \cite{jaganathan2000introduction}. For $q \to 1$, we would get back the classical commuting variables. 
Let us choose $q = e^{i\theta}$. Consider an example when the operators $T_\alpha$ and $G_{\theta/\alpha}$ are acting on a function $\psi(x)$ of real variable $x$, such that 
\eq{T_\alpha \psi(x) = \psi(x+\alpha), \qquad G_{\theta/\alpha} \psi(x) = e^{i \theta x / \alpha} \psi(x).}
When we apply both $T_\alpha$ and $G_{\theta/\alpha}$ operators, we get
\eq{T_\alpha G_{\theta/\alpha} \psi(x) = e^{i\theta(x+\alpha)} \psi(x+\alpha) = e^{i\theta} G_{\theta/\alpha} T_\alpha\psi(x).}
With the fixed value of variables $\theta$ and $\alpha$, $T_\alpha$ and $G_{\theta/\alpha}$ become noncommuting variables that can be written as
\eq{T_\alpha G_{\theta/\alpha} = e^{i\theta}G_{\theta/\alpha} T_\alpha.} 

\subsubsection{q-Analogs}

An anyon or a pair of anyons interacts through braiding in the plane deformed by the existence of the fields of the other anyons in an abstract way. This braiding interaction may cause the twist factor that is related to the topological spin of an anyon and involves the parameter $q$. Algebraically, we can think of $q$ as a small perturbation of the usual mathematical objects. The $q$ can be generic or a \textit{root of unity}. The root of unity is defined as when a complex number $q$ is raised to some power $n$ so that it equals 1 for that power, then we say that this complex number is $n$th root of unity. 
Now the $q$-analog quantities give the corresponding classical quantities for the limiting case when $q \rightarrow 1$.
For $n\in \mathbb{Z}$ we define what is called a \textbf{$q$-integer}
\eq{[n]_q= \frac{q^n - q^{-n}}{q-q^{-1}},}
which is the Laurent polynomial equal to $n \in \mathbb{Z}$ for $q \rightarrow 1$. 
It is done by taking the limit $q \to 1$ and applying L'Hospital's rule.
The $q$-analog of natural numbers is as follows, $[0]_q =0,\ [1]_q = 1,\ [2]_q = \frac{q^2-q^{-2}}{q-q^{-1}}$, and so on.
We can get \textbf{$q$-factorial} $[n]!$ that can be written as
\eq{[n]! &= [n][n-1]...[1] \\
	&=\frac{q^n - q^{-n}}{q-q^{-1}}\cdot \frac{q^{n-1} - q^{-(n-1)}}{q-q^{-1}}...\frac{q^2 - q^{-2}}{q-q^{-1}}\cdot 1}
Sometimes, we take $[n]_q = \frac{1-q^n}{1-q}$, as in \cite{jaganathan2000introduction,le2015quantum}.
In that case the $q$-analog of natural numbers and $q$-factorials are written as $[0]_q =0,\ [1]_q = 1,\ [2]_q = 1+q,\ [3]_q = 1+q+q^2$,
\eq{[n]!=1\cdot (1+q)(1+q+q^2)...(1+q+...+q^{n-1}).}

\subsection{$SU(2)_k$ Anyon Model}
The F-symbols in topological quantum computation can be computed using the $SU(2)_k$ model \cite{bonderson2007non} and Temperley-Lieb recoupling theory \cite{kauffman1994temperley} also called the $JK_k$ anyon model \cite{levaillant2015universal}. Where the $k$ is called the level of the theory. It is the coupling constant of Chern-Simons theory and is related to the number of particles present as we discussed in Section \ref{TQC}. These theories are the quantum analog of theory of addition of angular momentum. The $SU(2)_k$ is the q-deformed version of $SU(2)$ with $q=\exp(i2\pi/k+2)$, for $q$ at the root of unity. The detailed derivation of the parameter $q$ in this form is given in \cite{witten1989quantum}. The anyon's fusion amplitudes would be written as the recoupling coefficients. The $F$-symbols and $R$-symbols are obtained by using these two models at level $k=4$ gave identical values. For $F$ and $R$ symbols in the $JK_4$ model, see \cite{levaillant2015universal}. In our work, we will use the $SU(2)_4$ model. The topological data for this model are given as
\eq{[n]_q = \frac{q^{n/2}-q^{-n/2}}{q^{1/2} - q^{-1/2}}, \qquad {\mathcal{C}} = \left\{0, 1/2,..., k/2 \right\}, \qquad j_1 \times j_2 = \sum_{j = |{j_1-j_2}|}^{\min{j_1+j_2, k-j_1-j_2}},}

\eq{[F_j^{j_1,j_2,j_3}]_{j_{12},j_{23}} = (-1)^{j_1 + j_2 + j_3 +j} \sqrt{[2j_{12}+1]_q [2j_{23}+1]_q}
	\begin{Bmatrix} 
		j_1 & j_2 & j_{12}\\ 
		j_3 & j & j_{23}\\ 
	\end{Bmatrix}_q,}
where
\eq{\begin{Bmatrix} 
		j_1 & j_2 & j_{12}\\ 
		j_3 & j & j_{23}\\ 
	\end{Bmatrix}=\Delta(j_1,j_2,j_3)\Delta(j_{12},j_3,j)\Delta(j_2,j_3,j_{23})\Delta(j_1,j_{23},j) \nonumber\\ \times \sum_z \frac{(-1)^z[z+1]_q !}{[z-j_1-j_2 -j_{12}]_q![z-j_{12}-j_3 -j]_q![z-j_2-j_3 -j_{23}]_q![z-j_1-j_{23} -j]_q!} \nonumber\\ \times \frac{1}{[j_1 +j_2 +j_3 +j -z]_q![j_1 +j_{12} +j_3 +j_{23} -z]_q![j_2 +j_{12} +j +j_{23} -z]_q!},}

$$ \Delta(j_1,j_2,j_3) = \sqrt{\frac{[-j_1 +j_2 +j_3]_q![j_1-j_2+j_3]_q! [j_1 +j_2 -j_3]_q!}{[j_1 + j_2 +j_3 +1]_q!}}, \qquad [n]_q! \equiv \prod_{m=1}^n [m]_q,$$

$$ R_j^{j_1,j_2} = (-1)^{j-j_1-j_2} q^{\frac{1}{2}[j(j+1) - j_1(j_1+1) -j_2(j_2+1)]},$$

$$ d_j = [2j+1]_q = \frac{\sin \big[\frac{(2j+1)\pi}{k+2}\big]}{\sin \big(\frac{\pi}{k+2}\big)}, \quad {\mathcal{D}} = \frac{\sqrt{\frac{k+2}{2}}}{\sin \big(\frac{\pi}{k+2}\big)},$$


$$ \theta_j = q^{j(j+1)} = e^{i 2\pi \frac{j(j+1)}{k+2}}, \qquad S_{j_1j_2} = \sqrt{\frac{2}{k+2}} \sin \big[ \frac{(2j_1 +1)(2j_2 +1) \pi}{k+2} \big].$$

The data in $SU(2)_k$ theory are used to compute the $[F_j^{j_1,j_2,j_3}]_{j_{12},j_{23}}$ and $R^{j_1j_2}_j$ matrices \cite{cui2015universal}. The $F$ and $R$ symbols are calculated using these fusion rules and from the $F$ and $R$ matrices, the $\sigma$ matrices are obtained in Section \ref{Meta} \cite{cui2015universal,levaillant2015universal}.

\section{Ternary Logic Design with Metaplectic Anyons}\label{Meta}
Quantum computation is performed with metaplectic anyons which are simple objects in weakly integral categories. The term metaplectic is for a braid group that is in the metaplectic representation. These representations are the symplectic analog of spinor representation. \cite{cui2015universal,bocharov2016efficient}.

The metaplectic anyons can be studied from the category theory. Anyons are simple objects in a unitary modular category. See Ref. \cite{ilyas2021topological,bakalov2001lectures} for the introduction to category theory and \cite{rowell2018mathematics} for the use of category theory in topological quantum computation. A category is integral when the quantum dimension or Frobenius-Perron dimension of a simple object is an integer, whereas a category is called weakly integral if the squares of the quantum dimensions of all the simple objects are integer. Weakly integral categories are a class of metaplectic categories \cite{bruillard2016classification,hastings2014metaplectic}.
There are five anyons $\left\{1,Z,X,X^{'},Y\right\}$ in the theory of metaplectic anyons with fusion rules, quantum dimensions, and topological twists given as
\eq{X \otimes X &= 1 +Y, \nonumber\\
	Y \otimes Y &= 1 + Z + Y,\nonumber\\
	X \otimes Z &= X',\nonumber\\
	X \otimes X' &= Z+Y,} 
\eq{d_1 = d_Z = 1, \ d_X = d_{X'}= \sqrt{3}, \ d_{Y} = 2,}
\eq{\theta_0 = \theta_4 = 1, \ \theta_1 = \theta_3 = e^{i\pi/4}, \ \theta_2 = e^{i2\pi/3}.}

There is a non-Abelian boson quasiparticle $Z$. These theories also have a fundamental particle $X$. This particle is also a non-Abelian. It is a vortex for the $Z$ boson. This $X$ particle is fused with another $X$ particle to give $Y_i$ or vacuum, where $i = 1,2,...,r$ and $r= (m-1)/2$. These non-Abelian particles $Y_i$ have the quantum dimension $2$. When $X$ and $Z$ are fused, the result is the particle $X'$ \cite{hastings2013metaplectic,hastings2014metaplectic}.
A collection of $N$ quasiparticles $X$ at a fixed position has an $n_N$-dimensional degenerate subspace with $n_N \sim m^{N/2}$.
The proposed metaplectic anyon systems are the quantum Hall effect and Majorana zero modes \cite{barkeshli2012topological,clarke2013exotic,lindner2012fractionalizing,cheng2012superconducting,vaezi2013fractional}. 

\subsection{One-Qutrit Gates}
Let us consider four $X$ anyons. The first two of the four are fused to $c_{12}$ and the last two are fused to $c_{34}$ as shown in Fig. \ref{OneTwoQutrit} (a).
With the constraint $c_{14} = Y$, we get three fusion trees \cite{cui2015universal,bocharov2016efficient}
\eq{(c_{12},c_{34}) \in \left\{-(YY),(\textbf{1}Y),(Y\textbf{1})\right\}.}
These are corresponding to the three states of qutrit $\ket{0},\ket{1},\ket{2}$. The minus sign is just to make the algebra nicer later. 
Let $\sigma_1$ be a braid matrix for the first two particles, and $\sigma_2$ corresponds to a braid of the second with the third, and $\sigma_3$ is a braid matrix for the third and fourth as shown in Fig. \ref{Sigma13} and \ref{Sigma2}.
The associated Hilbert space is represented by $V_y^{\epsilon \epsilon \epsilon \epsilon}$, for $\epsilon=X$. Under the basis $\left\{-\ket{YY},\ket{\bm{1}Y}, \ket{Y\bm{1}}\right\}$, the generators of the braid group ${\mathcal{B}}_4$ for the representation $V_y^{\epsilon \epsilon \epsilon \epsilon}$ are
\eq{&\sigma_1 = \gamma \begin{pmatrix}
		1 & 0 & 0\\
		0 & \omega & 0\\
		0 & 0 & 1
	\end{pmatrix}, \qquad \sigma_3 = \gamma \begin{pmatrix}
		1 & 0 & 0\\
		0 & 1 & 0\\
		0 & 0 & \omega
	\end{pmatrix},}
\eq{\sigma_2 &= \frac{\gamma^3}{\sqrt{3}} \begin{pmatrix}
		1 & \omega & \omega\\
		\omega & 1 & \omega\\
		\omega & \omega & 1
	\end{pmatrix} = \gamma \begin{pmatrix}
		\frac{1}{2} + \frac{\sqrt{3}i}{6} & -\frac{1}{2} + \frac{\sqrt{3}i}{6} & -\frac{1}{2} + \frac{\sqrt{3}i}{6} \\
		-\frac{1}{2} + \frac{\sqrt{3}i}{6} & \frac{1}{2} + \frac{\sqrt{3}i}{6} & -\frac{1}{2} + \frac{\sqrt{3}i}{6} \\
		-\frac{1}{2} + \frac{\sqrt{3}i}{6} & -\frac{1}{2} + \frac{\sqrt{3}i}{6} & \frac{1}{2} + \frac{\sqrt{3}i}{6}
	\end{pmatrix},}
where $\omega = e^{2 \pi i/3}$ and $\gamma = e^{\pi i/12}$.
Ignoring the $\gamma$ in front, let us define \cite{cui2015universal}
$p = \sigma_1 \sigma_2 \sigma_1$, $q = \sigma_2 \sigma_3 \sigma_2$, 
\eq{p^2 = -\begin{pmatrix}
		0 & 1 & 0\\
		1 & 0 & 0\\
		0 & 0 & 1
	\end{pmatrix}, \qquad
	q^2 = -\begin{pmatrix}
		0 & 0 & 1\\
		0 & 1 & 0\\
		1 & 0 & 0
	\end{pmatrix}, \qquad
	-(q^2pq^2)^2 = \begin{pmatrix}
		1 & 0 & 0\\
		0 & 0 & 1\\
		0 & 1 & 0
	\end{pmatrix}, \nonumber \\
	(q^2pq^2)^2Z^*((q^2pq^2)^2)^* = \begin{pmatrix}
		0 & 0 & 1\\
		1 & 0 & 0\\
		0 & 1 & 0
	\end{pmatrix}, \qquad 
	(q^2pq^2)^2Z((q^2pq^2)^2)^* = \begin{pmatrix}
		0 & 1 & 0\\
		0 & 0 & 1\\
		1 & 0 & 0
	\end{pmatrix}.\label{ZGates}}
These gates correspond to one-qutrit gates $Z_3(+1)$, $Z_3(+2)$, $Z_3(01)$, $Z_3(12)$, $Z_3(02)$ in conventional quantum computing discussed in \ref{QC}. The phase gate $Z=\sigma_1\sigma_3^{-1}=\sigma_1\sigma_3^2$ and the ternary Hadamard gate $H=q^2pq^2$ can be written in matrix form as
\eq{Z = \begin{pmatrix}
		1 & 0 & 0\\
		0 & \omega & 0\\
		0 & 0 & \omega^2
	\end{pmatrix}, \
	H = \frac{1}{\sqrt{3}i}\begin{pmatrix}
		1 & 1 & 1\\
		1 & \omega & \omega^2\\
		1 & \omega^2 & \omega
	\end{pmatrix}.}

\subsection{Two-Qutrit Gates}
The two-qutrit model would consist of eight $X$ anyons with the final fusion outcome $Y$ as shown in Fig. \ref{OneTwoQutrit} (b). 
The braid matrices for two qutrits are written as $\sigma_1,\sigma_2,\sigma_3,\sigma_4,\sigma_5,\sigma_6,\sigma_7$.
Let us define
\eq{s_1 = \sigma_2\sigma_1\sigma_3\sigma_2, \ s_2 = \sigma_4\sigma_3\sigma_5\sigma_4, \ s_3 = \sigma_6\sigma_5\sigma_7\sigma_6.}
From these matrices, we can calculate a matrix
\eq{\Lambda (Z) = s_1^{-1}s_2^2 s_1 s_3^{-1}s_2^2s_3.}
The two-qutrit encoding is obtained when restricting the vector space $V_y^{\epsilon \epsilon \epsilon \epsilon \epsilon \epsilon \epsilon \epsilon}$ to nine dimensional subspace $V_y^{\epsilon \epsilon \epsilon \epsilon}\otimes V_y^{\epsilon \epsilon \epsilon \epsilon} \subset V_y^{\epsilon \epsilon \epsilon \epsilon \epsilon \epsilon \epsilon \epsilon}$
with $c_{14}=c_{58}= Y$. This nine-dimensional restriction of the $\Lambda (Z)$ is the Controlled-Z gate.
The SUM gate is a generalization of the CNOT gate \cite{cui2015universal,bocharov2016efficient}. It is related to CZ as
\eq{SUM = (I\otimes H)\Lambda (Z)(I \otimes H^{-1}).\label{SumGate}}
This $SUM$ gate will be combined with the topological charge measurement to build arithmetic circuits in the next section.
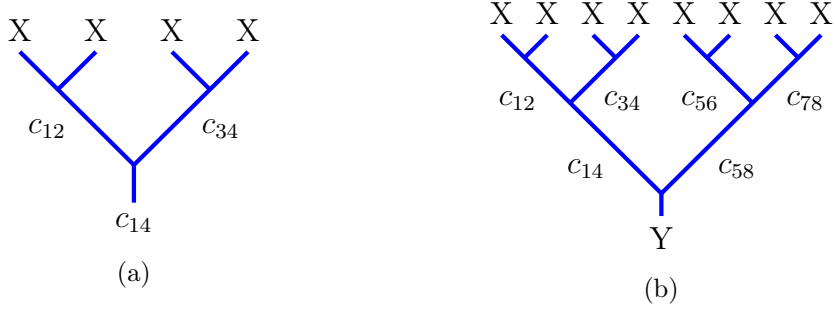
\begin{figure}[t!]
	\centering
	\begin{subfigure}{0.3\textwidth}
		\centering
		\begin{tikzpicture}[scale=0.5]
			\draw [blue, ultra thick] (3,0) -- (3,1); 
			\draw [blue, ultra thick] (3,1) -- (0,4);
			\draw [blue, ultra thick] (3,1) -- (6,4);
			\draw [blue, ultra thick] (1,3) -- (2,4);
			\draw [blue, ultra thick] (5,3) --(4,4);
			\node [below left] at (1.5,2.5) {$c_{12}$};
			\node [below right] at (4.5,2.5) {$c_{34}$};
			\node [below] at (3,0) {$c_{14}$};
			\node [above] at (0,4) {X};
			\node [above] at (2,4) {X};
			\node [above] at (4,4) {X};
			\node [above] at (6,4) {X};
		\end{tikzpicture}
		\caption{}
	\end{subfigure}
	\begin{subfigure}{0.5\textwidth}
		\centering
		\begin{tikzpicture}[scale = 0.3]
			\draw [blue, ultra thick] (7,0) -- (7,1);
			\draw [blue, ultra thick] (7,1) -- (0,8);
			\draw [blue, ultra thick] (7,1) -- (14,8);%
			\draw [blue, ultra thick] (3,5) --(6,8);
			\draw [blue, ultra thick] (11,5) --(8,8);%
			\draw [blue, ultra thick] (1,7) --(2,8);
			\draw [blue, ultra thick] (5,7) --(4,8);
			\draw [blue, ultra thick] (9,7) --(10,8);
			\draw [blue, ultra thick] (13,7) --(12,8);%
			\node [above] at (0,8) {X};
			\node [above] at (2,8) {X};
			\node [above] at (4,8) {X};
			\node [above] at (6,8) {X};%
			\node [above] at (8,8) {X};
			\node [above] at (10,8) {X};
			\node [above] at (12,8) {X};
			\node [above] at (14,8) {X};%
			\node [below left] at (2,6) {$c_{12}$};
			\node [below right] at (4,6) {$c_{34}$};%
			\node [below left] at (10,6) {$c_{56}$};
			\node [below right] at (12,6) {$c_{78}$};%
			\node [below left] at (5,3) {$c_{14}$};
			\node [below right] at (9,3) {$c_{58}$};%
			\node [below] at (7,0) {Y};
		\end{tikzpicture}
		\caption{}
	\end{subfigure}
	\caption{(a) One-qutrit fusion tree (b) Two-qutrit fusion tree \cite{bocharov2016efficient}.}
	\label{OneTwoQutrit}
\end{figure}
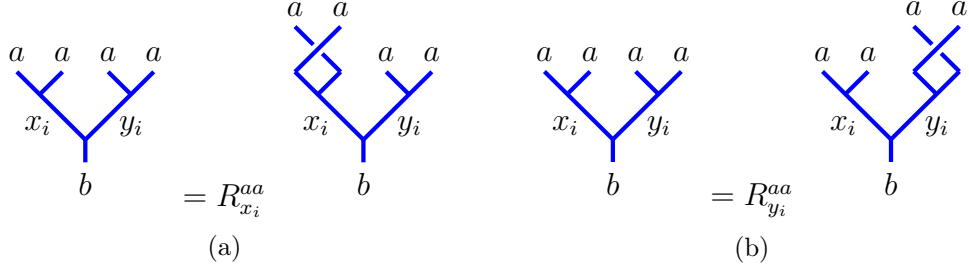
\begin{figure}[t!]
	\centering
	\begin{subfigure}{0.4\textwidth}
		\centering
		\begin{tikzpicture}[scale = 0.3]
			\draw [blue, ultra thick] (3,0) -- (3,1); 
			\draw [blue, ultra thick] (3,1) -- (6,4);
			\draw [blue, ultra thick] (3,1) -- (0,4);
			\draw [blue, ultra thick] (1,3) --(2,4);
			\draw [blue, ultra thick] (5,3) --(4,4);
			\node [below] at (3,0) {$b$};
			\node [above] at (0,4) {$a$};
			\node [above] at (2,4) {$a$};
			\node [above] at (4,4) {$a$};
			\node [above] at (6,4) {$a$};
			\node [below left] at (2,2.5) {$x_i$};
			\node [below right] at (4,2.5) {$y_i$};
		\end{tikzpicture}
		$=R^{aa}_{x_i}$
		\begin{tikzpicture}[scale = 0.3]
			\draw [knot=blue, ultra thick] (2,4) --(0,6);
			\draw [knot=blue, ultra thick] (0,4) -- (2,6);
			\draw [blue, ultra thick] (3,0) -- (3,1); 
			\draw [blue, ultra thick] (3,1) -- (6,4);
			\draw [blue, ultra thick] (3,1) -- (0,4);	
			\draw [blue, ultra thick] (1,3) --(2,4);	
			\draw [blue, ultra thick] (5,3) --(4,4);
			\node [below] at (3,0) {$b$};
			\node [above] at (2,6) {$a$};
			\node [above] at (0,6) {$a$};
			\node [above] at (4,4) {$a$};
			\node [above] at (6,4) {$a$};
			\node [below left] at (2,2.5) {$x_i$};
			\node [below right] at (4,2.5) {$y_i$};
		\end{tikzpicture}
		\caption{}
	\end{subfigure}
	\begin{subfigure}{0.4\textwidth}
		\centering
		\begin{tikzpicture}[scale = 0.3]
			\draw [blue, ultra thick] (3,0) -- (3,1); 
			\draw [blue, ultra thick] (3,1) -- (6,4);
			\draw [blue, ultra thick] (3,1) -- (0,4);
			\draw [blue, ultra thick] (1,3) --(2,4);
			\draw [blue, ultra thick] (5,3) --(4,4);
			\node [below] at (3,0) {$b$};
			\node [above] at (0,4) {$a$};
			\node [above] at (2,4) {$a$};
			\node [above] at (4,4) {$a$};
			\node [above] at (6,4) {$a$};
			\node [below left] at (2,2.5) {$x_i$};
			\node [below right] at (4,2.5) {$y_i$};
		\end{tikzpicture}
		$=R^{aa}_{y_i}$
		\begin{tikzpicture}[scale = 0.3]
			\draw [knot=blue, ultra thick] (6,4) --(4,6);
			\draw [knot=blue, ultra thick] (4,4) -- (6,6);
			\draw [blue, ultra thick] (3,0) -- (3,1); 
			\draw [blue, ultra thick] (3,1) -- (6,4);
			\draw [blue, ultra thick] (3,1) -- (0,4);	
			\draw [blue, ultra thick] (1,3) --(2,4);	
			\draw [blue, ultra thick] (5,3) --(4,4);
			\node [below] at (3,0) {$b$};
			\node [above] at (2,4) {$a$};
			\node [above] at (0,4) {$a$};
			\node [above] at (4,6) {$a$};
			\node [above] at (6,6) {$a$};
			\node [below left] at (2,2.5) {$x_i$};
			\node [below right] at (4,2.5) {$y_i$};
		\end{tikzpicture}
		\caption{}
	\end{subfigure}
	\caption{One-qutrit braid matrices (a) $\sigma_1$ and (b) $\sigma_3$.}
	\label{Sigma13}
\end{figure}
\begin{figure*}[h!]
	\centering
	\begin{tikzpicture}[scale = 0.3]
		\draw [blue, ultra thick] (3,0) -- (3,1); 
		\draw [blue, ultra thick] (3,1) -- (6,4);
		\draw [blue, ultra thick] (3,1) -- (0,4);
		\draw [blue, ultra thick] (1,3) --(2,4);
		\draw [blue, ultra thick] (5,3) --(4,4);
		\node [below] at (3,0) {$b$};
		\node [above] at (0,4) {$a$};
		\node [above] at (2,4) {$a$};
		\node [above] at (4,4) {$a$};
		\node [above] at (6,4) {$a$};
		\node [below left] at (2,2.5) {$x_i$};
		\node [below right] at (4,2.5) {$y_i$};
	\end{tikzpicture}
	$=\sum_c[F^{aay_i}_b]_{x_ic}$	
	\begin{tikzpicture}[scale = 0.3]	
		\draw [blue, ultra thick] (3,0) -- (3,1); 
		\draw [blue, ultra thick] (3,1) -- (6,4);
		\draw [blue, ultra thick] (3,1) -- (0,4);
		\draw [blue, ultra thick] (4,2) --(2,4);
		\draw [blue, ultra thick] (5,3) --(4,4);
		\node [below] at (3,0) {$b$};
		\node [above] at (0,4) {$a$};
		\node [above] at (2,4) {$a$};
		\node [above] at (4,4) {$a$};
		\node [above] at (6,4) {$a$};
		\node [below right] at (3,1.5) {$c$};
		\node [below right] at (4,2.5) {$y_i$};
	\end{tikzpicture}\\
	$=\sum_{cd}[F^{aay_i}_b]_{x_ic}[F^{aaa}_c]^{-1}_{dy_i}$
	\begin{tikzpicture}[scale = 0.3]	
		\draw [blue, ultra thick] (3,0) -- (3,1); 
		\draw [blue, ultra thick] (3,1) -- (6,4);
		\draw [blue, ultra thick] (3,1) -- (0,4);
		\draw [blue, ultra thick] (4,2) --(2,4);
		\draw [blue, ultra thick] (3,3) --(4,4);
		\node [below] at (3,0) {$b$};
		\node [above] at (0,4) {$a$};
		\node [above] at (2,4) {$a$};
		\node [above] at (4,4) {$a$};
		\node [above] at (6,4) {$a$};
		\node [below right] at (3,1.5) {$c$};
		\node [below left] at (3.45,3.3) {$d$};
	\end{tikzpicture}
	$=\sum_{cd}[F^{aay_i}_b]_{x_ic}[F^{aaa}_c]^{-1}_{dy_i}R^{aa}_d$
	\begin{tikzpicture}[scale = 0.3]
		\draw [blue, ultra thick] (3,0) -- (3,1); 
		\draw [blue, ultra thick] (3,1) -- (6,4);
		\draw [blue, ultra thick] (3,1) -- (0,4);
		\draw [blue, ultra thick] (4,2) --(2,4);
		\draw [blue, ultra thick] (3,3) --(4,4);
		\draw [blue, ultra thick] (2,4) --(4,6);
		\draw [blue, ultra thick] (4,4) --(3.2,4.8);
		\draw [blue, ultra thick] (2.8,5.2) --(2,6);
		\node [below] at (3,0) {$b$};
		\node [above] at (0,4) {$a$};
		\node [above] at (2,6) {$a$};
		\node [above] at (4,6) {$a$};
		\node [above] at (6,4) {$a$};
		\node [below right] at (3,1.5) {$c$};
		\node [below left] at (3.45,3.3) {$d$};
	\end{tikzpicture}
	$=\sum_{cde}[F^{aay_i}_b]_{x_ic}[F^{aaa}_c]^{-1}_{dy_i}R^{aa}_d [F^{aaa}_c]_{de}$
	\begin{tikzpicture}[scale = 0.3]
		\draw [blue, ultra thick] (3,0) -- (3,1); 
		\draw [blue, ultra thick] (3,1) -- (6,4);
		\draw [blue, ultra thick] (3,1) -- (0,4);
		\draw [blue, ultra thick] (4,2) --(2,4);
		\draw [blue, ultra thick] (5,3) --(4,4);
		\draw [blue, ultra thick] (2,4) --(4,6);
		\draw [blue, ultra thick] (4,4) --(3.2,4.8);
		\draw [blue, ultra thick] (2.8,5.2) --(2,6);
		\node [below] at (3,0) {$b$};
		\node [above] at (0,4) {$a$};
		\node [above] at (2,6) {$a$};
		\node [above] at (4,6) {$a$};
		\node [above] at (6,4) {$a$};
		\node [below right] at (4.5,3) {$e$};
		\node [below right] at (3,1.5) {$c$};
	\end{tikzpicture}\\
	$=\sum_{cdef}[F^{aay_i}_b]_{x_ic}[F^{aaa}_c]^{-1}_{dy_i}R^{aa}_d [F^{aaa}_c]_{de}
	[F^{aae}_b]^{-1}_{fc}$
	\begin{tikzpicture}[scale = 0.3]
		\draw [blue, ultra thick] (3,0) -- (3,1); 
		\draw [blue, ultra thick] (3,1) -- (6,4);
		\draw [blue, ultra thick] (3,1) -- (0,4);
		\draw [blue, ultra thick] (1,3) --(2,4);
		\draw [blue, ultra thick] (5,3) --(4,4);
		\draw [blue, ultra thick] (2,4) --(4,6);
		\draw [blue, ultra thick] (4,4) --(3.2,4.8);
		\draw [blue, ultra thick] (2.8,5.2) --(2,6);
		\node [below] at (3,0) {$b$};
		\node [above] at (0,4) {$a$};
		\node [above] at (2,6) {$a$};
		\node [above] at (4,6) {$a$};
		\node [above] at (6,4) {$a$};
		\node [below right] at (4,2.5) {$e$};
		\node [below left] at (2,2.5) {$f$};
	\end{tikzpicture}
	\caption{One-qutrit braid matrix $\sigma_2$.}
	\label{Sigma2}
\end{figure*}

\section{Ternary Arithmetic Circuits}\label{Arith}

The most important challenge in circuit design is reducing the number of gates. The more the gates, the harder it is to implement the circuit.
The MS gates in conventional quantum computing are designed by keeping the controlling value 2. Since, in topological quantum computation, any anyon can be braided to another anyon at any stage of the implementation, we can have a controlling value of $0,1,2$. Therefore, we can create a more general methodology of designing topological circuits presented here. We redesigned the qutrit arithmetic circuits that can be implemented with one-qutrit and two-qutrit gates made by metaplectic anyons \cite{bocharov2015improved} described in the last section. The universal set of gates cannot be made by braiding alone, it is to be combined with the topological charge measurement \cite{cui2015universal}.

The gates that can be obtained by braiding alone are the Clifford gates, whereas the non-Clifford gates cannot be implemented by braiding alone. In this work, we have the Clifford gate $SUM$ that can be implemented by braiding alone, whereas $C_c(X)$ is a non-Clifford gate that is implemented by the measurement of the topological charge.
As in \ref{QC}, one-qutrit ternary gates are represented as $Z_3(+1)$, $Z_3(+2)$, $Z_3(01)$, $Z_3(12)$, and $Z_3(02)$, where the first two are \textit{increment gates} and the last three are \textit{permutation gates}.
The non-Clifford gate, that is $C_c(X)$, applies $X$ when the controlling value is $c=0,1,2$, where $X$ can be a permutation or increment gate. The gates $C_c(X)$ are $9\cross 9$ matrices can be written as $diag(I_3,I_3,X)$ for control $\ket{2}$, $diag(I_3,X,I_3)$ for control $\ket{1}$, and $diag(X,I_3,I_3)$ for control $\ket{0}$, where $I_3$ is $3\cross 3$ identity matrix. In Ref. \cite{bocharov2015improved}, the $SUM$ gates are called soft-controlled whereas the $C_c(U)$ are called hard-controlled gates.

The measurement can be projective or based on interference \cite{cui2015universal,nayak2008non,stern2008anyons}.
In case of interferometric measurement, a probe charge is sent through the paths around some region. The interference between different paths is related to the total topological charge in that region. The charge of the region can be found by measuring the charge of the probe. This kind of measurement can distinguish charge of fusion channel from an overall collection. This is non-demolition measurement but the fusion channels evolve in non-universal manner.
Local measurement for topological charge of a single quasiparticle is performed by bringing two charges close to each other and find their charge from their fusion outcome.
Let the measurement $\mathcal{M}_1 = \left\{\Pi_1,\Pi'_1\right\}$ correspond to the topological charge measurement of the first pair of anyons spanned by $\ket{\textbf{1}Y}$ and its orthogonal complements $\ket{-YY},\ket{Y\textbf{1}}$. The topological charge of the first pair of anyons is found by this measurement. If it is $\textbf{1}$ or $Y$ then the second pair is still in a coherent superposition of $\textbf{1}$ and $Y$. This measurement allows us to find whether an anyon is trivial or not.
For the hard-controlled gates $C_c(U)$, braiding gate is applied only when the controlling value of topological charge is 'c', the one mentioned on the gate, otherwise go to previous step or start over. The process is repeated several times until we get the required result.
The braiding supplemented with the projective measurement provides the universal set of gates for anyonic quantum computation \cite{cui2015universal}.

\begin{figure}[h!]
	\centering
	\begin{tikzpicture}
		\draw[ultra thick,cyan] (0,0)--(0,1);
		\qnode{0}{-0.8}{$SUM_1$}
		\qgateCNC*[ibmqxD]{b}{0}{1}
		\qgateU[ibmqxD]{0}{0}{+1}
	\end{tikzpicture}\qquad
	\begin{tikzpicture}
		\draw[ultra thick,cyan] (0,0)--(0,1);
		\qnode{0}{-0.8}{$SUM_2$}
		\qgateCNC*[ibmqxD]{b}{0}{1}
		\qgateU[ibmqxD]{0}{0}{+2}
	\end{tikzpicture}\qquad
	\begin{tikzpicture}
		\draw[ultra thick,cyan] (0,0)--(0,1);
		\qnode{0}{-0.8}{$C_c(X)$}
		\qgateCNC[ibmqx]{b}{0}{1}
		\qgateU[ibmqxA]{0}{0}{X}
		\qnode{-0.2}{0.8}{$c$}
	\end{tikzpicture}\qquad
	\begin{tikzpicture}
		\draw[ultra thick,cyan] (0,0)--(0,2);
		\qnode{0}{-0.8}{$C(SUM)$}
		\qgateCNC[ibmqxD]{b}{0}{2}
		\qgateCNC*[ibmqxD]{b}{0}{1}
		\qgateU[ibmqxA]{0}{0}{X}
		\qnode{-0.2}{1.8}{$c$}
	\end{tikzpicture}
	\caption[Graphical representation of two-qutrit ternary gates.]{The graphical representation of two-qutrit ternary gates \cite{bocharov2015improved}.}
	\label{GTGates}
\end{figure}
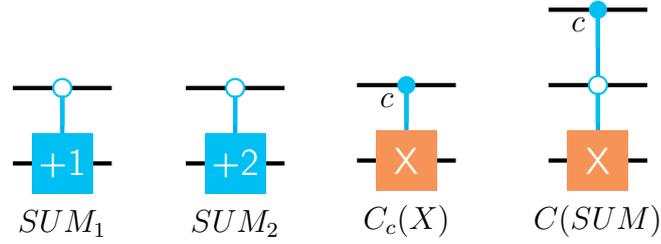

As discussed in previous section, the Clifford gate $SUM = (I\otimes H)\Lambda (Z)(I \otimes H^{-1})$ in Eq. \ref{SumGate} is a generalization of the CNOT gate. It can also be written as $SUM= \ket{0}\bra{0} \otimes I + \ket{1}\bra{1} \otimes X + \ket{2}\bra{2} \otimes X^2$. Here, $X$ is an increment gate $Z_3(+1)$ or $Z_3(+2)$. The two-qutrit $9\cross 9$ matrix for the $SUM$ is written as $diag(I_3, X, X^2)$. Let us call this gate $SUM_1$. But if we use $\Lambda (Z^{-1})$ in Eq. \ref{SumGate}, we get the matrix form as $diag(I_3, X^2, X)$. Let us represent this form as $SUM_2$. The braiding implementation of $SUM_1$ and $SUM_2$ is equivalent. The $SUM_1$ and $SUM_2$ are used for designing the two-qutrit braiding gates $Z_3(+1)$ and $Z_3(+2)$. These gates are shown in Fig. \ref{GTGates} and their matrices are shown in Eq. \ref{SUMMatrix}. We can also note that the matrix $diag(I_3, X^2, X)$ is the square of the matrix $diag(I_3, X, X^2)$. When control is the second qutrit and target is the first qutrit, then the $SUM$ gates are written in the form $SUM=I \otimes \ket{0}\bra{0} + X \otimes \ket{1}\bra{1} + X^2 \otimes \ket{2}\bra{2}= (H\otimes I)\Lambda (Z)(H^{-1} \otimes I)$.

Two qutrits can be swapped by the $SWAP$ gate, as discussed in \ref{QC}. This gate can be formed by braiding alone \cite{bocharov2016efficient} with the use of the permutation gates. One of the realizations of the $SWAP$ gate $SWAP: \ket{i,j} \to \ket{j,i}$ is obtained by using the gate $Z_3(12)$ \cite{bocharov2016efficient} and can be written as
\eq{SWAP = (Z_3(12) \otimes I)SUM_{1,2}SUM_{2,1}SUM_{2,1}SUM_{1,2}, \label{Swap}}
where $SUM_{j,k}$ is a two-qutrit $SUM$ gate applied to $k$th qutrit when $j$th qutrit is the control qutrit. Two other non-Clifford gates used in this paper are Honer gate and controlled-SUM gate given as
\eq{&Horner = \Lambda(\Lambda(X)): \ket{i,j,k} \to \ket{i,j,ij+k}\nonumber\\
	&C(SUM) = C_c(SUM): \ket{i,j,k} \to \ket{i,j,j\delta_{i,c}+k}}
These gates are generalization of Toffoli gate \cite{bocharov2015improved,bocharov2017factoring}.

In Fig. \ref{GTGates}, the non-Clifford gates are represented by the filled circles at the controlling values with $c=0,1,2$, whereas for the $SUM$ gates, the hollow circles are drawn at the controlling values. To avoid cluttering, labels of the control is omitted when its value is 2.
The increment gates will be represented by $+1$ and $+2$ and the permutation gates will be represented by $01$, $12$, and $02$. The gates in the boxes are related to the sum or product column of truth tables. Blue and orange colors of gates correspond to Clifford gates and non-Clifford gates respectively. A circuit is read from left to right but when it is written as matrices it is read from right to left, the same as the matrix multiplication.

\eq{SUM_1=\begin{pmatrix}
		1 & 0 & 0 & & & & & & \\
		0 & 1 & 0 & & & & & &\\
		0 & 0 & 1 & & & & & &\\
		& & & 0 & 0 & 1 & & &\\
		& & & 1 & 0 & 0 & & &\\
		& & & 0 & 1 & 0 & & &\\
		& & & & & & 0 & 1 & 0\\
		& & & & & & 0 & 0 & 1\\
		& & & & & & 1 & 0 & 0\end{pmatrix}, \ SUM_2=\begin{pmatrix}
		1 & 0 & 0 & & & & & & \\
		0 & 1 & 0 & & & & & &\\
		0 & 0 & 1 & & & & & &\\
		& & & 0 & 1 & 0 & & &\\
		& & & 0 & 0 & 1 & & &\\
		& & & 1 & 0 & 0 & & &\\
		& & & & & & 0 & 0 & 1\\
		& & & & & & 1 & 0 & 0\\
		& & & & & & 0 & 1 & 0\end{pmatrix}\label{SUMMatrix}}

\subsection{Ternary Adder}
The adder circuit is the most important arithmetic circuit used in almost all circuits, especially in algorithms such as Grover, Shor, and HHL algorithms.
Binary adder circuits are proposed by  Ref. \cite{draper2004logarithmic,cuccaro2004new,vedral1996quantum} and their ternary counterparts are given in Ref. \cite{khan2007quantum,khan2004quantum,monfared2017design}. 
The adder circuit of Ref. \cite{haghparast2017towards} consists of 14 MS and shift gates and the circuit from Ref. \cite{deibuk2015design} obtained using the genetic algorithm, has 13 MS and shift gates. In Ref. \cite{asadi2020efficient}, half adder is designed by 5 MS gates and one shift gate. An output that remains unused and thrown out is called the garbage output. Most of these designs used the Toffoli gate for their implementation, but the Toffoli gate cannot be built by braiding alone \cite{cui2015universal}. Our circuit design for half adder consist of four gates. Only one braiding gate is used to implement the sum and three non-Clifford gates are used for the implementation of carry. The constant inputs and the garbage outputs are the same for our designs as in the existing designs.

When we add two one-digit numbers, then we get the half adder, whereas the full adder circuit adds three one-digit numbers. The third digit can be a carry from the previous half adder.   
The truth table for a half adder is shown in Table \ref{TruthHalfAdd} and the circuit realization is shown in Fig. \ref{Half}. Let us discuss the cases when there is a nonzero carry. For example, the case when the input $A$ has value 1 and input $B$ has value 2. For the first gate, the control value is not 2 so it would not be applied. For the second gate, control value is 2 but third input is zero. The second gate will also remain ineffective. Third gate will be applied and it will give the carry 1. At the fourth gate, withing the box, $S$ will be zero as $B=2$ will add 2 to $A=1$.

When $A=2$ and $B=2$, the first gate will change the third qutrit from 0 to 2. At the second gate, third qutrit will be changed to 1. The third gate will not be applied. The fourth gate will add 2 to $A=2$ and give the sum $S=1$. The garbage bits at the end of the computation will be ignored.

\begin{figure}[h!]
	\centering
	\begin{minipage}{0.4\textwidth}
		\centering
		\begin{tikzpicture}[scale=1]
			\draw[fill=red!20,opacity=0.5] (4.6,0.6) rectangle (5.8,2.6);
			\qnode{0.2}{2}{$A$}
			\qnode{0.2}{1}{$B$}
			\qnode{0.2}{0}{$0$}
			\qwire[ibmqx]{1}{1}
			\draw[ultra thick,cyan] (1.3,0)--(1.3,2);			
			\qgateU[ibmqxA]{1}{0}{+1}
			\qgateCNC[ibmqx]{b}{1}{2}
			\qgateCNC*[ibmqx]{b}{1}{1}
			\qwire[ibmqx]{2}{2}
			\draw[ultra thick,cyan] (2.6,0)--(2.6,1);			
			\qgateU[ibmqxA]{2}{0}{12}
			\qgateCNC[ibmqx]{b}{2}{1}
			\draw[ultra thick,cyan] (3.9,0)--(3.9,2);			
			\qgateU[ibmqxA]{3}{0}{+1}
			\qgateCNC[ibmqx]{b}{3}{2}
			\qgateCNC[ibmqx]{b}{3}{1}
			\qnode{2.8}{2.3}{$1$}
			\qwire[ibmqx]{4}{0}		
			\draw[ultra thick,cyan] (5.2,1)--(5.2,2);
			\qgateU[ibmqxD]{4}{2}{+1}
			\qgateCNC*[ibmqx]{t}{4}{1}			
			\qnode{4.7}{2}{$S$}
			\qnode{4.7}{1}{$g$}
			\qnode{4.85}{0}{$c_{out}$}
		\end{tikzpicture}
		\captionof{figure}{Ternary half adder circuit realization.}
		\label{Half}
	\end{minipage}
	\begin{minipage}{0.4\textwidth}
		\centering			
		\begin{tabular}{|>{\columncolor{green!20}}c|>{\columncolor{green!20}}c||>{\columncolor{orange!40}}c|>{\columncolor{orange!20}}c|}
			\hline
			\rowcolor{cyan!50}
			$A$ & $B$ & $S$ & $c_{out}$\\
			\hline\hline
			0 & 0 & 0 & 0\\
			0 & 1 & 1 & 0\\
			0 & 2 & 2 & 0\\
			1 & 0 & 1 & 0\\
			1 & 1 & 2 & 0\\
			1 & 2 & 0 & 1\\
			2 & 0 & 2 & 0\\
			2 & 1 & 0 & 1\\
			2 & 2 & 1 & 1\\
			\hline
		\end{tabular}
		\captionof{table}{Truth table of the ternary half adder.}
		\label{TruthHalfAdd}
	\end{minipage}	
\end{figure}

The full adder adds three qutrits $A$, $B$, $C$ as shown in Fig. \ref{Full}. The truth table for the ternary full adder is given in Table \ref{TruthFullAdd}. The sum of $A$ and $B$ is obtained that is added to the third input qutrit $C$ to get the output $\bm{S}$. The input $C$ can be a carry from the previous sum of two qutrits. The garbage outputs $g_1$ and $g_2$ are ignored.

The addition of two-qutrit numbers and its circuit realization are shown in Fig. \ref{2-adder} (a), (b). A half adder and a full adder can be used. 
The first qutrit $A_0$ of $A_0B_0$ is added by the first half adder and the first digit of the sum $\bm{S}_0$ and $g_1$ are obtained. Their carry $c_0$ is to be added with the sum of the second qutrits $A_1$ and $B_1$. This $c_0$ corresponds to the input $C$ of the full adder. The addition of $A_1+B_1 +c_0$ gives the second digits of the output as $\bm{S}_1$ and a carry $\bm{c}_{out}$. The $SWAP$ gate in Eq. \ref{Swap} is used to exchange the qutrits $S$ and $c_0$, and the garbage qutrits are thrown out.

\begin{figure}[h!]
	\centering
	\begin{minipage}{0.65\textwidth}
		\centering
		\begin{tikzpicture}[scale=1]
			\draw[fill=red!20,opacity=0.5] (3.3,1.7) rectangle (4.5,3.6);
			\draw[fill=red!20,opacity=0.5] (8.5,0.7) rectangle (9.7,3.6);
			\qnode{-0.7}{3}{$A$}
			\qnode{-0.7}{2}{$B$}
			\qnode{-0.7}{1}{$C$}
			\qnode{-0.7}{0}{$0$}
			\qwire[ibmqx]{0}{3}
			\qwire[ibmqx]{0}{1}
			\draw[ultra thick,cyan] (0,0)--(0,3);
			\qgateU[ibmqxA]{0}{0}{+1}
			\qgateCNC[ibmqx]{b}{0}{3}
			\qgateCNC*[ibmqx]{b}{0}{2}
			\qwire[ibmqx]{1}{3}				
			\qwire[ibmqx]{1}{1}			
			\draw[ultra thick,cyan] (1.3,0)--(1.3,2);				
			\qgateCNC[ibmqx]{b}{1}{2}
			\qgateU[ibmqxA]{1}{0}{12}
			\qwire[ibmqx]{2}{1}
			\draw[ultra thick,cyan] (2.6,0)--(2.6,3);			
			\qgateCNC[ibmqx]{b}{2}{3}
			\qgateCNC[ibmqx]{b}{2}{2}
			\qgateU[ibmqxA]{2}{0}{+1}
			\qnode{1.8}{3.3}{$1$}
			\qwire[ibmqx]{3}{0}
			\qwire[ibmqx]{3}{1}
			\draw[ultra thick,cyan] (3.9,2)--(3.9,3);
			\qgateU[ibmqxD]{3}{3}{+1}
			\qgateCNC*[ibmqx]{t}{3}{2}
			\qwire[ibmqx]{4}{2}
			\draw[ultra thick,cyan] (5.2,0)--(5.2,3);			
			\qgateCNC[ibmqx]{b}{4}{3}
			\qgateCNC*[ibmqx]{b}{4}{1}
			\qgateU[ibmqxA]{4}{0}{+1}
			\qwire[ibmqx]{5}{2}
			\draw[ultra thick,cyan] (6.5,0)--(6.5,3);			
			\qgateCNC[ibmqx]{b}{5}{3}
			\qgateCNC[ibmqx]{b}{5}{1}
			\qgateU[ibmqxA]{5}{0}{+2}
			\qwire[ibmqx]{6}{2}
			\draw[ultra thick,cyan] (7.8,0)--(7.8,3);			
			\qgateCNC[ibmqx]{b}{6}{3}
			\qgateCNC[ibmqx]{b}{6}{1}
			\qgateU[ibmqxA]{6}{0}{+1}
			\qnode{5.8}{3.3}{$1$}
			\qwire[ibmqx]{7}{0}
			\qwire[ibmqx]{7}{2}
			\draw[ultra thick,cyan] (9.1,1)--(9.1,3);
			\qgateU[ibmqxD]{7}{3}{+1}
			\qgateCNC*[ibmqx]{t}{7}{1}
			\qnode{7.7}{3}{$S$}
			\qnode{7.7}{2}{$B$}
			\qnode{7.7}{1}{$C$}
			\qnode{7.9}{0}{$c_{out}$}
		\end{tikzpicture}
		\captionof{figure}[Ternary full adder circuit realization.]{Ternary full adder circuit realization.}
		\label{Full}
	\end{minipage}
	\begin{minipage}{0.3\textwidth}
		\centering
		\scalebox{0.8}{
			\begin{tabular}{|>{\columncolor{green!20}}c|>{\columncolor{green!20}}c|>{\columncolor{green!20}}c||>{\columncolor{orange!40}}c|>{\columncolor{orange!20}}c|}
				\hline
				\rowcolor{cyan!50}
				$A$ & $B$ & $C$ & $S$ & $c_{out}$\\
				\hline\hline
				0 & 0 & 0 & 0 & 0\\
				0 & 0 & 1 & 1 & 0\\
				0 & 0 & 2 & 2 & 0\\
				0 & 1 & 0 & 1 & 0\\
				0 & 1 & 1 & 2 & 0\\
				0 & 1 & 2 & 0 & 1\\
				0 & 2 & 0 & 2 & 0\\
				0 & 2 & 1 & 0 & 1\\
				0 & 2 & 2 & 1 & 1\\
				1 & 0 & 0 & 1 & 0\\
				1 & 0 & 1 & 2 & 0\\
				1 & 0 & 2 & 0 & 1\\
				1 & 1 & 0 & 2 & 0\\
				1 & 1 & 1 & 0 & 1\\
				1 & 1 & 2 & 1 & 1\\
				1 & 2 & 0 & 0 & 1\\
				1 & 2 & 1 & 1 & 1\\
				1 & 2 & 2 & 2 & 1\\
				2 & 0 & 0 & 2 & 0\\
				2 & 0 & 1 & 0 & 1\\
				2 & 0 & 2 & 1 & 1\\
				2 & 1 & 0 & 0 & 1\\
				2 & 1 & 1 & 1 & 1\\
				2 & 1 & 2 & 2 & 1\\
				2 & 2 & 0 & 1 & 1\\
				2 & 2 & 1 & 2 & 1\\
				2 & 2 & 2 & 0 & 2\\
				\hline
		\end{tabular}}
		\captionof{table}{Truth table for ternary full adder.}
		\label{TruthFullAdd}
	\end{minipage}	
\end{figure}

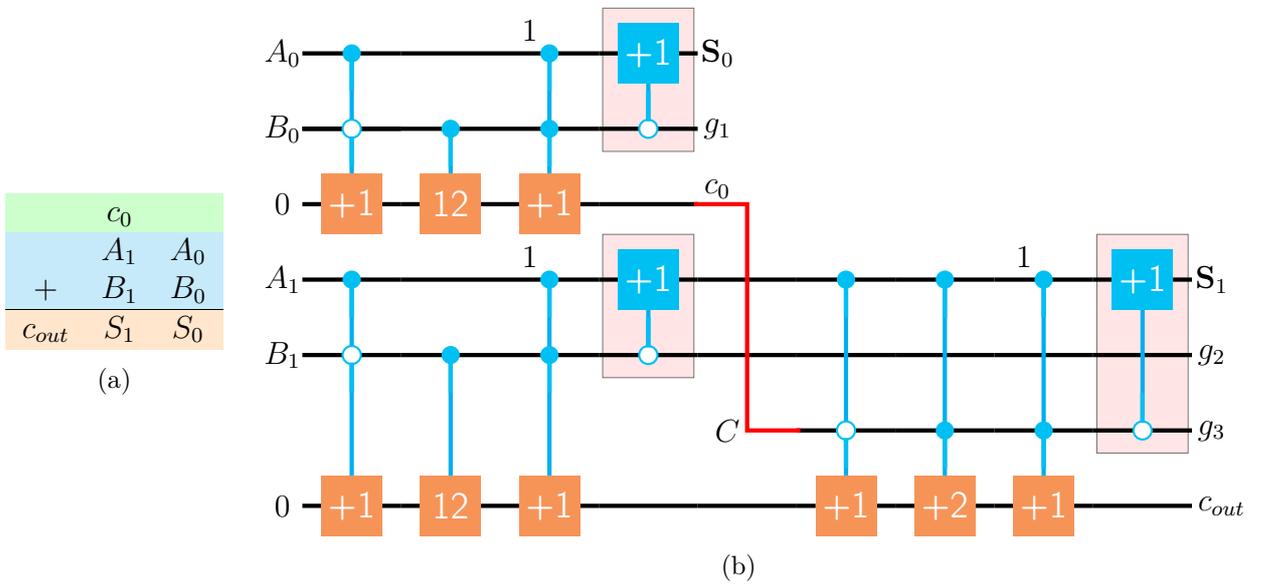
\begin{figure}[h!]
	\centering
	\begin{subfigure}{0.2\textwidth}
		\centering
		\begin{tabular}{ccc}
			\rowcolor{green!20}
			&$c_0$&\\
			\rowcolor{cyan!20}
			&$A_1$&$A_0$\\
			\rowcolor{cyan!20}
			$+$&$B_1$&$B_0$\\
			\hline
			\rowcolor{orange!20}
			$c_{out}$&$S_1$&$S_0$
		\end{tabular}
		\caption{}
	\end{subfigure}
	\begin{subfigure}{0.75\textwidth}
		\centering
		\begin{tikzpicture}[scale=1]
			\draw[fill=red!20,opacity=0.5] (3.3,4.7) rectangle (4.5,6.6);			
			\qnode{-0.7}{6}{$A_0$}
			\qnode{-0.7}{5}{$B_0$}
			\qnode{-0.7}{4}{$0$}
			\qwire[ibmqx]{0}{5}
			\draw[ultra thick,cyan] (0,4)--(0,6);			
			\qgateU[ibmqxA]{0}{4}{+1}
			\qgateCNC[ibmqx]{b}{0}{6}
			\qgateCNC*[ibmqx]{b}{0}{5}
			\qwire[ibmqx]{1}{6}
			\draw[ultra thick,cyan] (1.3,4)--(1.3,5);			
			\qgateU[ibmqxA]{1}{4}{12}
			\qgateCNC[ibmqx]{b}{1}{5}
			\draw[ultra thick,cyan] (2.6,4)--(2.6,6);			
			\qgateU[ibmqxA]{2}{4}{+1}
			\qgateCNC[ibmqx]{b}{2}{6}
			\qgateCNC[ibmqx]{b}{2}{5}
			\qnode{1.8}{6.3}{$1$}
			\qwire[ibmqx]{3}{4}
			\draw[ultra thick,cyan] (3.9,5)--(3.9,6);
			\qgateU[ibmqxD]{3}{6}{+1}
			\qgateCNC*[ibmqx]{t}{3}{5}			
			\qnode{3.7}{6}{$\textbf{S}_0$}
			\qnode{3.7}{5}{$g_1$}
			\qnode{3.7}{4.2}{$c_0$}
			
			===================================
			
			\draw[fill=red!20,opacity=0.5] (3.3,1.7) rectangle (4.5,3.6);
			\draw[fill=red!20,opacity=0.5] (9.8,0.7) rectangle (11,3.6);
			\qnode{-0.7}{3}{$A_1$}
			\qnode{-0.7}{2}{$B_1$}
			\qnode{3.8}{1}{$C$}
			\qnode{-0.7}{0}{$0$}
			
			\qwire[ibmqx]{4}{3}
			\qwire[ibmqx]{4}{2}
			\qwire[ibmqx]{4}{0}		
			
			\draw[ultra thick,cyan] (0,0)--(0,3);
			\qgateU[ibmqxA]{0}{0}{+1}
			\qgateCNC[ibmqx]{b}{0}{3}
			\qgateCNC*[ibmqx]{b}{0}{2}
			\qwire[ibmqx]{1}{3}				
			\draw[ultra thick,cyan] (1.3,0)--(1.3,2);				
			\qgateCNC[ibmqx]{b}{1}{2}
			\qgateU[ibmqxA]{1}{0}{12}
			\draw[ultra thick,cyan] (2.6,0)--(2.6,3);			
			\qgateCNC[ibmqx]{b}{2}{3}
			\qgateCNC[ibmqx]{b}{2}{2}
			\qgateU[ibmqxA]{2}{0}{+1}
			\qnode{1.8}{3.3}{$1$}
			\qwire[ibmqx]{3}{0}
			
			\draw[ultra thick,cyan] (3.9,2)--(3.9,3);
			\qgateU[ibmqxD]{3}{3}{+1}
			\qgateCNC*[ibmqx]{t}{3}{2}
			\qwire[ibmqx]{5}{2}
			\draw[ultra thick,cyan] (6.5,0)--(6.5,3);			
			\qgateCNC[ibmqx]{b}{5}{3}
			\qgateCNC*[ibmqx]{b}{5}{1}
			\qgateU[ibmqxA]{5}{0}{+1}
			\qwire[ibmqx]{6}{2}
			\draw[ultra thick,cyan] (7.8,0)--(7.8,3);			
			\qgateCNC[ibmqx]{b}{6}{3}
			\qgateCNC[ibmqx]{b}{6}{1}
			\qgateU[ibmqxA]{6}{0}{+2}
			\qwire[ibmqx]{7}{2}
			\draw[ultra thick,cyan] (9.1,0)--(9.1,3);			
			\qgateCNC[ibmqx]{b}{7}{3}
			\qgateCNC[ibmqx]{b}{7}{1}
			\qgateU[ibmqxA]{7}{0}{+1}
			\qnode{6.8}{3.3}{$1$}
			\qwire[ibmqx]{8}{0}
			\qwire[ibmqx]{8}{2}
			\draw[ultra thick,cyan] (10.4,1)--(10.4,3);
			\qgateU[ibmqxD]{8}{3}{+1}
			\qgateCNC*[ibmqx]{t}{8}{1}
			\qnode{8.7}{3}{$\textbf{S}_1$}
			\qnode{8.7}{2}{$g_2$}
			\qnode{8.7}{1}{$g_3$}
			\qnode{8.8}{0}{$c_{out}$}
			\draw[ultra thick, red] (4.5,4)--(5.2,4)--(5.2,1)--(5.9,1);
		\end{tikzpicture}
		\caption{}
	\end{subfigure}
	\caption[Ternary two-qutrit addition and its circuit realization.]{Ternary two-qutrit (a) addition and (b) circuit realization by using one half adder and one full adder.}
	\label{2-adder}
\end{figure}

\begin{figure*}[h!]
	\centering
	
\end{figure*}

\subsection{Ternary Subtractor}
A ternary subtractor gives an output as a difference between two inputs. The subtractor circuit takes two inputs $A$ and $B$ and one ancilla. The difference between two inputs and the borrow is obtained at the output. For ternary subtractor in conventional quantum computing, see Ref. \cite{monfared2017design}. The half subtractor truth table is shown in Table \ref{TruthHalfSub} and the circuit realization is shown in Fig. \ref{HalfSub}. Let us discuss the case when $A=1,B=2$. Since the controlling, value needs to be at 0, but we have $B=2$, the first $C(SUM)$ gate would not be applied. For the second gate, the controlling value is 2, therefore the MS permutation gate would be applied, but the value at the third qutrit is 0 so this gate will remain ineffective. The gate for the carry will give 1, because the controlling values are 2. The fourth gate changes $A$ from 1 to 2. Therefore, at the output, we get a difference $D=2$ and the borrow $b_{out} = 1$. This design of half subtractor circuit consists of only one braiding gate and three non-Clifford gates.

\begin{figure}[h!]
	\centering
	\begin{minipage}{0.4\textwidth}
		\centering
		\begin{tikzpicture}[scale=1]
			\draw[fill=red!20,opacity=0.6] (3.3,0.5) rectangle (4.5,2.6);
			\qnode{-0.7}{2}{$A$}
			\qnode{-0.7}{1}{$B$}
			\qnode{-0.7}{0}{$0$}
			\draw[ultra thick,cyan] (0,0)--(0,2);			
			\qgateU[ibmqxA]{0}{0}{+1}
			\qgateCNC[ibmqx]{b}{0}{2}
			\qgateCNC*[ibmqx]{b}{0}{1}
			\qnode{-0.2}{2.3}{$0$}
			\draw[ultra thick,cyan] (1.3,0)--(1.3,1);			
			\qwire[ibmqx]{1}{2}
			\qwire[ibmqx]{1}{1}
			\qgateCNC[ibmqx]{b}{1}{1}
			\qgateU[ibmqxA]{1}{0}{12}
			\draw[ultra thick,cyan] (2.6,0)--(2.6,2);			
			\qwire[ibmqx]{2}{2}
			\qwire[ibmqx]{2}{1}
			\qgateCNC[ibmqx]{b}{2}{2}
			\qgateCNC[ibmqx]{b}{2}{1}
			\qgateU[ibmqxA]{2}{0}{+1}
			\qnode{1.8}{2.3}{$1$}
			\draw[ultra thick,cyan] (3.9,1)--(3.9,2);
			\qgateU[ibmqxD]{3}{2}{+2}
			\qgateCNC*[ibmqx]{t}{3}{1}
			\qwire[ibmqx]{3}{0}
			\qnode{3.7}{2}{$D$}
			\qnode{3.7}{1}{$g$}
			\qnode{3.85}{0}{$b_{out}$}
		\end{tikzpicture}
		\captionof{figure}{Ternary half subtractor circuit realization.}
		\label{HalfSub}
	\end{minipage}
	\begin{minipage}{0.4\textwidth}
		\centering		
		\begin{tabular}{|>{\columncolor{green!20}}c|>{\columncolor{green!20}}c||>{\columncolor{orange!40}}c|>{\columncolor{orange!20}}c|}
			\hline
			\rowcolor{cyan!50}
			A & B & D & $b_{out}$ \\
			\hline\hline
			0 & 0 & 0 & 0\\
			0 & 1 & 2 & 1\\
			0 & 2 & 1 & 1\\
			1 & 0 & 1 & 0\\
			1 & 1 & 0 & 0\\
			1 & 2 & 2 & 1\\
			2 & 0 & 2 & 0\\
			2 & 1 & 1 & 0\\
			2 & 2 & 0 & 0\\
			\hline
		\end{tabular}
		\captionof{table}{Truth table for ternary half subtractor.}
		\label{TruthHalfSub}
	\end{minipage}	
\end{figure}

When taking the difference of two numbers and a borrow is needed, then have a full subtractor. It has three inputs $A$, $B$, and $C$, where $C$ is the borrow-in. The truth table for full subtractor is shown in Table \ref{TruthFullSub} and the circuit realization is shown in Fig. \ref{FullSub}. 

\begin{figure}[h!]
	\centering
	\begin{minipage}{0.7\textwidth}
		\centering
		\begin{tikzpicture}[scale=1]
			\draw[fill=red!20,opacity=0.5] (3.3,0.4) rectangle (5.8,3.6);
			\qnode{-0.7}{3}{$A$}
			\qnode{-0.7}{2}{$B$}
			\qnode{-0.7}{1}{$C$}
			\qnode{-0.7}{0}{$0$}
			\qwire[ibmqx]{0}{3}
			\qwire[ibmqx]{0}{1}
			\draw[ultra thick,cyan] (0,0)--(0,2);
			\qgateU[ibmqxA]{0}{0}{+1}
			\qgateCNC[ibmqx]{b}{0}{2}
			\qwire[ibmqx]{1}{1}			
			\draw[ultra thick,cyan] (1.3,0)--(1.3,3);				
			\qgateCNC[ibmqx]{b}{1}{3}
			\qgateCNC[ibmqx]{b}{1}{2}
			\qgateU[ibmqxA]{1}{0}{+1}
			\qnode{0.8}{3.3}{$0$}
			\qnode{0.8}{2.3}{$1$}
			\qwire[ibmqx]{2}{1}
			\draw[ultra thick,cyan] (2.6,0)--(2.6,3);			
			\qgateCNC[ibmqx]{b}{2}{3}
			\qgateCNC[ibmqx]{b}{2}{2}
			\qgateU[ibmqxA]{2}{0}{+2}
			
			\qwire[ibmqx]{3}{0}
			\qwire[ibmqx]{3}{1}
			\draw[ultra thick,cyan] (3.9,2)--(3.9,3);
			\qgateU[ibmqxD]{3}{3}{+2}
			\qgateCNC*[ibmqx]{t}{3}{2}			
			\qwire[ibmqx]{4}{0}
			\qwire[ibmqx]{4}{2}
			\draw[ultra thick,cyan] (5.2,1)--(5.2,3);
			\qgateU[ibmqxD]{4}{3}{+2}
			\qgateCNC*[ibmqx]{t}{4}{1}			
			
			\qwire[ibmqx]{5}{2}
			\draw[ultra thick,cyan] (6.5,0)--(6.5,3);			
			\qgateCNC[ibmqx]{b}{5}{3}
			\qgateCNC[ibmqx]{b}{5}{1}
			\qgateU[ibmqxA]{5}{0}{+1}
			\qwire[ibmqx]{6}{2}
			\draw[ultra thick,cyan] (7.8,0)--(7.8,3);			
			\qgateCNC[ibmqx]{b}{6}{3}
			\qgateCNC[ibmqx]{b}{6}{1}
			\qgateU[ibmqxA]{6}{0}{+1}
			\qnode{5.8}{1.3}{$1$}
			\qwire[ibmqx]{7}{2}
			\draw[ultra thick,cyan] (9.1,0)--(9.1,3);			
			\qgateCNC[ibmqx]{b}{7}{3}
			\qgateCNC[ibmqx]{b}{7}{1}
			\qgateU[ibmqxA]{7}{0}{+1}
			\qnode{6.8}{3.3}{$1$}
			\qnode{7.7}{3}{$D$}
			\qnode{7.7}{2}{$g_1$}
			\qnode{7.7}{1}{$g_2$}
			\qnode{7.85}{0}{$b_{out}$}
		\end{tikzpicture}
		\captionof{figure}{Ternary full subtractor circuit realization.}
		\label{FullSub}
	\end{minipage}
	\begin{minipage}{0.25\textwidth}
		\scalebox{0.8}[0.8]{
			\begin{tabular}{|>{\columncolor{green!20}}c|>{\columncolor{green!20}}c|>{\columncolor{green!20}}c||>{\columncolor{orange!40}}c|>{\columncolor{orange!20}}c|}
				\hline
				\rowcolor{cyan!50}
				A & B & C & D & $b_{out}$\\
				\hline\hline
				0 & 0 & 0 & 0 & 0\\
				0 & 0 & 1 & 2 & 1\\
				0 & 0 & 2 & 1 & 1\\
				0 & 1 & 0 & 2 & 1\\
				0 & 1 & 1 & 1 & 1\\
				0 & 1 & 2 & 0 & 1\\
				0 & 2 & 0 & 1 & 1\\
				0 & 2 & 1 & 0 & 1\\
				0 & 2 & 2 & 2 & 2\\
				
				1 & 0 & 0 & 1 & 0\\
				1 & 0 & 1 & 0 & 0\\
				1 & 0 & 2 & 2 & 1\\
				1 & 1 & 0 & 0 & 0\\
				1 & 1 & 1 & 2 & 1\\
				1 & 1 & 2 & 1 & 1\\
				1 & 2 & 0 & 2 & 1\\
				1 & 2 & 1 & 1 & 1\\
				1 & 2 & 2 & 0 & 1\\
				
				2 & 0 & 0 & 2 & 0\\
				2 & 0 & 1 & 1 & 0\\
				2 & 0 & 2 & 0 & 0\\
				2 & 1 & 0 & 1 & 0\\
				2 & 1 & 1 & 0 & 0\\
				2 & 1 & 2 & 2 & 1\\
				2 & 2 & 0 & 0 & 0\\
				2 & 2 & 1 & 2 & 1\\
				2 & 2 & 2 & 1 & 1\\
				\hline
		\end{tabular}}
		\captionof{table}{Truth table for ternary full subtractor.}
		\label{TruthFullSub}
	\end{minipage}	
\end{figure}

\subsection{Ternary Multiplier}
In a two-qutrit multiplier $A_0B_0 \cross A_1B_1$, each digit of the first number is multiplied by each digit of the second number. Then all the partial products are added in the way shown in Fig. \ref{2-Multiply}. Therefore, we need ternary partial product generation (TPPG) circuits and adder circuits for the two-qutrit multiplier. This kind of multiplier is discussed in Ref. \cite{panahi2019novel}. The outputs of the two numbers are $P_0$, $P_1$, $P_2$, and $P_3$ and the carry is $c_{out}$.
\begin{figure}[h!]
	\centering
	\begin{tabular}{ccccc}
		\rowcolor{green!20}
		&&&$A_1$&$A_0$\\
		\rowcolor{green!20}
		&&$\cross$&$B_1$&$B_0$\\
		\hline
		\rowcolor{cyan!20}
		&&$c_0$&&\\
		\rowcolor{cyan!20}
		&$c_2$&$cp_1$&$cp_0$&\\
		\rowcolor{cyan!20}
		&$c_1$&$cp_2$&$A_1B_0$&$A_0B_0$\\
		\rowcolor{cyan!20}
		+&$cp_3$&$A_1B_1$&$A_0B_1$&0\\
		\hline
		\rowcolor{orange!20}
		$c_{out}$&$P_3$&$P_2$&$P_1$&$P_0$
	\end{tabular}
	\caption{Two-qutrit multiplication.}
	\label{2-Multiply}
\end{figure}

The carries produced by addition are represented as $c_i$, whereas the $cp_i$ are the carries we get as a result of one-digit multiplication. The first digit of multiplication is $P_0=A_0B_0$ and its carry is represented by $cp_0$. To compute $P_1,P_2,P_3$ and $c_{out}$, adder circuits are needed. As each of the digits $A_0,B_0,A_1,B_1$ can have values $0,1,2$, not all the partial products produce carries in multiplication. We do not need extra input lines and gates corresponding to the carries, and therefore, it is cheaper to implement the circuit without full adders. Instead, we are using the Adder Blocks as in Panahi \cite{panahi2019novel}. The one-digit TPPG circuit is shown in Fig. \ref{TPPG} and the truth table is shown in Fig. \ref{TruthTPPG}. The carry-out appears only when both the input values are at value $2$ in the one-digit multiplication. The existing realization in Ref. \cite{panahi2019novel} has 13 MS and shift gates.
\eq{P_0&=A_0B_0\nonumber\\
	P_1&=cp_0+A_1B_0+A_0B_1\nonumber\\
	P_2&=c_0+cp_1+cp_2+A_1B_1\nonumber\\
	P_3&=c_2+c_1+cp_3}

\begin{figure}[h!]
	\centering
	\begin{minipage}{0.5\textwidth}
		\centering
		\begin{tikzpicture}[scale=1]
			\draw[fill=red!20,opacity=0.5] (2,0.4) rectangle (5.8,3.6);
			\qnode{1.2}{3}{$A$}
			\qnode{1.2}{2}{$B$}
			\qnode{1.2}{1}{$0$}
			\qnode{1.2}{0}{$0$}
			\draw[ultra thick,cyan] (2.6,1)--(2.6,2);
			\qwire[ibmqx]{2}{3}
			\qgateU[ibmqxD]{2}{1}{+1}
			\qgateCNC*[ibmqx]{b}{2}{2}
			
			\qwire[ibmqx]{3}{2}
			\draw[ultra thick,cyan] (3.9,1)--(3.9,3);
			\qgateCNC[ibmqx]{b}{3}{3}
			\qgateU[ibmqxA]{3}{1}{12}
			\qwire[ibmqx]{4}{2}
			\draw[ultra thick,cyan] (5.2,1)--(5.2,3);		
			\qgateCNC[ibmqx]{b}{4}{3}
			\qgateCNC*[ibmqx]{b}{4}{2}
			\qgateU[ibmqxA]{4}{1}{+2}
			\qnode{3.8}{3.3}{$0$}
			
			\qwire[ibmqx]{5}{1}
			\draw[ultra thick,cyan] (6.5,0)--(6.5,3);					
			\qgateCNC[ibmqx]{b}{5}{3}
			\qgateCNC[ibmqx]{b}{5}{2}
			\qgateU[ibmqxA]{5}{0}{+1}
			\qnode{5.8}{3}{$A$}
			\qnode{5.8}{2}{$B$}
			\qnode{5.8}{1}{$P$}
			\draw[ultra thick] (1.95,0)--(6,0);
			\qnode{5.8}{0}{$cp$}
		\end{tikzpicture}
		\captionof{figure}{Circuit design of the TPPG component.}
		\label{TPPG}
	\end{minipage}
	\begin{minipage}{0.3\textwidth}
		\centering
		\begin{tabular}{|>{\columncolor{green!20}}c|>{\columncolor{green!20}}c||>{\columncolor{orange!40}}c|>{\columncolor{orange!20}}c|}
			\hline
			\rowcolor{cyan!50}
			A & B & $P$ & $cp$\\
			\hline\hline
			0 & 0 & 0 & 0\\
			1 & 0 & 0 & 0\\
			2 & 0 & 0 & 0\\
			0 & 1 & 0 & 0\\
			1 & 1 & 1 & 0\\
			2 & 1 & 2 & 0\\
			0 & 2 & 0 & 0\\
			1 & 2 & 2 & 0\\
			2 & 2 & 1 & 1\\
			\hline
		\end{tabular}
		\captionof{table}{Truth table for a two-qutrit partial product.}
		\label{TruthTPPG}
	\end{minipage}	
\end{figure}

From the Fig. \ref{2-Multiply}, we have $P_1=cp_0+A_1B_0+A_0B_1$. Adder Block 1 adds the partial products $A_1B_0,A_0B_1$ and the carry $cp_0$ of the partial products. As we can see from the truth table \ref{TruthTPPG} of the partial product, the carry in the first partial product is never 2. The input values can be $0$, $1$ and $2$ but carry is only $0$ or $1$. That is, the $cp_0$, $cp_1$, $cp_2$, and $cp_3$ are 1 and the carry is only for the case when inputs are at values 2. The truth table for the Adder Block 1 is shown in Table \ref{TruthBlock1}, and the implementation is shown in Fig. \ref{Block1}. The input lines with label $0$ are ancilla lines, and $g$ are the garbage outputs.
Adder Block 2 adds three qutrits $A_1B_1$, $cp_1$ and $cp_2$ to get $P_2$. The sum of these three qutrits will be added to $c_0$ using Block 4. The first input has values $0,1,2$, but the second and third inputs have values $0,1$. The partial product $A_1B_1$ is added to the two input carry numbers, which have values $0,1$. The truth table and implementation of Block 2 is shown in Table \ref{TruthBlock2} and Fig. \ref{Block2}.
Adder Block 3 adds qutrits $c_1$, $c_2$ and $cp_3$ to get $P_3$. The truth table is shown in Table \ref{TruthBlock3} and the implementation is shown in Fig. \ref{Block3}. All the inputs have values $0$ or $1$. The output carry appears only when all the inputs have values $1$.
Adder Block 4 is used to add the sum of Adder Block 2 and $c_0$. The half adder is used for its implementation. The output carry is only for one of the cases. The input carry $c_{in}$ is either $0$ or $1$ and input $A=0,1,2$. The truth table of Block 4 is shown in Table \ref{TruthBlock4} and its implementation is shown in Fig. \ref{Block4}. The Block 1 is designed with two Clifford and 4 non-Clifford gates, Block 2 has 2 Clifford and 2 non-Clifford gates, Block 3 consists of 2 Clifford and one non-Clifford gates, whereas Block 4 is implemented with only one Clifford gate and 1 non Clifford gate.

\begin{figure}[h!]
	\centering
	\begin{minipage}{0.6\textwidth}
		\centering
		\begin{tikzpicture}[scale=1]
			\draw[fill=red!20,opacity=0.5] (4.6,0.5) rectangle (7.1,3.6);
			\qnode{0.2}{3}{$A$}
			\qnode{0.2}{2}{$B$}
			\qnode{0.2}{1}{$c_{in}$}	
			\qnode{0.2}{0}{$0$}
			\qwire[ibmqx]{1}{3}
			\qwire[ibmqx]{1}{1}
			\draw[ultra thick,cyan] (1.3,0)--(1.3,2);			
			\qgateCNC[ibmqx]{b}{1}{2}
			\qgateU[ibmqxA]{1}{0}{+1}
			\qwire[ibmqx]{2}{1}
			\draw[ultra thick,cyan] (2.6,0)--(2.6,3);			
			\qgateCNC[ibmqx]{b}{2}{3}
			\qgateCNC[ibmqx]{b}{2}{2}
			\qgateU[ibmqxA]{2}{0}{+1}			
			\qnode{1.8}{2.3}{$1$}
			\qwire[ibmqx]{3}{1}
			\draw[ultra thick,cyan] (3.9,0)--(3.9,3);			
			\qgateCNC[ibmqx]{b}{3}{3}
			\qgateCNC[ibmqx]{b}{3}{2}
			\qgateU[ibmqxA]{3}{0}{01}			
			\qnode{2.8}{3.3}{$0$}
			\draw[ultra thick,cyan] (5.2,2)--(5.2,3);
			\qgateU[ibmqxD]{4}{3}{+1}
			\qgateCNC*[ibmqx]{t}{4}{2}
			\qwire[ibmqx]{4}{1}
			\qwire[ibmqx]{4}{0}
			\qwire[ibmqx]{5}{2}
			\qwire[ibmqx]{5}{0}
			\draw[ultra thick,cyan] (6.5,1)--(6.5,3);
			\qgateU[ibmqxD]{5}{3}{+1}
			\qgateCNC*[ibmqx]{t}{5}{1}
			\qwire[ibmqx]{6}{2}
			\draw[ultra thick,cyan] (7.8,0)--(7.8,3);			
			\qgateCNC[ibmqx]{b}{6}{3}
			\qgateCNC[ibmqx]{b}{6}{1}
			\qgateU[ibmqxA]{6}{0}{+1}			
			\qnode{5.8}{1.3}{$1$}
			\qnode{5.8}{3.3}{$0$}
			\qnode{6.9}{3}{$Sum$}
			\qnode{6.8}{2}{$g_0$}
			\qnode{6.8}{1}{$g_1$}	
			\qnode{6.85}{0}{$c_{out}$}
		\end{tikzpicture}
		\captionof{figure}{Implementation for the ternary Block 1.}
		\label{Block1}		
	\end{minipage}
	\begin{minipage}{0.3\textwidth}
		\centering
		\scalebox{0.9}[0.9]{
			\begin{tabular}{|>{\columncolor{green!20}}c|>{\columncolor{green!20}}c|>{\columncolor{green!20}}c||>{\columncolor{orange!40}}c|>{\columncolor{orange!20}}c|}
				\hline
				\rowcolor{cyan!50}
				A & B & $c_{in}$ & $Sum$ & $c_{out}$\\
				\hline\hline
				0 & 0 & 0 & 0 & 0\\
				1 & 0 & 0 & 1 & 0\\
				2 & 0 & 0 & 2 & 0\\
				0 & 1 & 0 & 1 & 0\\
				1 & 1 & 0 & 2 & 0\\
				2 & 1 & 0 & 0 & 1\\
				0 & 2 & 0 & 2 & 0\\
				1 & 2 & 0 & 0 & 1\\
				2 & 2 & 0 & 1 & 1\\
				0 & 0 & 1 & 1 & 0\\
				1 & 0 & 1 & 2 & 0\\
				2 & 0 & 1 & 0 & 1\\
				0 & 1 & 1 & 2 & 0\\
				1 & 1 & 1 & 0 & 1\\
				2 & 1 & 1 & 1 & 1\\
				0 & 2 & 1 & 0 & 1\\
				1 & 2 & 1 & 1 & 1\\
				2 & 2 & 1 & 2 & 1\\
				\hline
		\end{tabular}}
		\captionof{table}{Truth table for ternary Adder Block 1.}
		\label{TruthBlock1}
	\end{minipage}	
\end{figure}


\begin{figure}[h!]
	\centering
	\begin{minipage}{0.5\textwidth}
		\centering
		\begin{tikzpicture}[scale=1]
			\draw[fill=red!20,opacity=0.5] (4.6,0.5) rectangle (7,3.7);
			\qnode{2.2}{3}{$A$}
			\qnode{2.2}{2}{$B$}
			\qnode{2.2}{1}{$c_{in}$}	
			\qnode{2.2}{0}{$0$}
			\qwire[ibmqx]{3}{1}
			\draw[ultra thick,cyan] (3.9,0)--(3.9,3);			
			\qgateCNC[ibmqx]{b}{3}{3}
			\qgateCNC[ibmqx]{b}{3}{2}
			\qgateU[ibmqxA]{3}{0}{+1}			
			\qnode{2.8}{2.3}{$1$}
			\draw[ultra thick,cyan] (5.2,2)--(5.2,3);
			\qgateU[ibmqxD]{4}{3}{+1}
			\qgateCNC*[ibmqx]{t}{4}{2}
			\qwire[ibmqx]{4}{1}
			\qwire[ibmqx]{4}{0}
			\qwire[ibmqx]{5}{2}
			\qwire[ibmqx]{5}{0}
			\draw[ultra thick,cyan] (6.5,1)--(6.5,3);
			\qgateU[ibmqxD]{5}{3}{+1}
			\qgateCNC*[ibmqx]{t}{5}{1}
			\qwire[ibmqx]{6}{2}
			\draw[ultra thick,cyan] (7.8,0)--(7.8,3);			
			\qgateCNC[ibmqx]{b}{6}{3}
			\qgateCNC[ibmqx]{b}{6}{1}
			\qgateU[ibmqxA]{6}{0}{+1}			
			\qnode{5.8}{3.3}{$0$}
			\qnode{5.8}{1.3}{$1$}
			\qnode{6.9}{3}{$Sum$}
			\qnode{6.8}{2}{$g_2$}
			\qnode{6.8}{1}{$g_3$}	
			\qnode{6.9}{0}{$c_{out}$}
		\end{tikzpicture}
		\captionof{figure}{Implementation of Block 2.}
		\label{Block2}		
	\end{minipage}
	\begin{minipage}{0.4\textwidth}
		\centering
		\scalebox{0.9}[0.9]{
			\begin{tabular}{|>{\columncolor{green!20}}c|>{\columncolor{green!20}}c|>{\columncolor{green!20}}c||>{\columncolor{orange!40}}c|>{\columncolor{orange!20}}c|}
				\hline
				\rowcolor{cyan!50}
				A & B & $c_{in}$ & $Sum$ & $c_{out}$\\
				\hline\hline
				0 & 0 & 0 & 0 & 0\\
				1 & 0 & 0 & 1 & 0\\
				2 & 0 & 0 & 2 & 0\\
				0 & 1 & 0 & 1 & 0\\
				1 & 1 & 0 & 2 & 0\\
				2 & 1 & 0 & 0 & 1\\
				0 & 0 & 1 & 1 & 0\\
				1 & 0 & 1 & 2 & 0\\
				2 & 0 & 1 & 0 & 1\\
				0 & 1 & 1 & 2 & 0\\
				1 & 1 & 1 & 0 & 1\\
				2 & 1 & 1 & 1 & 1\\
				\hline
		\end{tabular}}
		\captionof{table}{Truth table for ternary Adder Block 2.}
		\label{TruthBlock2}
	\end{minipage}	
\end{figure}

\begin{figure}[h!]
	\centering
	\begin{minipage}{0.5\textwidth}
		\centering
		\begin{tikzpicture}[scale=1]
			\draw[fill=red!20,opacity=0.5] (0.6,0.5) rectangle (3.2,3.6);
			\qnode{0.2}{3}{$A$}
			\qnode{0.2}{2}{$B$}
			\qnode{0.2}{1}{$c_{in}$}	
			\qnode{0.2}{0}{$0$}
			\draw[ultra thick,cyan] (1.3,2)--(1.3,3);
			\qgateU[ibmqxD]{1}{3}{+1}
			\qgateCNC*[ibmqx]{t}{1}{2}
			\qwire[ibmqx]{1}{1}
			\qwire[ibmqx]{1}{0}
			
			\qwire[ibmqx]{2}{2}
			\qwire[ibmqx]{2}{0}
			\draw[ultra thick,cyan] (2.6,1)--(2.6,3);
			\qgateU[ibmqxD]{2}{3}{+1}
			\qgateCNC*[ibmqx]{t}{2}{1}
			
			\qwire[ibmqx]{3}{2}
			\draw[ultra thick,cyan] (3.9,0)--(3.9,3);			
			\qgateCNC[ibmqx]{b}{3}{3}
			\qgateCNC[ibmqx]{b}{3}{1}
			\qgateU[ibmqxA]{3}{0}{+1}			
			\qnode{2.8}{1.3}{$1$}
			\qnode{2.8}{3.3}{$0$}
			
			\qnode{3.9}{3}{$Sum$}
			\qnode{3.8}{2}{$g_4$}
			\qnode{3.8}{1}{$g_5$}	
			\qnode{3.85}{0}{$c_{out}$}
		\end{tikzpicture}
		\captionof{figure}{Implementation of ternary Adder Block 3.}
		\label{Block3}
	\end{minipage}
	\begin{minipage}{0.4\textwidth}		
		\centering	
		\begin{tabular}{|>{\columncolor{green!20}}c|>{\columncolor{green!20}}c|>{\columncolor{green!20}}c||>{\columncolor{orange!40}}c|>{\columncolor{orange!20}}c|}
			\hline
			\rowcolor{cyan!50}
			A & B & $c_{in}$ & $Sum$ & $c_{out}$\\
			\hline\hline
			0 & 0 & 0 & 0 & 0\\
			1 & 0 & 0 & 1 & 0\\
			0 & 1 & 0 & 1 & 0\\
			1 & 1 & 0 & 2 & 0\\
			0 & 0 & 1 & 1 & 0\\
			1 & 0 & 1 & 2 & 0\\
			0 & 1 & 1 & 2 & 0\\
			1 & 1 & 1 & 0 & 1\\
			\hline
		\end{tabular}
		\captionof{table}{Truth table for ternary Adder Block 3.}
		\label{TruthBlock3}
	\end{minipage}	
\end{figure}


\begin{figure}[h!]
	\centering
	\begin{minipage}{0.4\textwidth}
		\centering
		\begin{tikzpicture}[scale=1]
			\draw[fill=red!20,opacity=0.6] (2,0.5) rectangle (3.2,2.6);
			\qnode{0.2}{2}{$A$}
			\qnode{0.2}{1}{$c_{in}$}	
			\qnode{0.2}{0}{$0$}
			\draw[ultra thick,cyan] (1.3,0)--(1.3,2);
			\qwire[ibmqx]{1}{2}
			\qgateCNC[ibmqx]{b}{1}{1}
			\qgateCNC[ibmqx]{b}{1}{2}
			\qgateU[ibmqxA]{1}{0}{+1}
			\qnode{0.8}{1.3}{$1$}
			\draw[ultra thick,cyan] (2.6,1)--(2.6,2);	
			\qwire[ibmqx]{2}{0}
			\qgateCNC*[ibmqx]{t}{2}{1}
			\qgateU[ibmqxD]{2}{2}{+1}
			\qnode{2.9}{2}{$Sum$}
			\qnode{2.8}{1}{$g_6$}
			\qnode{2.9}{0}{$c_{out}$}
		\end{tikzpicture}
		\captionof{figure}{Implementation of Block 4.}
		\label{Block4}
	\end{minipage}
	\begin{minipage}{0.4\textwidth}		
		\centering
		\scalebox{1}[1]{
			\begin{tabular}{|>{\columncolor{green!20}}c|>{\columncolor{green!20}}c||>{\columncolor{orange!40}}c|>{\columncolor{orange!20}}c|}
				\hline
				\rowcolor{cyan!50}
				A & $c_{in}$ & $Sum$ & $c_{out}$\\
				\hline\hline
				0 & 0 & 0 & 0\\
				1 & 0 & 1 & 0\\
				2 & 0 & 2 & 0\\
				0 & 1 & 1 & 0\\
				1 & 1 & 2 & 0\\
				2 & 1 & 0 & 1\\
				\hline
		\end{tabular}}
		\captionof{table}{Truth table for ternary Adder Block 4.}
		\label{TruthBlock4}
	\end{minipage}	
\end{figure}

Now we will combine the TPPG circuits and the Adder Blocks to get a full two-digit multiplier circuit. From the two-digit qutrit multiplication in Fig. \ref{2-Multiply}, the first digits of two numbers give the first partial product and first multiplication digit $P_0$. The second multiplication digit is the addition of partial products $A_0B_1$ and $A_1B_0$ and added to the carry from the first partial product. These partial products also create the carries. The carries from these partial products and the carries of additions are added to the partial product $A_1B_1$. These give the third digit of multiplication. The additions generate two carry digits.

\begin{figure}[t!]
	\centering
	\scalebox{0.9}{
		\begin{tikzpicture}[scale=1]
			\draw[dashed,fill=cyan!10] (1.8,9.7) rectangle (3.35,10.5);
			\draw[dashed,fill=cyan!10] (4.45,9.7) rectangle (7.2,10.5);
			\draw[dashed,fill=cyan!10] (8.35,9.7) rectangle (11.1,10.5);
			\draw[dashed,fill=cyan!10] (12.25,9.7) rectangle (13.7,10.5);
			\draw[dashed,fill=cyan!10] (14.85,9.7) rectangle (16.3,10.5);
			\node at (2.5,10.1) {$1$};
			\node at (5.8,10.1) {$2$};
			\node at (9.7,10.1) {$3$};
			\node at (12.95,10.1) {$4$};
			\node at (15.6,10.1) {$5$};
			\draw[dashed,fill=cyan!10] (1.8,-0.5) rectangle (3.35,9.5);
			\draw[dashed,fill=cyan!10] (4.45,-2.5) rectangle (7.2,9.5);
			\draw[dashed,fill=cyan!10] (8.35,-4.5) rectangle (11.1,9.5);
			\draw[dashed,fill=cyan!10] (12.25,-5.5) rectangle (13.7,9.5);
			\draw[dashed,fill=cyan!10] (14.85,-6.5) rectangle (16.3,9.5);
			\qnode{1}{9}{$A_0$}
			\qnode{1}{8}{$0$}	
			\qnode{1}{7}{$0$}
			\qnode{1}{6}{$B_0$}
			\qnode{1}{3}{$A_1$}
			\qnode{1}{2}{$0$}
			\qnode{1}{1}{$0$}
			\qnode{1}{0}{$B_1$}
			\qgateUu[ibmqxE]{2}{8}{}
			\qgateUuu[ibmqxE]{2}{7}{{\small \begin{turn}{-90}$TPPG$\end{turn}}}
			\qgateUu[ibmqxE]{2}{2}{}
			\qgateUuu[ibmqxE]{2}{1}{{\small \begin{turn}{-90}$TPPG$\end{turn}}}
			\qnode{3}{9.23}{$A_0$}
			\qnode{3}{8.23}{$A_0B_0$}	
			\qnode{3}{7.2}{$cp_0$}
			\qnode{3}{6.23}{$B_0$}
			\qnode{3}{5}{$0$}
			\qnode{3}{4}{$0$}
			\qnode{3}{3.23}{$A_1$}
			\qnode{3}{2.23}{$A_1B_1$}
			\qnode{3}{1.2}{$cp_3$}
			\qnode{3}{0.23}{$B_1$}
			\qwire[ibmqx]{3}{9}
			\qwire[ibmqx]{3}{8}
			\qwire[ibmqx]{3}{7}
			\qwire[ibmqx]{3}{6}
			\qwire[ibmqx]{3}{3}
			\qwire[ibmqx]{3}{2}
			\qwire[ibmqx]{3}{1}
			\qwire[ibmqx]{3}{0}
			\qwire[ibmqx]{4}{9}
			\qwire[ibmqx]{4}{8}
			\qwire[ibmqx]{4}{7}
			\qwire[ibmqx]{4}{6}
			\qwire[ibmqx]{4}{3}
			\qwire[ibmqx]{4}{2}
			\qwire[ibmqx]{4}{1}
			\qwire[ibmqx]{4}{0}
			\qnode{4}{-1}{$0$}
			\qnode{4}{-2}{$0$}
			\qgateUu[ibmqxD]{4}{5}{}
			\qgateUuu[ibmqxD]{4}{4}{{\small \begin{turn}{-90}$TPPG$\end{turn}}}
			\qgateUuu[ibmqxD]{5}{8}{}
			\qgateUuu[ibmqxD]{5}{6}{}
			\qgateUuu[ibmqxD]{5}{4}{}
			\qgateUuu[ibmqxD]{5}{2}{}
			\qgateUu[ibmqxD]{5}{0}{}
			\qgateUuu[ibmqxD]{5}{-1}{{\small \begin{turn}{-90}$TPPG$\end{turn}}}
			\qwire[ibmqx]{5}{8}
			\qwire[ibmqx]{5}{7}
			\qwire[ibmqx]{5}{6}
			\qwire[ibmqx]{5}{5}
			\qwire[ibmqx]{5}{4}
			\qwire[ibmqx]{5}{3}
			\qwire[ibmqx]{5}{2}
			\qwire[ibmqx]{5}{1}
			\qwire[ibmqx]{6}{9}
			\qwire[ibmqx]{6}{8}
			\qwire[ibmqx]{6}{7}
			\qwire[ibmqx]{6}{6}
			\qwire[ibmqx]{6}{5}
			\qwire[ibmqx]{6}{4}
			\qwire[ibmqx]{6}{3}
			\qwire[ibmqx]{6}{2}
			\qwire[ibmqx]{6}{1}
			\qwire[ibmqx]{6}{0}
			\qwire[ibmqx]{6}{-1}
			\qwire[ibmqx]{6}{-2}
			\qnode{6}{7.23}{$cp_0$}
			\qnode{6}{6.23}{$B_0$}
			\qnode{6}{5.2}{$A_1B_0$}
			\qnode{6}{4.2}{$cp_1$}
			\qnode{6}{3.23}{$A_1$}
			\qnode{6}{2.23}{$A_1B_1$}
			\qnode{6}{1.23}{$cp_3$}
			\qnode{6}{0.23}{$B_1$}
			\qnode{6}{-0.77}{$A_0B_1$}
			\qnode{6}{-1.77}{$cp_2$}
			\qnode{6}{-3}{$0$}
			\qgateUuu[ibmqxC]{7}{6}{}
			\qgateUuu[ibmqxC]{7}{4}{}
			\qgateUuu[ibmqxC]{7}{2}{}
			\qgateUuu[ibmqxC]{7}{1}{}
			\qgateUu[ibmqxC]{7}{-3}{}		
			\qgateUuu[ibmqxC]{7}{-1}{{\small \begin{turn}{-90}$\ \ Block \ 1$\end{turn}}}
			\qwire[ibmqx]{7}{9}
			\qwire[ibmqx]{7}{8}
			\qwire[ibmqx]{7}{6}
			\qwire[ibmqx]{7}{4}
			\qwire[ibmqx]{7}{3}
			\qwire[ibmqx]{7}{2}
			\qwire[ibmqx]{7}{1}
			\qwire[ibmqx]{7}{0}
			\qwire[ibmqx]{7}{-2}
			\qnode{7}{-4}{$0$}
			\qgateUu[ibmqxC]{8}{3}{}
			\qgateUuu[ibmqxC]{8}{2}{}
			\qgateUuu[ibmqxC]{8}{0}{}
			\qgateUu[ibmqxC]{8}{-4}{}
			\qgateUuu[ibmqxC]{8}{-2}{{\small \begin{turn}{-90}$\ \ Block \ 2$\end{turn}}}
			\qwire[ibmqx]{8}{9}
			\qwire[ibmqx]{8}{8}
			\qwire[ibmqx]{8}{7}
			\qwire[ibmqx]{8}{6}
			\qwire[ibmqx]{8}{5}
			\qwire[ibmqx]{8}{3}
			\qwire[ibmqx]{8}{1}
			\qwire[ibmqx]{8}{0}
			\qwire[ibmqx]{8}{-1}
			\qwire[ibmqx]{8}{-3}
			\qwire[ibmqx]{9}{9}
			\qwire[ibmqx]{9}{8}
			\qwire[ibmqx]{9}{7}
			\qwire[ibmqx]{9}{6}
			\qwire[ibmqx]{9}{5}
			\qwire[ibmqx]{9}{4}
			\qwire[ibmqx]{9}{3}
			\qwire[ibmqx]{9}{2}
			\qwire[ibmqx]{9}{1}
			\qwire[ibmqx]{9}{0}
			\qwire[ibmqx]{9}{-1}
			\qwire[ibmqx]{9}{-2}
			\qwire[ibmqx]{9}{-3}
			\qwire[ibmqx]{9}{-4}
			\qnode{9}{-0.77}{$P_1$}
			\qnode{9}{-1.77}{$s_1$}
			\qnode{9}{-2.77}{$c_0$}
			\qnode{9}{-3.77}{$c_1$}
			\qnode{9}{-5}{$0$}
			\qgateUu[ibmqxF]{10}{-5}{}
			\qgateUuu[ibmqxF]{10}{-3}{{\small \begin{turn}{-90}$\ Block \ 4$\end{turn}}}
			\qwire[ibmqx]{10}{9}
			\qwire[ibmqx]{10}{8}
			\qwire[ibmqx]{10}{7}
			\qwire[ibmqx]{10}{6}
			\qwire[ibmqx]{10}{5}
			\qwire[ibmqx]{10}{4}
			\qwire[ibmqx]{10}{3}
			\qwire[ibmqx]{10}{2}
			\qwire[ibmqx]{10}{1}
			\qwire[ibmqx]{10}{0}
			\qwire[ibmqx]{10}{-1}
			\qwire[ibmqx]{10}{-4}
			\qwire[ibmqx]{11}{9}
			\qwire[ibmqx]{11}{8}
			\qwire[ibmqx]{11}{7}
			\qwire[ibmqx]{11}{6}
			\qwire[ibmqx]{11}{5}
			\qwire[ibmqx]{11}{4}
			\qwire[ibmqx]{11}{3}
			\qwire[ibmqx]{11}{2}
			\qwire[ibmqx]{11}{1}
			\qwire[ibmqx]{11}{0}
			\qwire[ibmqx]{11}{-1}
			\qwire[ibmqx]{11}{-2}
			\qwire[ibmqx]{11}{-3}
			\qwire[ibmqx]{11}{-4}
			\qwire[ibmqx]{11}{-5}
			\qnode{11}{1.23}{$cp_3$}
			\qnode{11}{-1.77}{$P_2$}
			\qnode{11}{-2.77}{$g_5$}
			\qnode{11}{-3.77}{$c_1$}
			\qnode{11}{-4.77}{$c_2$}
			\qnode{11}{-6}{$0$}
			\qgateUu[ibmqxB]{12}{0}{}
			\qgateUuu[ibmqxB]{12}{-1}{}
			\qgateUuu[ibmqxB]{12}{-3}{}
			\qgateUuu[ibmqxB]{12}{-5}{{\small \begin{turn}{-90} $ Block \ 3$\end{turn}}}
			\qwire[ibmqx]{12}{9}
			\qwire[ibmqx]{12}{8}
			\qwire[ibmqx]{12}{7}
			\qwire[ibmqx]{12}{6}
			\qwire[ibmqx]{12}{5}
			\qwire[ibmqx]{12}{4}
			\qwire[ibmqx]{12}{3}
			\qwire[ibmqx]{12}{2}
			\qwire[ibmqx]{12}{0}
			\qwire[ibmqx]{12}{-1}
			\qwire[ibmqx]{12}{-2}
			\qwire[ibmqx]{12}{-3}
			\qnode{13}{9}{$A_0$}
			\qnode{13}{8}{$\bm{P_0}$}
			\qnode{13}{7}{$g_0$}
			\qnode{13}{6}{$B_0$}
			\qnode{13}{5}{$g_1$}
			\qnode{13}{4}{$g_2$}
			\qnode{13}{3}{$A_1$}
			\qnode{13}{2}{$g_3$}
			\qnode{13}{1}{$g_4$}
			\qnode{13}{0}{$B_1$}
			\qnode{13}{-1}{$\bm{P_1}$}
			\qnode{13}{-2}{$\bm{P_2}$}
			\qnode{13}{-3}{$g_6$}
			\qnode{13}{-4}{$g_5$}
			\qnode{13}{-5}{$\bm{P_3}$}
			\qnode{13}{-6}{$\bm{c_{out}}$}
	\end{tikzpicture}}
	\caption[Ternary two-qutrit multiplier is implemented using the Adder Blocks and TPPG circuits.]{Ternary two-qutrit multiplier is implemented using the Adder Blocks and TPPG circuits \cite{panahi2019novel}.}
	\label{MultiFull}
\end{figure}
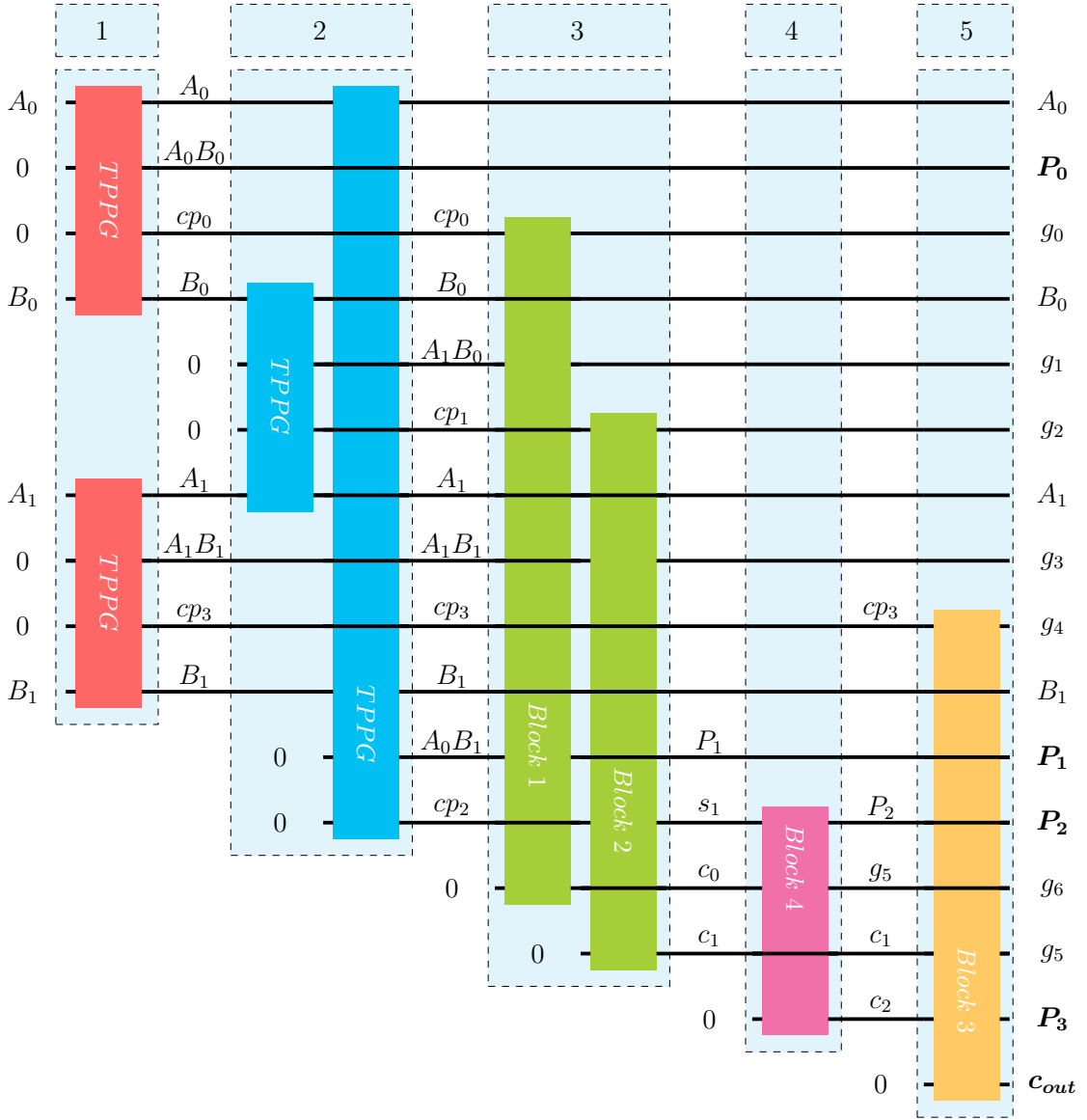

The full two-digit qutrit multiplication circuit is shown in Fig. \ref{MultiFull}. When the line goes on the top of a Block, then it is non-interacting, but when a line goes below the Block, then it is given to that Block as the input line.
In the Fig., the upper TPPG at stage 1 gives a partial product $P_0=A_0B_0$ and the carry $cp_0$.
The lower TPPG at the stage 1 computes $A_1B_1$, and we have the carry $cp_3$.
At stage 2, the first TPPG computes partial product $A_1B_0$ and produces the carry $cp_1$, whereas the second TPPG computes $A_0B_1$ and produces the carry $cp_2$. Inputs of the first TPPG are $B_0$ and $A_1$ and two ancillae, whereas the inputs of the second TPPG are $A_0$ and $B_1$.
The stages 3, 4 and 5 consist of Adder Blocks. As we can see in Fig. \ref{2-Multiply}, we need to add $cp_0$, $A_0B_1$ and $A_1B_0$ to get $P_1$ and the carry $c_0$. This is done by Block 1 at stage 3. At this stage, the Adder Block 2 adds $cp_1,A_1B_1$ and $cp_2$. The output of this block is $s_1$ and the carry is $c_1$. At stage 4, the sum $s_1$ is added to $c_0$ by using the Block 4 and the output $P_2$ is obtained. The Block 4 has a carry $c_2$. At stage 5, $cp_3$, $c_1$ and $c_2$ are added using the Block 3 to get an output qutrit $P_3$ and the carry $c_{out}$.

\section{Conclusion}\label{Conclusion}
The topological quantum computation is a promising candidate for fault-tolerant quantum computation. Our main focus in this article is to find the ternary arithmetic circuits implementation in topological quantum computation.
We explained how to build the fault-tolerant quantum gates by using the ideas based on topology and the braiding of non-Abelian anyons. The Hilbert space, fusion, and braiding matrices are computed for a topological qubit. 
Metaplectic anyons are discussed and fusion and braiding matrices for ternary logic are obtained by the quantum deformation of the recoupling theory. The universal set of gates cannot be obtained by braiding alone, it needs to be combined with the measurement gates. The quantum ternary arithmetic circuits in this paper can be realized by braiding the metaplectic anyons and the topological charge measurement gates.

\appendix
\section{Quantum Computing}\label{QC}
The time evolution of a state in quantum mechanics is represented by the unitary time evolution operator $U(t)$ such that $U^{-1}=U^\dagger$ and $U(t)U^\dagger(t)= U^\dagger(t)U(t)=I$. It is written as $U(t)=\exp(-iHt)$, where $H$ is the Hamiltonian operator corresponds to the energy eigenvalues. The unitary evolution relates the changes in the state to the energy of the system. These changes are reversible so that the inner product is preserved. That means, the phase can be changed but the amplitude remains the same with time.
When the initial state $\ket{\psi_i}$ evolves unitarily to the final state $\ket{\psi_f}$, it is written as 
\eq{\ket{\psi_f}=U(t)\ket{\psi_i}.}
\textit{A quantum computation model involves three steps; initialization, unitary evolution, and measurement \cite{divincenzo2000physical}. The initial state is an input state and the final state is an output state \cite{Nielsen2002quantum} of a quantum gate. The evolution operator $U(t)$ corresponds to a quantum gate. The readout is a measurement in certain bases that gives a classical result.} For basic study on quantum computing, see the books \cite{yanofsky2008quantum,mcmahon2007quantum,mermin2007quantum}, and for technical details, see \cite{Nielsen2002quantum,marinescu2011classical}.
\subsection*{A.1. Binary Quantum Gates}
A classical bit has a value 0 or 1. A qubit is a superposition of 0 and 1, written as
\eq{\ket{\psi}= \alpha\ket{0} + \beta\ket{1}.} 
A qubit can be written in a matrix form as
\eq{\ket{\psi} = \alpha \begin{pmatrix}
		1 \\
		0
	\end{pmatrix}+ \beta \begin{pmatrix}
		0 \\
		1
	\end{pmatrix} = \begin{pmatrix}
		\alpha \\
		\beta
	\end{pmatrix},}
where $\alpha$ and $\beta$ are complex numbers. The sum of their squares is one, $\abs{\alpha}^2+\abs{\beta}^2 =1$, which means that the sum of probabilities is equal to one. A qubit can be made by any two-level quantum mechanical system. For example, a spin-half particle can be in a superposition state of spin-up state $\ket{0}$ and spin down spin state $\ket{1}$, a photon can be in a superposition of two polarization states, or an atom can be in a superposition of the ground state and the excited state. The states $\ket{0}$ and $\ket{1}$ are the eigenvectors of a system. These eigenvectors $\ket{0}$ and $\ket{1}$ provide the bases for a qubit state. These bases are orthonormal, that is $\bra{i}\ket{j}=\delta_{ij}$, where $i,j=\{0,1\}$. When $\ket{i}$ is a column vector, $\bra{i}$ is a row vector. The superposition allows us to do many calculations in parallel. For $n$ qubits, a state is written as a $2^n$-dimensional vector in a Hilbert space $\cal H$ and the qubits can be entangled.

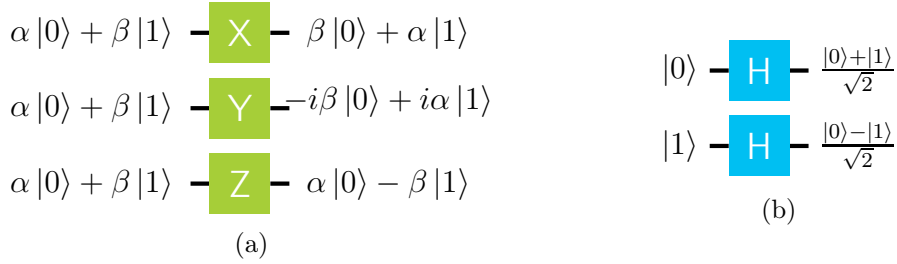
\begin{figure}[h!]
	\centering
	\begin{subfigure}{0.4\textwidth}
		\centering
		\begin{tikzpicture}
			\qnode{0.5}{2}{$\alpha\ket{0}+\beta \ket{1}$}
			\qgateX[ibmqxA]{2}{2}
			\qnode{3.5}{2}{$\beta \ket{0} + \alpha \ket{1}$}
			\qnode{0.5}{1}{$\alpha\ket{0}+\beta \ket{1}$}
			\qgateY[ibmqxE]{2}{1}
			\qnode{3.5}{1.1}{$-i\beta \ket{0} + i\alpha \ket{1}$}
			\qnode{0.5}{0}{$\alpha\ket{0}+\beta \ket{1}$}
			\qgateZ[ibmqxC]{2}{0}
			\qnode{3.5}{0}{$\alpha\ket{0} - \beta\ket{1}$}
		\end{tikzpicture}
		\caption{}
	\end{subfigure}
	\begin{subfigure}{0.4\textwidth}
		\centering
		\begin{tikzpicture}
			\qnode{0.2}{1}{$\ket{0}$}
			\qgateH[ibmqxD]{1}{1}
			\qnode{2}{1}{$\frac{\ket{0} + \ket{1}}{\sqrt{2}}$}
			\qnode{0.2}{0}{$\ket{1}$}
			\qgateH[ibmqxA]{1}{0}
			\qnode{2}{0}{$\frac{\ket{0} - \ket{1}}{\sqrt{2}}$}
		\end{tikzpicture}
		\caption{}
	\end{subfigure}	
	\caption[Implementation of Pauli gates, Hadamard gate, and CNOT gate.]{The implementation of (a) Pauli gates, (b) Hadamard gate.}
	\label{PauliH}
\end{figure}

The purpose of quantum gates and circuits is to get the required output with maximum probability. Mathematically, a gate is represented by a matrix that must be unitary. The matrix elements correspond to the probabilities of getting the respective basis state. Two matrices do not commute in general.
Some elementary gates are represented by symbols $I, X, Y, Z$. These are called \textit{Pauli matrices} in physics and denoted as $\sigma_I,\sigma_x,\sigma_y, \sigma_z$. These matrices have the effect of rotating the qubit about the z-axis by an angle $\theta$.
\eq{I = \begin{pmatrix}
		1 & 0\\
		0 & 1
	\end{pmatrix} \ , \ X =
	\begin{pmatrix}
		0 & 1\\
		1 & 0
	\end{pmatrix} \ , \ Y =
	\begin{pmatrix}
		0 & -i\\
		i & 0
	\end{pmatrix} \ , \ Z =
	\begin{pmatrix}
		1 & 0\\
		0 & -1
	\end{pmatrix},}
where $X$ is the NOT gate, $Z$ is the phase gate and $Y$ is the phase and NOT gate together, that is $Y = iXZ$. These matrices have properties that $X^2=Y^2=Z^2=I$ and $XY=iZ, \ YZ=iX, \ ZX=iY$. The superposition is created by the \textit{Hadamard gate},
\eq{H\ket{\psi} = \frac{1}{\sqrt{2}}\begin{pmatrix}
		1 & 1 \\
		1 & -1
	\end{pmatrix} \begin{pmatrix}
		\alpha \\
		\beta
	\end{pmatrix} = \frac{1}{\sqrt{2}} \begin{pmatrix}
		\alpha + \beta \\
		\alpha - \beta
	\end{pmatrix}.}
The implementation of these gates is shown in Fig. \ref{PauliH} (a) and (b).
Another example of one-qubit gates is \textit{phase gate} that can be written as
\eq{P(\phi)= \begin{pmatrix}
		1 & 0\\
		0 & e^{i\phi}
	\end{pmatrix}.}
Applying on the state ket $\ket{\psi}$, we get
\eq{P\ket{\psi} = \ket{0}+e^{i\phi} \ket{1}=\begin{pmatrix}
		\alpha\\
		e^{i\phi}\beta
	\end{pmatrix}.}
When $\phi=\pi$ we have the Pauli matrix $Z$, that is $P(\pi)=Z$. Other examples of phase gates are $S$ and $T$ gates written as
\eq{S=P(\pi/2)=\begin{pmatrix}
		1 & 0\\
		0 & i
	\end{pmatrix}, \qquad T = e^{i\pi/8}\begin{pmatrix}
		e^{-i\pi/8} & 0\\
		0 & e^{i\pi/8}
	\end{pmatrix}=\begin{pmatrix}
		1 & 0\\
		0 & e^{i\pi/4}
	\end{pmatrix}.}
We can see $T^2=S$. $T$ is also known as $\pi/8$ gate.

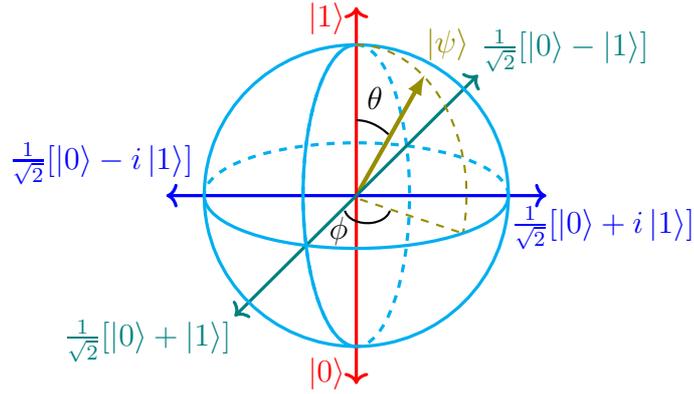
\begin{figure}
	\centering
	\begin{tikzpicture}
		\draw [very thick,cyan,dashed] (2,0) arc(0:180:2cm and 0.7cm);
		\draw [very thick,cyan,dashed] (0,2) arc(90:-270:0.7cm and 2cm);
		\draw[ultra thick,-{latex[scale=3.0]},olive] (0,0)--(0.9,1.6);
		
		\draw[very thick,->,blue] (0,0)--(2.5,0);
		\draw[very thick,->,blue] (0,0)--(-2.5,0);
		\draw[very thick,->,red] (0,0)--(0,2.5);
		\draw[very thick,->,red] (0,0)--(0,-2.5);
		
		\draw[very thick,->,teal] (0,0)--(-1.6,-1.6);
		\draw [very thick,cyan] (0,2) arc(90:270:0.7cm and 2cm);
		\draw [very thick,cyan] (2,0) arc(0:-180:2cm and 0.7cm);
		\draw[very thick,->,teal] (0,0)--(1.6,1.6);
		\draw [very thick,cyan] (0,0) ellipse (2 and 2);
		
		\node [above left,red] at (0,2) {$\ket{1}$};
		\node [below left,red] at (0,-2) {$\ket{0}$};
		\node [below right,blue] at (1.9,0) {$\frac{1}{\sqrt{2}}[\ket{0}+i\ket{1}]$};
		\node [above left,blue] at (-2,0) {$\frac{1}{\sqrt{2}}[\ket{0}-i\ket{1}]$};
		\node [above right,teal] at (1.5,1.5) {$\frac{1}{\sqrt{2}}[\ket{0}-\ket{1}]$};
		\node [below left,teal] at (-1.5,-1.5) {$\frac{1}{\sqrt{2}}[\ket{0}+\ket{1}]$};
		
		\draw[thick,dashed,olive] (0.05,-0.05)--(1.4,-0.5);
		\draw [thick,dashed,olive] (0,2) arc(90:-14:1.45cm and 2cm);
		\draw [thick] (0,1) arc(90:65:1cm and 2cm);
		\node [above right] at (0,1) {$\theta$};
		\draw [thick] (-0.15,-0.2) arc(190:350:0.3cm and 0.2cm);
		\node [above right] at (-0.5,-0.8) {$\phi$};
		\node [olive] at (1.2,2) {$\ket{\psi}$};	
	\end{tikzpicture}
	\caption{One-qubit state can be represented by a Bloch sphere.}
	\label{Bloch}	
\end{figure}

The one-qubit state can be represented by the Bloch sphere as shown in Fig. \ref{Bloch}. The general state on the Bloch sphere is given by \cite{Nielsen2002quantum}
\eq{\ket{\psi} = \cos \frac{\theta}{2} \ket{0} + e^{i\phi}\sin\frac{\theta}{2} \ket{1} = \begin{pmatrix}
		\cos\frac{\theta}{2} \\
		e^{i\phi}\sin\frac{\theta}{2}
	\end{pmatrix}, \ \text{where} \ 0 \le \theta \le \pi, \ \ 0 \le \phi \le 2\pi .}
The operators that rotate the state on the Bloch sphere can be written as
\eq{&R_x(\theta) \equiv e^{-i\frac{\theta}{2} X} = \cos \frac{\theta}{2} I + i\sin\frac{\theta}{2} X = \begin{pmatrix}
		\cos \frac{\theta}{2} & -i\sin\frac{\theta}{2}\\
		-i\sin\frac{\theta}{2} & \cos \frac{\theta}{2}
	\end{pmatrix},\nonumber \\
	&R_y(\theta) \equiv e^{-i\frac{\theta}{2} Y} = \cos \frac{\theta}{2} I + i\sin\frac{\theta}{2} Y = \begin{pmatrix}
		\cos \frac{\theta}{2} & -\sin\frac{\theta}{2}\\
		\sin\frac{\theta}{2} & \cos \frac{\theta}{2}
	\end{pmatrix},\nonumber \\
	&R_z(\theta) \equiv e^{-i\frac{\theta}{2} Z} = \cos \frac{\theta}{2} I + i\sin\frac{\theta}{2} Z = \begin{pmatrix}
		e^{-i \frac{\theta}{2}} & 0\\
		0 & e^{i \frac{\theta}{2}}
	\end{pmatrix}.}
These three rotations on a Bloch sphere are combined into a general rotation as
\eq{R_{\hat{n}}(\theta) = \exp(-i \frac{\theta}{2} \hat{n}\cdot \vec{\sigma}) = \cos \frac{\theta}{2} I - i\sin\frac{\theta}{2}\Big(n_xX + n_yY + n_zZ\Big),}
where $\hat{n}$ is a unit vector in three dimensions and $\vec{\sigma}$ is the three-component vector of Pauli matrices. $R_{\hat{n}}(\theta)$ is the effect of rotation on the state around the unit vector $\hat{n}$. Let there exist real numbers $\alpha$, $\beta$, $\gamma$ and $\delta$ such that a general unitary operation on a qubit \cite{Nielsen2002quantum} can be written as
\eq{U = e^{i\alpha}R_z(\beta)R_y(\gamma)R_z(\delta) = \begin{pmatrix}
		e^{i(\alpha - \beta/2 - \delta/2)}\cos\frac{\gamma}{2} & -e^{i(\alpha - \beta/2 + \delta/2)}\sin\frac{\gamma}{2} \\
		e^{i(\alpha + \beta/2 - \delta/2)}\sin\frac{\gamma}{2} & e^{i(\alpha + \beta/2 + \delta/2)}\cos\frac{\gamma}{2}
	\end{pmatrix}.}
The Bloch sphere representation is limited to a single qubit state only.

Two-qubit states can be \textit{separable or inseparable}. The separable states are written for independent composite systems, whereas the inseparable states are entangled state. The separable state, also called product state, can be factorized into two separate states and written as the \textit{tensor product} of the two states as
\eq{\ket{\psi} =& \ket{\phi_1} \otimes \ket{{\phi_2}}  
	= \Big(\alpha_1 \ket{0} + \beta_1 \ket{1} \Big)\otimes \Big(\alpha_2 \ket{0} + \beta_2 \ket{1}\Big) \nonumber \\
	=& \alpha_1 \alpha_2 \ket{0}\ket{0} + \alpha_1 \beta_2\ket{0}\ket{1} 
	+\beta_1 \alpha_2 \ket{1}\ket{0} + \beta_1 \beta_2 \ket{1}\ket{2} \nonumber \\
	=& \alpha \ket{00} + \beta\ket{01} + \gamma\ket{10} + \delta \ket{11},}
where $\abs{\alpha}^2 + \abs{\beta}^2 + \abs{\gamma}^2 + \abs{\delta}^2 = 1$. The tensor product notation $\otimes$ means that we multiply each term of the first vector to each term of the second vector. The dimension of the Hilbert space of the product state is a product of the dimensions of the two systems, that is $\mathcal{H}=\mathcal{H}_1\otimes \mathcal{H}_2$. Since the gates are operators and we write them as matrices, the tensor product of two operators is expressed in such a way that multiply each entry of the first matrix to all entries of the second matrix. As an example, let us consider two matrices $A$ and $B$ as
$$A=\begin{pmatrix}
	a_1 & a_2\\
	a_3 & a_4
\end{pmatrix}, \qquad B=\begin{pmatrix}
	b_1 & b_2\\
	b_3 & b_4
\end{pmatrix}.$$
The tensor product of $A$ and $B$ is given as
\eq{A\otimes B = \begin{pmatrix}
		a_1\begin{pmatrix}
			b_1 & b_2\\
			b_3 & b_4
		\end{pmatrix} & a_2 \begin{pmatrix}
			b_1 & b_2\\
			b_3 & b_4
		\end{pmatrix}\\
		a_3\begin{pmatrix}
			b_1 & b_2\\
			b_3 & b_4
		\end{pmatrix} & a_4 \begin{pmatrix}
			b_1 & b_2\\
			b_3 & b_4
		\end{pmatrix}
	\end{pmatrix}=\begin{pmatrix}
		a_1b_1 & a_1b_2 & a_2b_1 & a_2b_2\\
		a_1b_3 & a_1b_4 & a_2b_3 & a_2b_4\\
		a_3b_1 & a_3b_2 & a_4b_1 & a_4b_2\\
		a_3b_3 & a_3b_4 & a_4b_3 & a_4b_4
	\end{pmatrix}.}
When $A\ket{a}=\alpha \ket{a}$ and $B\ket{b}=\beta \ket{b}$ then the following rules are defined for the tensor product
\eq{&(A\otimes B)(\ket{a}\otimes \ket{b}) = A\ket{a} \otimes B\ket{b},\nonumber\\	
	&(\ket{a}+\ket{b})\otimes \ket{c}= \ket{a}\otimes \ket{c}+\ket{b}\otimes \ket{c},\nonumber\\
	&\ket{a}\otimes(\ket{b}+ \ket{c})= \ket{a}\otimes \ket{b}+\ket{a}\otimes \ket{c}.}
We also have the notations $\ket{a}\otimes \ket{b}\equiv \ket{a}\ket{b} \equiv \ket{ab}$. If $\ket{a}$ and $\ket{b}$ are column vectors of two elements each, then $\ket{ab}$ is a column vector of four elements. Corresponding operators $A$ and $B$ would become a four by four matrix $A\otimes B$. It can be generalized to $n$-dimensional Hilbert space. \textit{The number of states is increased exponentially with the increase of the number of qubits.}

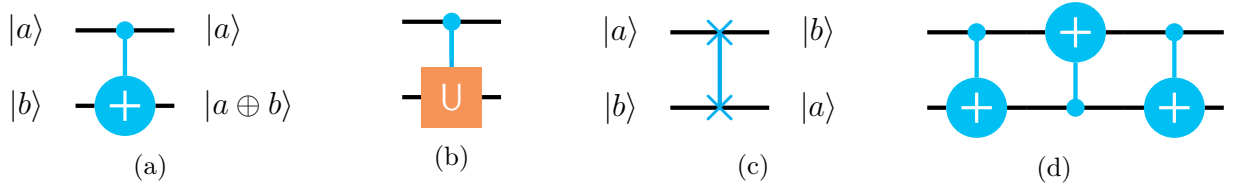
\begin{figure}[h!]
	\begin{subfigure}{0.25 \textwidth}
		\centering
		\begin{tikzpicture}
			\qnode{0}{1}{$\ket{a}$}
			\qnode{0}{0}{$\ket{b}$}
			\qgateCNC[ibmqx]{b}{1}{1}
			\qgateCNX[ibmqx]{t}{1}{0}
			\qnode{2}{1}{$\ket{a}$}
			\qnode{2.25}{0}{$\ket{a\oplus b}$}
		\end{tikzpicture}
		\caption{}
	\end{subfigure}
	\begin{subfigure}{0.2\textwidth}
		\centering
		\begin{tikzpicture}
			\draw[ultra thick,cyan] (0,0)--(0,1);
			\qgateCNC[ibmqx]{b}{0}{1}
			\qgateU[ibmqxA]{0}{0}{U}
		\end{tikzpicture}
		\caption{}
	\end{subfigure}
	\begin{subfigure}{0.25\textwidth}
		\begin{tikzpicture}
			\draw[ultra thick,cyan] (0,0)--(0,1);	
			\qwire[ibmqx]{0}{1}
			\qwire[ibmqx]{0}{0}
			\qnode{0}{1}{\Large\textcolor{cyan}{\textbf{$\cross$}}}
			\qnode{0}{0}{\Large\textcolor{cyan}{\textbf{$\cross$}}}
			\qnode{-1}{1}{$\ket{a}$}
			\qnode{-1}{0}{$\ket{b}$}
			\qnode{1}{1}{$\ket{b}$}
			\qnode{1}{0}{$\ket{a}$}
		\end{tikzpicture}
		\caption{}
	\end{subfigure}
	\begin{subfigure}{0.2\textwidth}
		\begin{tikzpicture}
			\qgateCNC[ibmqx]{b}{1}{1}
			\qgateCNX[ibmqx]{t}{1}{0}
			\qgateCNC[ibmqx]{t}{2}{0}
			\qgateCNX[ibmqx]{b}{2}{1}
			\qgateCNC[ibmqx]{b}{3}{1}
			\qgateCNX[ibmqx]{t}{3}{0}
		\end{tikzpicture}
		\caption{}
	\end{subfigure}
	\caption[Controlled-NOT, Controlled-U, SWAP gate, and its physical realization.]{(a) CNOT gate (b) Controlled-U gate and (c) SWAP gate (d) Physical realization of SWAP gate.}
	\label{TwoQubit}
\end{figure}

A typical example of a two-qubit gate is the \textit{controlled-NOT or CNOT gate} shown in Fig. \ref{TwoQubit} (a). This gate flips the second qubit when the first qubit state is $\ket{1}$. The first qubit is called \textit{control qubit} and the second qubit is called \textit{target qubit}. This gate is a classical analog of exclusive-OR gate based on exclusive-OR logic represented by $\oplus$. The symbol $\oplus$ is defined such that the output state $\ket{x\oplus y}$ will give a value 0 when both inputs are either zero or 1, but the output value will be 1 when one of the inputs is 1.
A two-qubit state
\eq{\ket{\psi} &= \alpha \ket{00} + \beta\ket{01} + \gamma\ket{10} + \delta \ket{11}}
is changed by the operation of CNOT gate as
\eq{CNOT\ket{\psi} &= \begin{pmatrix}
		1 & 0 & 0 & 0 \\
		0 & 1 & 0 & 0 \\
		0 & 0 & 0 & 1 \\
		0 & 0 & 1 & 0 
	\end{pmatrix}
	\begin{pmatrix}
		\alpha \\
		\beta \\
		\gamma \\
		\delta
	\end{pmatrix} = \begin{pmatrix}
		\alpha \\
		\beta \\
		\delta \\
		\gamma
	\end{pmatrix} \nonumber\\
	&= \alpha \ket{00} + \beta\ket{01} + \delta\ket{10} + \gamma \ket{11}.}
The second qubit remains the same when the first qubit is $\ket{0}$, whereas $X$ gate is applied to the target qubit when the control state is $\ket{1}$. But in general, there can be any one-qubit gate at the place of $X$ as shown in Fig. \ref{TwoQubit} (b). In that case, we can write the controlled-U (CU) gate as
\eq{CU =\begin{pmatrix}
		1 & 0 & 0 & 0\\
		0 & 1 & 0 & 0\\
		0 & 0 & u_1 & u_2\\
		0 & 0 & u_3 & u_4
	\end{pmatrix}.}

Another two-qubit gate is a SWAP gate that swaps the states of input qubits. The SWAP gate and its physical realization are shown in Fig. \ref{TwoQubit} (c) and (d). It can also be written in matrix notation as
\eq{SWAP=\begin{pmatrix}
		1 & 0 & 0 & 0\\
		0 & 0 & 1 & 0\\
		0 & 1 & 0 & 0\\
		0 & 0 & 0 & 1
	\end{pmatrix}.}

\begin{figure}[h!]
	\centering
	\begin{tikzpicture}
		\qnode{-1}{1}{$\ket{0}$}
		\qnode{-1}{0}{$\ket{0}$}
		\qgateH[ibmqxE]{0}{1}
		\qwire[ibmqx]{0}{0}
		\qgateCNC[ibmqx]{b}{1}{1}
		\qgateCNX[ibmqx]{t}{1}{0}
		\qnode{2}{0.5}{$\ket{\Psi^+}$}
	\end{tikzpicture}
	\caption{The entanglement generating gate.}
	\label{BellState}
\end{figure}
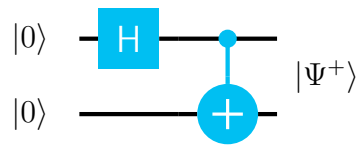

The computation cannot be performed by a single qubit only. A system should consist of several qubits and should have the capability to entangle these qubits. One qubit is a superposition of two basis states, but when there is a quantum correlation among two systems, then we say that these systems are entangled. The entangled state is non-separable and cannot mathematically be factorized into two separate superposition states. The Bell's state
\eq{\ket{\Psi^+} = \alpha_{00}\ket{00} + \alpha_{11}\ket{11},}
with $\abs{\alpha_{00}}^2 + \abs{\alpha_{11}}^2 =1$, is an example of the entangled state. It is non-local, that is, the information of only one qubit is not accessible locally when two states are far apart. Classically, the first qubit can be in state $\ket{0}$ or $\ket{1}$, so can be the second qubit. Therefore, there are four possibilities of values on measurement. But an entangled state like this Bell's state would give $\ket{00}$ or $\ket{11}$ with the probabilities $\abs{\alpha_{00}}^2$ and $\abs{\alpha_{11}}^2$. On measurement, if the first qubit collapse to the state $\ket{0}$ $(\ket{1})$ then the second qubit is forced to collapse to $\ket{0}$ $(\ket{1})$. The state is maximally entangled when $\abs{\alpha_{00}}^2 = \abs{\alpha_{11}}^2 = \frac{1}{2}$. The entangled state is created in a process that can be shown as in Fig. \ref{BellState}.

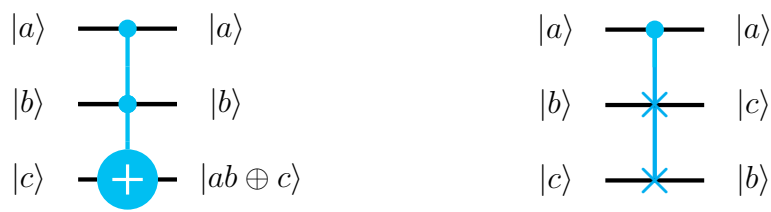
\begin{figure}[h!]
	\centering
	\begin{subfigure}{0.4\textwidth}
		\begin{tikzpicture}
			\qnode{0}{2}{$\ket{a}$}
			\qnode{0}{1}{$\ket{b}$}
			\qnode{0}{0}{$\ket{c}$}
			\qgateCNC[ibmqx]{b}{1}{2}
			\qgateCNC[ibmqx]{b}{1}{1}
			\qgateCNX[ibmqx]{t}{1}{0}
			\qgateCNC[ibmqx]{t}{1}{1}
			\qnode{2}{2}{$\ket{a}$}
			\qnode{2}{1}{$\ket{b}$}
			\qnode{2.25}{0}{$\ket{ab\oplus c}$}
		\end{tikzpicture}
	\end{subfigure}
	\begin{subfigure}{0.2\textwidth}
		\begin{tikzpicture}
			\draw[ultra thick,cyan] (0,0)--(0,2);
			\qgateCNC[ibmqx]{b}{0}{2}	
			\qwire[ibmqx]{0}{1}
			\qwire[ibmqx]{0}{0}
			\qnode{0}{1}{\Large\textcolor{cyan}{\textbf{$\cross$}}}
			\qnode{0}{0}{\Large\textcolor{cyan}{\textbf{$\cross$}}}
			\qnode{-1}{2}{$\ket{a}$}
			\qnode{-1}{1}{$\ket{b}$}
			\qnode{-1}{0}{$\ket{c}$}
			\qnode{1}{2}{$\ket{a}$}
			\qnode{1}{1}{$\ket{c}$}
			\qnode{1}{0}{$\ket{b}$}
		\end{tikzpicture}
	\end{subfigure}
	\caption[Toffoli and Fredkin gates.]{(a) Toffoli and (b) Fredkin gates.}
	\label{Toffoli}
\end{figure}

Toffoli and Fredkin gates shown in Fig. \ref{Toffoli} are examples of three-qubit gates. Toffoli gate, also known as controlled-controlled-NOT or CCNOT, consists of two control qubits and one target qubit. When the first and second qubits will be in state $\ket{1}$, then the NOT gate $X$ will be applied to the third gate, but nothing will happen in all other cases. For the Fredkin gate, a SWAP gate is applied on the second and third qubit when the first one is in state $\ket{1}$, nothing will happen otherwise. Therefore, this gate is also known as the controlled-SWAP gate.

Gates are drawn from left to right in a diagram but mathematically they appear in order from right to left. For example, the gate in Fig. \ref{BellState} is written as
\eq{\ket{\Psi^+} = CNOT(H\otimes I)\ket{00}.}

The set of elementary gates used to perform all kinds of computations is called the \textit{universal set of gates} \cite{barenco1995elementary}. There are several universal sets of gates. Hadamard, $\pi/8$, and a CNOT gate can make one of the sets of universal quantum gates \cite{Nielsen2002quantum}. The number of gates used to implement a circuit is called \textit{quantum cost} of the circuit.
Quantum algorithms are used to solve a particular problem. The most popular quantum algorithms are Shor's factoring algorithm \cite{lavor2003shor} and Grover search algorithm \cite{lavor2003grover}. The former is an exponential speedup over classical factoring algorithm, and the latter is a quadratic speedup.

The gates and circuits in quantum computing are made \textit{reversible}.
The reversible circuits are the ones with the same number of outputs as the number of inputs, and there is also a one-to-one correspondence between input and output states. Landauer \cite{landauer1961irreversibility} proved that the minimum energy dissipation for the processing of information is  $KT \ln 2$.  
Bennett et al. \cite{bennett1973logical} proposed that the energy dissipation can be avoided if the information processing is made reversible.

The main factors in designing the reversible circuits are the total quantum cost, total hardware complexity, number of shift gates, number of MS gates, and delay time. The total quantum cost refers to the number of ternary shift gates required to realize the circuit. The total hardware complexity is the complexity of the circuit in which $\varepsilon$ denotes a ternary one-qutrit shift gate and $\gamma$ denotes a two-qutrit MS gate. The number of constant inputs and the number of unutilized garbage outputs is also one of the factors sometimes taken into consideration. The delay time indicated by $\Delta$ is the logical depth of the circuit. It is 1 for the shift gates and MS gates. Shift gates and MS gates have quantum cost unity \cite{asadi2020efficient,monfared2017design}.

The conditions necessary for constructing a quantum computer are known as \textit{DiVincenzo' criteria} \cite{divincenzo2000physical}. Five requirements for quantum computations are scalability of the physical system, ability to initialize the qubits in a particular basis state, long decoherence time, universal set of gates, and measurement capability. Two more conditions are added for quantum communication, which are the ability to interconvert stationary and flying qubits, and the ability to transmit the qubits between two locations.
Some physical systems used to construct a quantum computer are: ion trap, neutral atoms trapped in an optical lattice, superconductors, quantum dots, nitrogen vacancy centers in diamond, optical, and topological.

\subsection*{A.2. Ternary Quantum Gates}
In quantum technologies, hybrid circuits are sometimes employed, which are a combination of binary and multivalued circuits. Such gates and corresponding circuits may be advantageous in some ways, such as the reduction in inputs and outputs, reduction in the quantum cost, and the complexity of interconnects. Ternary logic is the most popular multi-value logic. The basic unit of information for multivalued logic is called a \textit{qudit} and that of ternary logic is called a \textit{qutrit}.
Khan and Perkowski \cite{khan2007quantum} showed that the ternary logic needs fewer gates comparing with its respective binary system. A binary quantum system requires $n_2 = \log_2N$ qubits for a Hilbert space of dimensions $N$. On the other hand, an $m$-valued quantum system requires $n_m = \log_mN$ qudits, we have 
\eq{n_m = \log_mN = \frac{\log_2N}{\log_2m} = \frac{n_2}{\log_2m}.}
Therefore, an $m$-valued quantum system requires $1/\log_2m$ times the memory of its binary counterpart. Hence, a logarithmic reduction in the number of qudits for an $m$-valued logic. It is also shown in \cite{khan2007quantum} that $m=3$ is the most favorable choice. Haghparast et al. \cite{haghparast2017towards} proved that the ternary is $37\%$ more compact than binary. Therefore, by using ternary logic gates, we can reduce the cost of circuits and make them more efficient.

In ternary quantum logic, the state $\ket{0}$, $\ket{1}$ and $\ket{2}$ are computational bases. A state can be in a superposition of these three basis states, and is written as 
\eq{\phi = \alpha \ket{0} + \beta \ket{1} + \gamma\ket{2},}
with $\alpha$, $\beta$ and $\gamma$ being complex numbers such that $\abs{\alpha}^2 + \abs{\beta}^2 + \abs{\gamma}^2 = 1$. The state vector $\ket{\psi}$ is a three-dimensional column vector
\eq{\ket{\psi}=\begin{pmatrix}
		\alpha\\\beta\\\gamma
	\end{pmatrix} =\alpha\begin{pmatrix}
		1\\0\\0
	\end{pmatrix}+\beta\begin{pmatrix}
		0\\1\\0
	\end{pmatrix}+\gamma\begin{pmatrix}
		0\\0\\1
	\end{pmatrix}.} 

The elementary gates for ternary logic are $3\cross 3$ unitary matrices \cite{khan2006design,khan2007quantum,di2011elementary,shah2010design,giesecke2006ternary}. The qutrit gates and their symbols are shown in Fig. \ref{1QutritGates}, where $Z_3(+1)$ shifts the qutrit state by 1 and $Z_3(+2)$ gate shifts the qutrit state by 2. $Z_3(01)$, $Z_3(12)$ and $Z_3(02)$ permute the states $\ket{0}$ and $\ket{1}$, $\ket{1}$ and $\ket{2}$, and $\ket{0}$ and $\ket{2}$ respectively \cite{giesecke2006ternary,khan2007quantum}. 

\begin{figure}[h!]
	\centering
	\begin{tikzpicture}
		\qgateU[ibmqxA]{0}{0}{+1}
		\qnode{0}{-1.5}{$\begin{pmatrix}
				0 & 0 & 1\\
				1 & 0 & 0\\
				0 & 1 & 0
			\end{pmatrix}$}
	\end{tikzpicture}
	\begin{tikzpicture}
		\qgateU[ibmqxA]{0}{0}{+2}
		\qnode{0}{-1.5}{$\begin{pmatrix}
				0 & 1 & 0\\
				0 & 0 & 1\\
				1 & 0 & 0
			\end{pmatrix}$}
	\end{tikzpicture}
	\begin{tikzpicture}
		\qgateU[ibmqxD]{0}{0}{01}
		\qnode{0}{-1.5}{$\begin{pmatrix}
				0 & 1 & 0\\
				1 & 0 & 0\\
				0 & 0 & 1
			\end{pmatrix}$}
	\end{tikzpicture}
	\begin{tikzpicture}
		\qgateU[ibmqxD]{0}{0}{12}
		\qnode{0}{-1.5}{$\begin{pmatrix}
				1 & 0 & 0\\
				0 & 0 & 1\\
				0 & 1 & 0
			\end{pmatrix}$}
	\end{tikzpicture}
	\begin{tikzpicture}
		\qgateU[ibmqxD]{0}{0}{02}
		\qnode{0}{-1.5}{$\begin{pmatrix}
				0 & 0 & 1\\
				0 & 1 & 0\\
				1 & 0 & 0
			\end{pmatrix}$}
	\end{tikzpicture}	
	\caption{Quantum one-qutrit gates.}
	\label{1QutritGates}
\end{figure}

Analogous to the Hadamard in binary, there is a Chrestenson transform that creates a superposition state from the bases states and is written as \cite{al2002multiple,moraga2014some}
\eq{CH=\frac{1}{\sqrt{3}}\begin{pmatrix}
		1 &1 & 1\\
		1 & \omega & \omega^*\\
		1 & \omega^* & \omega 
	\end{pmatrix},}
where $\omega= \exp(2\pi i/3)$. The $\omega$ is a cube root of unity, that means that it is equal to unity if we raise it to the cubic power.

A two-qutrit state is written as 
\eq{\ket{\psi} =& \ket{\phi_1} \otimes \ket{\phi_2}
	= \big(\alpha_1 \ket{0} + \beta_1 \ket{1} + \gamma_1\ket{2} \big) \otimes \big(\alpha_2 \ket{0} + \beta_2 \ket{1} + \gamma_2\ket{2}\big) \nonumber \\
	=& \alpha_1 \alpha_2 \ket{00} + \alpha_1 \beta_2 \ket{01} + \alpha_1 \gamma_2 \ket{02} + \beta_1 \alpha_2 \ket{10} + \beta_1 \beta_2\ket{11} + \beta_1 \gamma_2\ket{12} +\nonumber\\
	& \gamma_1 \alpha_2\ket{20} + \gamma_1\beta_2\ket{21} + \gamma_1\gamma_2\ket{22}.}
The two-qutrit gates analogous to CNOT gate in Fig. \ref{TwoQubit} is such that the $U$ is applied when the controlling qutrit is at $\ket{2}$ otherwise the second qutrit does not change. Here, $U$ is one of the five one-qutrit gates in Fig. \ref{1QutritGates}. The Toffoli gate is implemented in such a way that $U$ is applied on the third qutrit when both the controlling qutrit at $\ket{2}$. The one-qutrit gates are called \textit{shift gates} and the two-qutrit gates are referred to as \textit{Muthukrishnan-Stroud (MS) gates} \cite{muthukrishnan2000multivalued}. The shift gates and MS gates have quantum cost unity \cite{asadi2020efficient,monfared2017design}. We will further discuss ternary gates and circuits in Section \ref{Meta}.

The physical realization of ternary logic was suggested by Ref. \cite{muthukrishnan2000multivalued} for an ion-trap quantum computer, by Ref. \cite{morisue1989novel,morisue1998josephson} for a Josephson junction, by Ref. \cite{smith2013quantum} for cold atoms, and by Ref. \cite{malik2016multi} for entangled photons. Some circuit architectures are better described by the multi-valued logic. In certain systems containing the non-Abelian anyons, called metaplectic anyons, qutrits naturally appear. We will discuss ternary gates with metaplectic anyons in Section \ref{Meta}.

\section{Topology and Knot Theory}\label{Knot}
The knot theory is of fundamental importance in topological quantum computing. Topological quantum gates are made up of knots, links, and braids. The path integral of the worldlines of quasiparticles in topological materials gives the knot invariants \cite{witten1989quantum}. The geometric phases are associated with such invariants. The knot theory is studied as a branch of topology.

\subsection*{B.1. Topology}
\begin{figure}[h!]
	\centering	
	\includegraphics[scale=0.4]{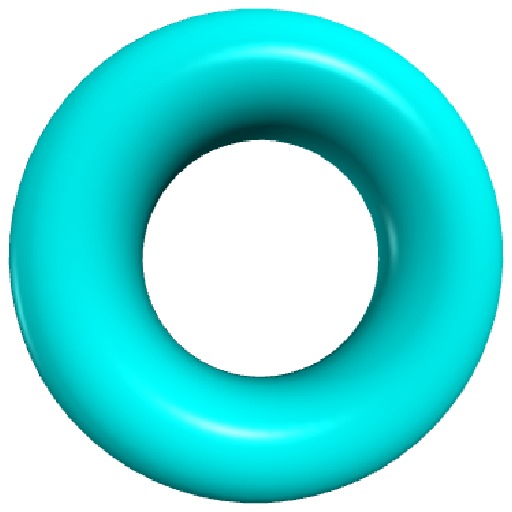}$=$
	\includegraphics[scale=0.45]{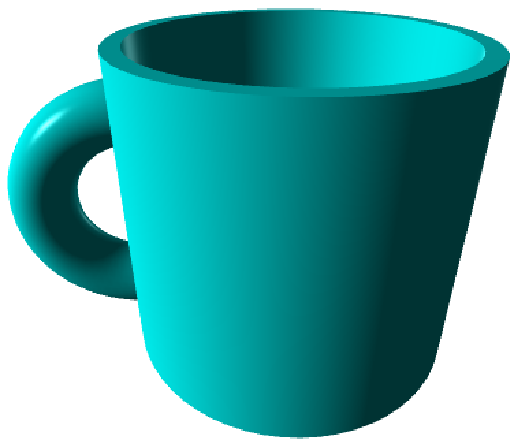}
	\caption{Equivalence of topological diagrams.}
	\label{TopDiagrams}
\end{figure}

The equivalence of two spaces in Euclidean geometry is shown by comparing their lengths and angles, but angles and lengths are irrelevant in topology. 
Instead, imagine that the spaces are made up of a stretchable and moldable material so that we can continuously deform one space to the other without tearing. For example, a sphere cannot be turned into a torus without tearing a hole, so they are topologically different. A hole or a handle in a topological space is called a \textit{genus}. From this point of view, a disk is equivalent to a rectangle or a square but different from an annulus. A curve and a straight line are equivalent shapes, but both are different from a closed curve. A closed curve is equivalent to a circle. A torus is equivalent to a coffee cup as both have one hole in it as shown in Fig. \ref{TopDiagrams}, and we can smoothly deform one into the other.
The spaces in topology are called the \textit{manifolds}. A manifold is a space that looks Euclidean if we take a small patch of the surface, but globally it may have a non-Euclidean structure. Topology is a study of properties that are preserved under continuous deformation in such a way that the dimension of the manifold should not change. The continuous deformation is called \textit{homeomorphism}. The topological properties that characterize the equivalence of two shapes under homeomorphism are called \textit{topological invariants}. These invariants can be numbers, or certain properties of the topological spaces like connectedness, compactness, homotopy group, homology group, or cohomology group \cite{nakahara2003geometry,isham1999modern,armstrong2013basic}. A genus is a topological invariant, but in some cases, it is not a very useful one. 

\begin{figure}[h!]
	\centering
	\includegraphics[scale=1]{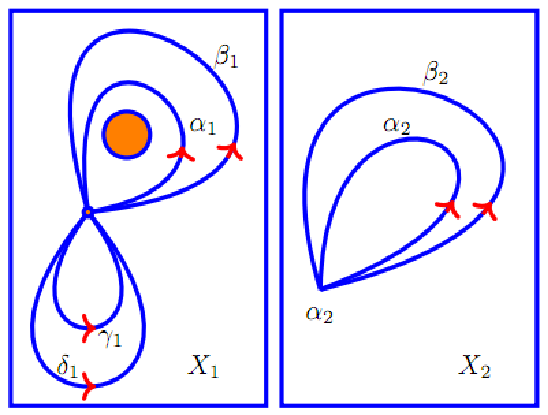}
	\caption{In the region $X_1$, the loops $\beta_1$ and $\alpha_1$ can be deformed to each other but they cannot be deformed to $\delta_1$ or $\gamma_1$. In the region $X_2$, both the loops can be shrunk to identity.}
	\label{Homotopy}
\end{figure}

The first homotopy group provides an intuition for anyonic statistics and braids.
Consider two regions $X_1$ and $X_2$ in Euclidean space as shown in Fig. \ref{Homotopy}. Imagine that any loop in $X_2$ can be shrunk to a point, but when there is a hole as in $X_1$, a loop cannot be shrunk to a point. From the Fig. \ref{Homotopy}, $\alpha _1$ can be deformed to $\beta _1$, and $\gamma_1$ can be deformed to $\delta_1$, but the loops $\alpha_1$ and $\beta_1$ cannot be deformed to $\gamma_1$ and $\beta_1$. 
When two loops can be continuously deformed to each other then they are in the same equivalence class or a homotopic class \cite{nash1988topology}. 
Spaces are distinguished by working with equivalent classes rather than loops. This suggests that the holes are determined by using these equivalence classes. The group structure on these equivalence classes is called the \textit{first homotopy group or a fundamental group} represented by $\pi _1(X)$. The group axioms of a homotopy group are described below.
The composition of two group elements corresponds to two loops that start at the same point and are combined to make the third one. In Fig. \ref{Homotopy}, $\gamma_1$ and $\beta_1$ loops are combined to make $\tilde {\gamma}$ that can be written as $\tilde {\gamma} = \gamma_1 \ \beta_1$. This loop first goes along the $\beta_1$ then along the $\gamma_1$. The inverse $\beta_1^{-1}$ is given by a loop that goes in the opposite direction. The identity loop is the one that stays at some point all the time. The loop $\varepsilon = \beta_1 \ \beta_1^{-1}$ is not an identity but homotopic to the identity \cite{nash1988topology}. When these loops are physically made by the motion of particles on a two-dimensional space then they make braids in the third dimension which is time.

\subsection*{B.2. Knot Invariants}
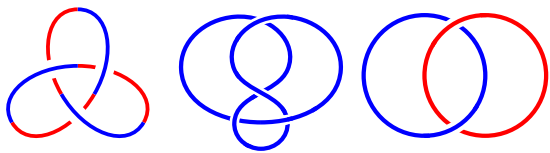
\begin{figure}[h!]
	\centering
	\begin{tikzpicture}[use Hobby shortcut,every path/.style={line width=1mm,white,double=red, double distance=.5mm},scale=0.5]
		\def\nfoil{3}
		\draw ([closed]0,2)
		foreach \k in {1,...,\nfoil} {
			.. ([blank=soft]90+360*\k/\nfoil-180/\nfoil:-.5) .. (90+360*\k/\nfoil:2)
		};
		\draw[use previous Hobby path={invert soft blanks,disjoint},double=blue];
	\end{tikzpicture}
	\begin{tikzpicture}[use Hobby shortcut,scale=0.7]
		\begin{knot}[consider self intersections=true,ignore endpoint intersections=false,flip crossing=3,only when rendering/.style={ultra thick,blue}]
			\strand ([closed,ultra thick,blue]0,0) .. (1.5,1) .. (.5,2) .. (-.5,1) .. (.5,0) .. (0,-.5) .. (-.5,0) .. (.5,1) .. (-.5,2) .. (-1.5,1) .. (0,0);
		\end{knot}
		\path (0,-.7);
	\end{tikzpicture}\ \
	\begin{tikzpicture}[scale=0.8]
		\begin{knot}[flip crossing=2]
			\strand [red,ultra thick](2,0) circle[radius=1cm];
			\strand[blue,ultra thick,->] (1,0) circle[radius=1cm];
		\end{knot}
	\end{tikzpicture}
	\caption{The trefoil knot, figure 8 knot, and the Hopf link.}
	\label{Knots}
\end{figure}

A knot is a closed loop embedded in the three-dimensional space. A link is a disjoint union of more than one loop. 
A knot diagram is a projection of a knot into the plane $\mathbb{R}^2$ such that the points are segments and double points are under-crossings and over-crossings.  A circle is an \textit{unknot or a trivial link}. The simplest non-trivial link is a \textit{Hopf link}. For example, the trefoil, figure 8 knot, and Hopf link are shown in Fig. \ref{Knots}. For further study on knots, see \cite{kauffman2001knots,adams2004knot,baez1994gauge}.

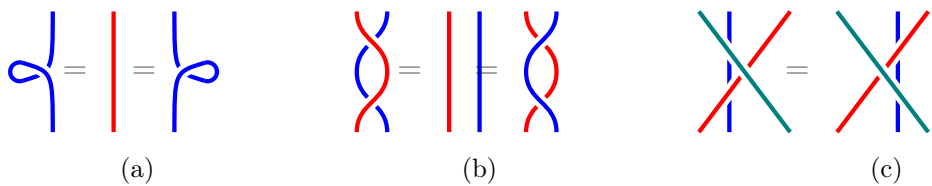
\begin{figure*}[h!]
	\centering
	\begin{subfigure}{0.2\textwidth}
		\begin{tikzpicture}[scale = 0.4]
			\draw[blue,knot,ultra thick] (1,4) .. controls +(0,-2) .. (0,1.7);
			\draw[knot=blue,ultra thick] (1,0) .. controls +(0,2) .. (0,2.3);
			\draw [blue,ultra thick] (0,1.7) to [curve through={(-0.4,2)}](0,2.3);
			\node [left] at (2.5,2) {=};
			\draw[red,knot,ultra thick] (3,0) -- (3,4);
			\node [] at (4,2) {=};
			\draw[blue,knot,ultra thick] (5,4) .. controls +(0,-2) .. (6,1.7);
			\draw[knot=blue,ultra thick] (5,0) .. controls +(0,2) .. (6,2.3);
			\draw [blue,ultra thick] (6,1.7) to [curve through={(6.4,2)}](6,2.3);
		\end{tikzpicture}
		\caption{}
	\end{subfigure} \ \qquad
	\begin{subfigure}{0.2\textwidth}
		\begin{tikzpicture}[scale = 0.4]
			\draw[knot=blue,ultra thick] (1,0) .. controls +(0,1) and +(0,-1) .. (0,2) .. controls +(0,1) and +(0,-1) .. (1,4);
			\draw[red,knot,ultra thick] (0,0) .. controls +(0,1) and +(0,-1) .. (1,2) .. controls +(0,1) and +(0,-1) .. (0,4);
			\node [left] at (2.5,2) {=};
		\end{tikzpicture}
		\begin{tikzpicture}[scale = 0.4]
			\draw[blue,ultra thick] (1,0) -- (1,4);
			\draw[red,knot,ultra thick] (0,0) -- (0,4);
			\node [left] at (2,2) {=};
		\end{tikzpicture}
		\begin{tikzpicture}[scale = 0.4]
			\draw[red,knot,ultra thick] (0,0) .. controls +(0,1) and +(0,-1) .. (1,2) .. controls +(0,1) and +(0,-1) .. (0,4);
			\draw[knot=blue,ultra thick] (1,0) .. controls +(0,1) and +(0,-1) .. (0,2) .. controls +(0,1) and +(0,-1) .. (1,4);
		\end{tikzpicture}
		\caption{}
	\end{subfigure} \ \qquad
	\begin{subfigure}{0.3\textwidth}
		\begin{tikzpicture}[scale = 0.4]
			\draw[blue,knot,ultra thick] (1,0) -- (1,4);
			\draw[red,knot,ultra thick] (0,0) -- (3,4);
			\draw[teal,knot,ultra thick] (3,0) -- (0,4);
			\node [left] at (4,2) {=};
		\end{tikzpicture}
		\begin{tikzpicture}[scale = 0.4]
			\draw[blue,knot,ultra thick] (2,0) -- (2,4);
			\draw[red,knot,ultra thick] (0,0) -- (3,4);
			\draw[teal,knot,ultra thick] (3,0) -- (0,4);
		\end{tikzpicture}
		\caption{}
	\end{subfigure}
	\caption{Reidemeister moves: (a) The move I is undoing a twist in the strand, (b) the move II separates two unbraided strands, and (c) the move III slides strand under a crossing.}
	\label{Reid}
\end{figure*}

Two knot diagrams are equivalent if we can bend, stretch and smoothly deform one to the other without cutting. In knot theory, the equivalence of two knots is called \textit{ambient isotopy}. Other than stretching and bending, a simple way of showing that two knots are isotopic to each other is by a finite number of the \textit{Reidemeister moves} shown in Fig. \ref{Reid}. These moves are always permitted but not always sufficient to show the isotopy of two knots. When a knot is modified by applying these moves on a small portion of the diagram while keeping the rest of the diagram fixed, we get the formulas called \textit{Skein Relations}.

The \textit{knot invariants} are a set of rules that give the same output for two equivalent knots. That is, these invariants should not change under an ambient isotopy. Therefore, different knots are distinguished by their knot invariants. The knot invariants have their merits and limitations. The \textit{knot polynomials} are among several knot invariants assigned to knots and relatively easy to calculate. 
The \textit{Jones polynomial} \cite{jones1997polynomial} is of particular interest to us because of its connection to physics. This connection was first explored by Edward Witten \cite{witten1989quantum}. Physically, the trajectories of anyons in spacetime make knots. The knot invariants of the trajectories are calculated by path integral approach to the Chern-Simons theory, see \cite{nayak2008non,ilyas2021topological}. Edward Witten made this connection between the knot theory and quantum physics in his seminal paper in 1989 \cite{witten1989quantum}. He won the field medal for this work with Vaughn Jones in 1990. The present form of the Jones polynomial is due to Kauffman who formulated it in a simpler way \cite{kauffman2001knots}.

\subsubsection*{B.2.1. Kauffman Bracket}
The Kauffman bracket is a polynomial invariant of unoriented link. The normalized version of Kauffman bracket yields the Jones polynomial when framing of a knot or a link is also considered. A Kaufman bracket $\langle L \rangle$ of a knot or a link $L$ assigns to each crossing a number that is either $A$ or $B$ as in diagrammatic Eq. \ref{SkeinKauff}. The values of variables $A$ and $B$ are to be computed in present section. This \textit{Kauffman skein relation} is used recursively until we get a resulting link diagram to have no crossings. Thus it consists of a finite set of unlinks or circles \cite{kauffman1990invariant,kauffman2001knots}. The Kauffman skein relation is written as
\eq{
	\begin{tikzpicture}[scale = 0.5]
		\draw [teal, ultra thick] (1,-1)--(0,0)--(1,1);
		\draw [teal, ultra thick] (2,-1)--(3,0)--(2,1);
		\draw[blue,knot,ultra thick] (1.9,-0.8) -- (1.1,0.8);
		\draw[red,knot,ultra thick] (1.1,-0.8) -- (1.9,0.8);
		\node [left] at (5,0) {$ = A $};
	\end{tikzpicture}
	\begin{tikzpicture}[scale = 0.5]
		\draw [teal, ultra thick] (1,-1)--(0,0)--(1,1);
		\draw [teal, ultra thick] (2,-1)--(3,0)--(2,1);
		\draw [red,ultra thick] (1.1,-0.8) to [curve through={(1.3,0)}](1.1,0.8);
		\draw [blue,ultra thick] (1.9,-0.8) to [curve through={(1.7,0)}](1.9,0.8);
		\node [left] at (5,0) {$ + B $};
	\end{tikzpicture}
	\begin{tikzpicture}[scale = 0.5]
		\draw [teal, ultra thick] (1,-1)--(0,0)--(1,1);
		\draw [teal, ultra thick] (2,-1)--(3,0)--(2,1);
		\draw [red,ultra thick] (0.8,0.5) to [curve through={(1.4,0.2)}](2.2,0.5);
		\draw [blue,ultra thick] (0.8,-0.5) to [curve through={(1.4,-0.2)}](2.2,-0.5);
	\end{tikzpicture}
	\label{SkeinKauff}}
Here the variables $A$ and $B$ are assigned according to the convention such that when the first strand goes over the second, that is an overcrossing, we call it the positive crossing. For a negative crossing, when the first strand goes below the second, the variables $A$ and $B$ will be exchanged. It is proved that $B=A^{-1}$ and $d=-(A^2+A^{-2})$. See \cite{kauffman2001knots,ilyas2021topological} for the detailed derivation. The number $d$ is assigned to a simple loop therefore, it is called the \textit{loop number}. The Kauffman bracket is not invariant under Reidemeister move I. 
Let us take a twist in a \textit{framed or ribbon} strand as in Fig. \ref{RibbonTwist}. The smoothing out of the twist gives a factor of $-A^3$.
Hence, a ribbon is related to the string with a factor $-A^{3}$ multiplied. In quantum theory, it is related to the phase accumulated by a particle with a spin when it does a $2\pi$ rotation. 
The skein relations for the Kauffman bracket are now written as 
\eq{&\langle L \rangle = A \langle L_A \rangle + A^{-1}\langle L_B \rangle\nonumber\\
	&\langle L \cup O \rangle = d \langle L \rangle = -(A^2+A^{-2}) \langle L \rangle \nonumber\\
	&\langle O \rangle = 1 \label{SkeinKauff2}}
where $O$ refers to unknot or a trivial link. The first relation is the same as Eq. \eqref{SkeinKauff}, the second relation tells that if a knot or a link is a union of a knot and an unknot then the resultant knot would be $d$ times that knot. The third relation implies that an isolated unknot is assigned a value 1.
\fig{[h!]
	\centering
	\begin{tikzpicture}[use Hobby shortcut,scale=1.5]
		\begin{knot}[
			consider self intersections=true,
			ignore endpoint intersections=false,
			flip crossing=2,
			only when rendering/.style={
			}
			]
			\strand [ultra thick,blue](0.34,0.9)..(0.5,0.8).. (0.7,0.9) .. (0.7,1.05) .. (0.33,1)..(0.2,0.6);
			\strand [ultra thick,blue](0.27,0.7).. (0.45,0.57).. (0.95,0.95) .. (0.5,1.37) .. (0.25,1.26)..(0,0.75);
			\strand [ultra thick,blue] (0,1) .. (0,1.8);
			\strand [ultra thick,blue] (0.2,1.3) .. (0.2,1.8);
			\strand [ultra thick,blue] (0,0) .. (0,0.8);
			\strand [ultra thick,blue] (0.2,0) .. (0.2,0.6);
			\draw [ultra thick,blue] (0,1.8) -- (0.2,1.8);
			\draw [ultra thick,blue] (0,0) -- (0.2,0);
		\end{knot}
		\path (0,0);
	\end{tikzpicture}
	\begin{tikzpicture}[scale=1.5]
		\node [] at (-0.2,1) {$ \rightarrow $};
		\draw[ultra thick,blue] (0,0) -- (0,0.2);
		\draw[ultra thick,blue] (0.2,0) -- (0.2,0.2);
		\draw[ultra thick,knot=blue] (0.2,0.2) -- (0,0.6);
		\draw[ultra thick,knot=blue] (0,0.2) -- (0.2,0.6);		
		\draw[ultra thick,blue] (0,0.6) -- (0,1.2);
		\draw[ultra thick,blue] (0.2,0.6) -- (0.2,1.2);
		\draw[ultra thick,knot=blue] (0.2,1.2) -- (0,1.6);
		\draw[ultra thick,knot=blue] (0,1.2) -- (0.2,1.6);		
		\draw[ultra thick,blue] (0,1.6) -- (0,1.8);
		\draw[ultra thick,blue] (0.2,1.6) -- (0.2,1.8);
		\draw[ultra thick,blue] (0,0) -- (0.2,0);
		\draw[ultra thick,blue] (0,1.8) -- (0.2,1.8);	
	\end{tikzpicture}
	\begin{tikzpicture}[scale=1.5]
		\node [] at (-0.5,1) {$ =-A^3 $};
		\draw[ultra thick,blue] (0,0) -- (0,1.8)--(0.2,1.8)--(0.2,0)--cycle;	
	\end{tikzpicture}
	\caption{Straightening a loop in a ribbon gives a twist factor.}
	\label{RibbonTwist}}

\subsubsection*{B.2.2. Jones Polynomial}
\begin{figure}[h!]
	\centering
	\begin{tikzpicture}[scale = 0.4]
		\node [left] at (0,2) {$L_+ = $};
		\draw[blue,knot,ultra thick] (3,0) -- (0,4);
		\draw[red,knot,ultra thick] (0,0) -- (3,4);
	\end{tikzpicture}
	\begin{tikzpicture}[scale = 0.4]
		\node [left] at (0,2) {$, \ L_0 = $};
		\draw[blue,ultra thick] (1,0) -- (1,4);
		\draw[red,knot,ultra thick] (0,0) -- (0,4);
	\end{tikzpicture}
	\begin{tikzpicture}[scale = 0.4]
		\node [left] at (0,2) {$, \ L_- = $};
		\draw[red,knot,ultra thick] (0,0) -- (3,4);
		\draw[blue,knot,ultra thick] (3,0) -- (0,4);
	\end{tikzpicture}
	\caption{The skein relations for Jones polynomial}
	\label{SkeinJones}
\end{figure}
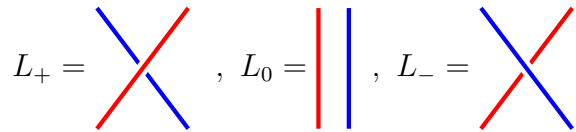

The Kauffman bracket is not an invariant under all the Reidemeister moves. We need to account for the twist factor or the self linking. This twist is also called a \textit{writhe}. Now we will also assign an orientation to the knot diagrams. If $w_+$ is an overcrossing and $w_-$ is an undercrossing for an oriented knot or link, then the writhe is given by $w(L) = w_+ - w_-$. We can construct a quantity as 
\eq{V_L(A) = (-A^3)^{-w (L)} \langle L \rangle.}
This is called \textit{Jones polynomial}. It is an invariant under all three Reidemeister moves. Two knots are equivalent if they have the same value of the Jones polynomial. A twist that contributes a factor $(-A^{-3})$ would get canceled with $(-A^3)^{w(L)}$. With the change of variable $t^{1/2} = A^{-2}$, the Jones polynomial agrees with the original form in Jones' paper \cite{jones1997polynomial}. The Jones polynomial is an invariant under orientation change.
The skein relations for Jones polynomial can be written as
\eq{-t^{-1} V(L_+) + (t^{1/2} - t^{1/2})V(L_0) + t V(L_-) = 0,}
where $L_0,L_-,L_+$ are shown in Fig. \ref{SkeinJones}. The writhe is $+1$ for $L_+$, $-1$ for $L_-$, and $0$ for $L_0$. The Jones polynomial is an invariant for a knot's mirror image when $t$ is replaced by $t^{-1}$.

\subsection*{B.3. Braid Group}
The trajectories of $N$ particles from their initial position at a time $t_i$ to the final position at a time $t_f$ are in one-to-one correspondence with the elements of the braid group $\cal{B}_N$. 
The time direction is taken vertically upward. The trajectories of particles are equivalence classes of all those trajectories which can be continuously deformed into each other. 
Assume that the particle number is fixed, which is the same as saying that there are no loops inside the braids.

\begin{figure}[h!]
	\centering
	\begin{subfigure}{0.2\textwidth}
		\centering
		\begin{tikzpicture}[rotate=180]
			\draw[ultra thick,red] (1,0) -- (1,-2.5);
			\draw[ultra thick,blue] (2,0) -- (2,-2.5);
			\node [left] at (1.2,-1.3) {$...$};
		\end{tikzpicture}
		\caption{Identity}
	\end{subfigure}
	\begin{subfigure}{0.25\textwidth}
		\centering
		\begin{tikzpicture}
			\braid[ultra thick, style strands={1}{blue},style strands={2}{red},style strands={3}{green}]
			s_1^{-1};
			\draw[ultra thick,blue] (1,0) -- (1,0.5);
			\draw[ultra thick,red] (2,0) -- (2,0.5);
			\draw[ultra thick,red] (1,-1.5) -- (1,-2);
			\draw[ultra thick,blue] (2,-1.5) -- (2,-2);
		\end{tikzpicture}
		\begin{tikzpicture}
			\braid[ultra thick, style strands={1}{red},style strands={2}{blue},style strands={3}{green}]
			s_1;
			\draw[ultra thick,red] (1,0) -- (1,0.5);
			\draw[ultra thick,blue] (2,0) -- (2,0.5);
			\draw[ultra thick,blue] (1,-1.5) -- (1,-2);
			\draw[ultra thick,red] (2,-1.5) -- (2,-2);
		\end{tikzpicture}
		\caption{$\sigma_i$ and $\sigma_i^{-1}$}
	\end{subfigure}
	\begin{subfigure}{0.4\textwidth}
		\centering
		\begin{tikzpicture}
			\braid[ultra thick, style strands={1}{red},style strands={2}{blue},style strands={3}{green}]
			s_1^{-1};
			\draw[ultra thick,red] (1,0) -- (1,1);
			\draw[ultra thick,blue] (2,0) -- (2,1);
		\end{tikzpicture}
		\begin{tikzpicture}
			\braid[ultra thick, style strands={1}{red},style strands={2}{blue},style strands={3}{green}]
			s_1^{-1};
			\draw[ultra thick,blue] (1,-2.5) -- (1,-1.5);
			\draw[ultra thick,red] (2,-2.5) -- (2,-1.5);
			\node [left] at (3,-1.5) {$=$};
		\end{tikzpicture}
		\begin{tikzpicture}
			\braid[ultra thick, style strands={1}{red},style strands={2}{blue},style strands={3}{green}]
			s_1^{-1};
			\draw[ultra thick,blue] (1,-2.5) -- (1,-1.5);
			\draw[ultra thick,red] (2,-2.5) -- (2,-1.5);
		\end{tikzpicture}
		\begin{tikzpicture}
			\braid[ultra thick, style strands={1}{red},style strands={2}{blue},style strands={3}{green}]
			s_1^{-1};
			\draw[ultra thick,red] (1,0) -- (1,1);
			\draw[ultra thick,blue] (2,0) -- (2,1);
		\end{tikzpicture}
		\caption{$\sigma_i\sigma_j = 
			\sigma_j \sigma_i$ for $\abs{i-j}>1$}
	\end{subfigure}
	\begin{subfigure}{0.3\textwidth}
		\centering
		\begin{tikzpicture}[scale=0.7]
			\braid[ultra thick, style strands={1}{red},style strands={2}{blue},style strands={3}{teal}]
			s_1^{-1} s_2^{-} s_1^{-1};
			\node [left] at (4,-2) {$=$};
		\end{tikzpicture}
		\begin{tikzpicture}[scale=0.7]
			\braid[ultra thick, style strands={1}{red},style strands={2}{blue},style strands={3}{teal}]
			S_2^{-1} s_1^{-1} s_2^{-1}; 
		\end{tikzpicture}
		\caption{$\sigma_i\sigma_{i+1}\sigma_i=\sigma_{i+1}\sigma_i\sigma_{i+1}$}
	\end{subfigure}
	\caption{The braid group; generators and their properties.}
	\label{Braiding1}
\end{figure}
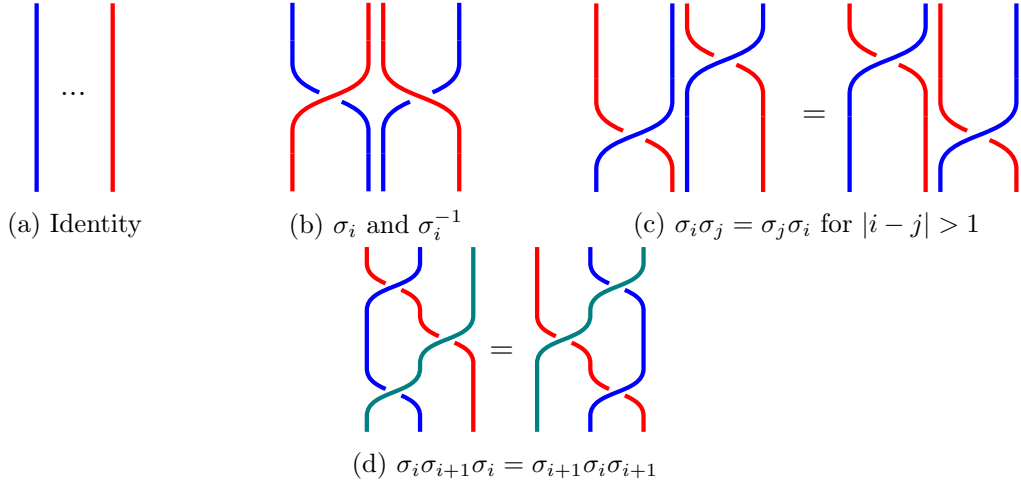

Let the braiding of the first and the second strand be represented by  $\sigma_1$ and braiding of the second and the third strands be represented by $\sigma_2$ and so on.
The braid group generators are shown in Fig. \ref{Braiding1}. (a) is the identity element of the braid group. It consists of all straight strands. As shown in (b), the clockwise exchange of strands $i$ and $i+1$ is represented by the generator $\sigma_i$, whereas counterclockwise exchange is represented by the inverse $\sigma_i^{-1}$. 
The group composition of two braids is given by stacking the strands on top of each other. For a non-Abelian group, the multiplication is noncommutative, and the order of stacking matters in this case. This is because of the degeneracy of the ground state, as discussed in \ref{GeoPhase}. The braid group generators also satisfy two conditions shown in Fig. \ref{Braiding1} (c) and (d).
\eq{
	\sigma_i\sigma_j &= \sigma_j\sigma_i \ \text{for} \ \abs{i-j} > 1,\\
	\sigma_i\sigma_{i+1}\sigma_i &= \sigma_{i+1}\sigma_i\sigma_{i+1}.} 
The second relation is the famous Yang-Baxter equation.

\section{Geometric Phases in Quantum Physics}\label{GeoPhase}
Anyons are charge-flux composites that arise in a highly correlated system. Interaction between these charges and fluxes is through braiding. Topological quantum gates are implemented by braiding of anyons on a two-dimensional manifold. The state-space of a topological quantum gate is the ground state degeneracy of the system. This braiding and the ground state degeneracy can be understood through the Aharonov-Bohm phase \cite{aharonov1959significance} and the Berry phase \cite{berry1984quantal}. For further details of the derivation of geometric phases, see \cite{bohm2013geometric}.

\subsection*{C.1. A Charged Particle in a Magnetic Field}
Classically, the motion of a nonrelativistic charge particle in a magnetic field is described by the Lorentz equation given as
\eq{m\bm{\ddot{x}}_i = q \big(\bm{E} + (\bm{v}\cross \bm{B}\big)).}
The electric field $\bm{E}$ and the magnetic field $\bm{B}$ can be written in terms of vector potential $\textbf{A} = (A_x,A_y,A_z)$ and scalar potential $\phi(x)$ as
\eq{\bm{E}= - \frac{\partial}{\partial t}\bm{A} - \nabla \phi , \ \ \bm{B}= \nabla \cross \bm{A}.\label{EM}}
In quantum mechanics, the momentum $p$ is used instead of $v$ and its status is raised to an operator $\bm{p}= -i\hbar \nabla$. 
The Schr\"odinger's equation for a charged particle moving in an electromagnetic field can be written as
\eq{i\hbar \frac{\partial \psi(x)}{\partial t} = \Big[\frac{1}{2m}\big(\bm{p} - q \bm{A}\big)^2 + q \phi\Big]\psi(x).}
Let $\psi_0(x)$ be an eigenstate of the Hamiltonian when there is no vector potential. The wave function of the particle in the presence of vector potential is related to $\psi_0$ as 
\eq{\psi(x) = \exp({i\frac{q}{\hbar} \int \bm{A}\cdot d\bm{x}}) \psi_0(x). \label{WaveFuncA}}
The wave function in a magnetic field will get a phase $\phi=\frac{q}{\hbar} \int \bm{A}\cdot d\bm{x}$ other than the dynamical phase. A dynamical phase is the one that a wave function gets during the time evolution. The momentum operator $(\bm{p}-q\bm{A})$ appears as a combination of $\bm{p} = -i\hbar \nabla$ and the vector potential $\bm{A}$.

\subsubsection*{C.1.1. Gauge Transformation}
Let us transform the vector potential as
\eq{\bm{A} \rightarrow \bm{A}^{'} = \bm{A} + \nabla \Lambda, \ \phi \to \phi' = \phi- \frac{\partial \Lambda}{\partial t}, \label{GaugeTransf}} 
where $\Lambda$ is a scalar function. The transformation in Eq. \ref{GaugeTransf} is called \textit{gauge transformation}. Now the Schr\"odinger equation will be written in terms of $\bm{A}',\psi'$, and $\phi'$. The wave function in Eq. \ref{WaveFuncA} can be written as
\eq{\psi(x)\to \psi'(x) &= \exp(i\frac{q}{\hbar}\int \bm{A}'\cdot d\bm{x})\psi_0(x)\nonumber\\
	&= \exp(i\frac{q}{\hbar}\Lambda(x))\psi(x).}
As a result of the gauge transformation, the wave function gets an additional phase of $\exp(i\frac{q}{\hbar} \Lambda(x))$. The fields $\bm{E}$ and $\bm{B}$ will remain invariant under this transformation. The physical quantities are modulus squared, so the complex phases do not appear.
This gauge transformation is \textit{local} as $\Lambda (x)$ is a function of $x$. The global gauge transformation is not as significant. It is independent of the position and is corresponding to the transformation of the whole system.

\subsection*{C.2. Aharonov-Bohm Effect}
In 1959, Yakir Aharonov and David Bohm \cite{aharonov1959significance} suggested that in quantum mechanics, the vector potential is not just a mathematical artifact, but it leads to detectable results. The effect of vector potential can be observed in a region where $\bm{B}= 0$ but $\bm{A} \ne 0$. They proposed an experiment shown in Fig. \ref{AB1}.
Suppose an infinitely long solenoid having a current through it produces a magnetic field along the z-axis. Since according to the right-hand rule the magnetic field is along the axis of the solenoid and zero outside, it can be taken as a tube of magnetic flux. The electron beam from the source $S$ is separated into two parts as shown in Fig. \ref{AB1}. The two parts of the beam combined at the screen make an interference pattern. 
The phase we get with the wave function of a charged particle in a magnetic field is given in the Eq. \ref{WaveFuncA}.
The phase acquired by the evolution of the wave function around a loop $C$ can be derived as
\eq{\phi = \frac{q}{\hbar} \oint\bm{A}\cdot d\bm{r} = \frac{q}{\hbar}\int_S \nabla \cross \bm{A} \cdot d\bm{s} = \frac{q}{\hbar} \int_S \bm{B}\cdot d\bm{s} = \frac{q}{\hbar}\Phi,}
where $d\bm{r}$ is a segment of the loop $C$ and $S$ is the surface enclosed by $C$, $d\bm{s}$ is the surface area element, and $\Phi$ is the total flux through $S$. This phase is gauge invariant, i.e. it is independent of the choice of $\bm{A}$ provided that it gives the same $\bm{B}$.
This phase is topological, as it does not depend on the shape of the path around the flux. Also, it remains invariant under the deformation of the surface that makes $\Phi$ fixed.
The two paths in Fig. \ref{AB1} are facing different relative vector potentials, hence interference fringes are modulated by the magnetic flux in the coil which is affected by the change of the electric current through the coil. Therefore, the choice of potentials instead of fields is not merely a convenience but a necessity. The electromagnetic field needs to be described in terms of an abstract four-dimensional vector $A_\mu = (\bm{A}, \phi)$.
\fig{[h!]
	\centering
	\tik{[scale=0.8]
		\draw[ultra thick,blue,-latex] (-3,-1) .. controls(0,2) and (1.5,1.5)..(3,1.1);
		\node [cylinder,draw=black,thick,aspect=1.5,minimum height=3cm,minimum width=1cm,shape border rotate=90,cylinder uses custom fill, cylinder body fill=green!30,cylinder end fill=red!5,opacity=0.5] at (0.5,0){};
		\draw[ultra thick,->,red] (0.5,-1.5)--(0.5,2);
		\draw[ultra thick,blue,-latex] (-3,-1) .. controls(-1,-1) and (1.5,-1.5)..(3,0.9);
		\draw [snake=coil, segment amplitude=12pt, segment length=4pt](0.5,-1.2)--(0.5,1.3);
		\draw[snake=coil,segment aspect=0,segment amplitude=15pt,ultra thick,red] (3.2,1) -- (4.8,1);
		\node [above] at (0.5,2){$\Phi$};
	}
	\caption{Aharonov-Bohm Effect: switching on and off the flux in the flux tube, causes a shift in the interference fringes.}
	\label{AB1}}

\subsubsection*{C.2.1. Anyon and Aharonov-Bohm Effect}
An anyon is a quasiparticle having fractional charge and fractional statistics. We can think of these particles as a composite of charge $q$ and flux $\Phi$. These composites arise in two-dimensional physical systems \cite{girvin1987quantum,wilczek1982quantum}. See \cite{ilyas2021topological} for a brief introduction to the quantum Hall effect and the effective field theory for the attachment of magnetic flux to the quasiparticle..
Let us exchange two anyons in a 2-dimensional space. The movement of anyons around each other in the $(2+1)$-dimensional space, is described by the braid group, see \ref{Knot}. The charge 1 going around the flux of 2 gets the Aharanov-Bohm phase $e^{iq \Phi}$. At the same time, the flux of 1 going around the charge of 2 and gets the phase $e^{iq \Phi}$, as in Fig. \ref{AnyonAB}. Therefore, the system gets a total phase $e^{2iq \Phi}$. This phase depends on the number of times one charge circulates the other, but it does not depend on the shape of the path, provided that the adiabaticity condition is satisfied. The \textit{adiabaticity condition} dictates that the charges must be moved slowly enough so that the system is not perturbed drastically from the ground state. The number of times a charge circulates another charge is called the \textit{winding number}. The phase will be written as $e^{imq\Phi}$ when the winding number is $m$. The statistical angle $\phi = q \Phi$ on an exchange corresponds to the phase shift of their wave function.
The $2\pi$ rotation of an anyon around itself gives a phase of $e^{iq \Phi}$ due to the charge ring around it. The spin-statistics theorem \cite{pauli1940connection} says that if $s$ is the effective spin of an anyon taken counterclockwise, we get a phase $e^{i2\pi s}$. Thus, we have a non-trivial spin $s= \frac{q \Phi}{2\pi}$ \cite{pachos2012introduction}. 
The \textit{topological spin} or a twist of an anyon is the rotation of the charge around its own flux. The phase due to the topological spin and the phase due to the braiding are related to each other as discussed in Section \ref{TQC}.
\fig{[h!]
	\centering\tik{
		\draw[fill=red!10] (-0.5,1.3) -- (6.5,1.3) -- (5.2,-1.3) -- (-2.1,-1.3)-- cycle;
		\draw [ultra thick,fill=cyan] (0,0) ellipse (0.8 and 0.4);
		\draw [ultra thick,blue] (2.5,0) ellipse (2.6 and 1);
		\draw [ultra thick,fill=cyan] (4,0) ellipse (0.8 and 0.4);
		\node [cylinder,draw=black,thick,aspect=1.5,minimum height=2cm,minimum width=1cm,shape border rotate=90,cylinder uses custom fill, cylinder body fill=green!30,cylinder end fill=red!5] at (0,0.6){};
		\draw[ultra thick,-latex,red] (0,1)--(0,2);
		\draw[ultra thick,-latex,red] (0.2,1)--(0.2,2);
		\draw[ultra thick,-latex,red] (-0.2,1)--(-0.2,2);
		\node [cylinder,draw=black,thick,aspect=1.5,minimum height=2cm,minimum width=1cm,shape border rotate=90,cylinder uses custom fill, cylinder body fill=green!30,cylinder end fill=red!5,opacity=0.8] at (4,0.6){};
		\draw[ultra thick,-latex,red] (4,1)--(4,2);
		\draw[ultra thick,-latex,red] (3.8,1)--(3.8,2);
		\draw[ultra thick,-latex,red] (4.2,1)--(4.2,2);
		\draw[->,bend left,ultra thick] (4.5,2) to (4.8,1);
		\node [above] at (0,2){$\Phi$};
		\node [above] at (4,2){$\Phi$};
		\node [] at (-0.8,-0.5){$q_1$};
		\node [] at (3.2,-0.5){$q_2$};
		\node [] at (4.2,-1){$C$};
	}\caption{Anyons moving around each other obtain Aharanov-Bohm Phase.}
	\label{AnyonAB}}
\subsection*{C.3. Berry Phase}
The Aharanov-Bohm phase is a special case of the geometric phases when the system has time reversal symmetry \cite{berry2010geometric,cohen2019geometric}. It appears when the underlying geometry is changed by the magnetic field in an abstract way. A more general geometric phase, acquired by a wave function during the evolution in parametric space, is called the Berry phase. As an example, consider a spin-1/2 particle in a magnetic field that is oriented in a particular direction. By slowly varying the magnetic field orientation and bringing it back to the initial value, the system will come back to the initial state up to an overall phase with the wave function of the particle \cite{pachos2012introduction}.

Let us compute the geometric phase in quantum mechanics for a general situation described by two variables $\bm{r}$ and $\bm{R}(t)$. Let $\bm{r}$ describe a fast motion and $\bm{R}(t)$ be a variable that describes a slow motion. The slow variable describes a parameter that varies slowly with time and modulates the fast variable. For example, the motion of electrons of the atoms in a diatomic molecule is described by the fast variable $r$ and the vibratory motion of atoms is described by a slower variable $\bm{R}(t)$.
We suppose that the system returns to the original state after completing a loop in parametric space.
The adiabaticity condition should be satisfied, which means that the motion of the system in the parametric space should be slow enough so that the system does not go to the excited state.
The Schr\"odinger equation with state vector $\psi(t)$ can be written as
\eq{ i\hbar \frac{\partial }{\partial t} \ket{\psi (t)} = H(\bm{R(t)}) \ket{\psi (t)}.} 
Let $\ket{n, \bm{R}}$ be an eigenstate of the Hamiltonian that has energy eigenvalue as $E_n(\bm{R})$.
As the $\bm{R}(t)$ is slowly varying, at an instant of time $t$ we can take $\ket{n, \bm{R}(t)}$ as a basis vector, therefore we can write
\eq{H(\bm{R}(t)) \ket{n, \bm{R}(t)} = E_n(\bm{R}(t))\ket{n, \bm{R}(t)}.}
The solution of Schr\"odinger equation is given by
\eq{\ket{\psi(t)} = e^{i\gamma_n (t)} \exp \Big[-\frac{i}{\hbar} \int_{0}^{t}dt^{'} E_n(\bm{R}(t^{'}))\Big] \ket{n,\bm{R}(t)}. \label{BerrySol}}
The phase in brackets is the dynamical phase that depends on time, whereas $\gamma_n(t)$ is the Berry phase that depends on the geometry of parametric space.
In 1984, Berry pointed out that $\gamma_n(t)$ has deep physical meaning and cannot be ignored \cite{berry1984quantal}. Consider a situation when the slow variable $\bm{R}(t)$ returns to the starting point $(\bm{R}(t)=\bm{R}(T))$ at a time $t=T$ after completing a turn in the parametric space in a closed path $C$. If we put the solution \ref{BerrySol} in the Schr\"odinger equation, take derivative, and cancel then exponentials on both sides, and then apply $\bra{n,\bm{R(t')}}$, we get
\eq{\gamma_n(C) = i \int_{0}^{t} dt'\bra{n, \bm{R}(t')} \frac{d}{dt'} \ket{n, \bm{R}(t')} = i\oint_{C} d\bm{R}\cdot\bra{n, \bm{R}} \nabla_R \ket{n, \bm{R}} \label{BerryPhase}.}
One of the examples of the Berry phase is the evolution of a system from one ground state to the other in topological materials.The degeneracy corresponds to the parametric space $\bm{R}(t)$. This idea is used in topological quantum computation in Section \ref{TQC}.

\subsection*{C.4. Anyons on a Torus}\label{AnyonOnTorus}
\begin{figure}[h!]
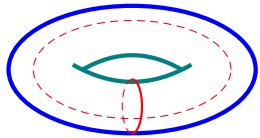

	\centering
	\tik{[scale=0.65]
		\draw[ultra thick,blue] (0,0) ellipse (2.5 and 1.3);
		\draw[bend left,ultra thick,teal] (-1,-0) to (1,-0);
		\draw[bend right,ultra thick,teal] (-1.2,0.1) to (1.2,0.1);	
		\draw[densely dashed,red] (0,0) ellipse (2 and 1);	
		\draw[densely dashed,red] (0,-1.3) arc (270:90:.2 and 0.550);
		\draw[thick,red] (0,-1.3) arc (-90:90:.2 and .550);}
	\caption{Two non-trivial paths in torus.}
	\label{PathsTorus}
\end{figure}

The system may have multiple types of anyons categorized according to their topological charge. As we discussed in \ref{Intro}, anyons are charge-flux composite found in topological materials. When two anyons are brought close to each other, their fusion may result in another anyon or a superposition of several anyons. These two anyons may annihilate to vacuum if they are antiparticles to each other. The ground state somehow knows what types of anyons can be created. Let there be two paths $C_1$ and $C_2$ on a torus along meridian and longitude, as shown in Fig. \ref{PathsTorus}.
Let $T_1$ and $T_2$ be operators correspond to the creation of anyon-antianyon pair from the vacuum and carrying around meridian and longitude respectively. 
$T_2^{-1}T_1^{-1}T_2T_1$ is two particles created, braided around each other and then re-annihilated. Since the operators $T_1$ and $T_2$ are implemented with some time-dependent Hamiltonian \cite{nayak2008non}, they are unitary. These two operators do not commute with each other, we have 
\eq{T_2T_1 = e^{-2i\theta} T_1T_2.}
Therefore, the system has ground state degeneracy. As $T_1$ is unitary, its eigenvalues must have a unit modulus, that is, they are just complex numbers. The operation of $T_1$ on a state $\alpha$ can be written as 
\eq{T_1\ket{\alpha} = e^{i\alpha}\ket{\alpha}.}
where $\alpha$ is the space of possible ground states. $T_2\ket{\alpha}$ must also be a ground state since $T_2$ commutes with $H$. Therefore, we can write
\eq{T_1(T_2 \ket{\alpha}) = e^{2i\theta}e^{i\alpha} (T_2 \ket{\alpha}).}
Let us call this new ground state $\ket{\alpha+2\theta} = T_2\ket{\alpha}$. 
On similar lines, we can generate more ground states. Consider a system where anyons have a statistical phase $\theta = \pi p/m$, where $p$ and $m$ are relatively prime so that $p/m$ is an irreducible fraction. The ground states can be written as
\eq{\ket{\alpha},\ket{\alpha+2\pi p/m}, \ket{\alpha+4\pi p/m},...,\ket{\alpha+2\pi (m-1)/m}.}
The phase $\alpha+2\pi = \alpha$ so that we are back to the original state. Now we have $m$ independent ground states. Since anyons have the statistical angle $\theta = \pi p/m$, the charge-flux composite will get $(q,\Phi) = (\pi p/m,1)$. When there is a fusion of $n$ elementary anyons then we have $\ket{n} = (q=n\pi p/m,\Phi =n) = (n\pi p/m,n)$. When there are $m$ anyons, we have $\ket{m} = (\pi p,m)$.
Now if we braid $\ket{n} = (n \pi p/m ,n)$ around one of these $\ket{m} = (\pi p/m)$, we obtain a net phase of $2\pi p$ which is equivalent to no phase at all. 
Hence, the cluster of $m$ elementary anyons is equivalent to a vacuum. In this way, we have $m$ species of anyon and $m$ different ground states on torus \cite{nayak2008non}. The subspace used to implement the topological gates depends on a particular model of anyons and also on the number of anyons present.

In the case of an annulus instead of a torus, the $T_1$ operator corresponds to a particle moving along a circular loop and $T_2$ to the particle moving from the inside edge to the outside edge. The degeneracy is $2$. On similar bases, the degeneracy for the higher genus space is $m^g$, where $g$ is \textit{genus}. The genus is a handle in a topological space. 

The total charge and flux of a fusion of two particles must be zero: that is, we should get the vacuum. Therefore, the antianyon must have charge $-q$ and phase $\Phi$. The phase of an anyon moving clockwise around another anyon is the same as an antianyon moving clockwise around another antianyon. However, the phase of an anyon around an antianyon is $-2\phi$. The fusion of a particle with its antiparticle gives the vacuum, but when two particles are pushed together, we get a charge $2q$ and flux $2\Phi$. Now, the phase of exchanging these two particles is $\phi= 4q\Phi/\hbar$.

\section{Recoupling Theory of Angular Momenta}\label{Rec}
We will briefly present the quantum theory of angular momentum to get the idea of the $SU(2)_k$ anyonic model will be discussed in the next section. This model is the quantum deformation of recoupling theory. For more detailed study, see \cite{biedenharn1981angular,varshalovich1988quantum,rose1995elementary}.
Let us consider two systems with angular momentum operators $J_1$ and $J_2$ with eigenvalues $j_1$ and $j_2$. These two systems may be the orbital angular momentum of two different particles or they may be the spin and orbital angular momentum of a single particle. The z-components of the angular momenta $J_z$ have allowed eigenvalues $-j_1 \le m_1 \le +j_1$ with $2j_1 + 1$ values, and $-j_2 \le m_2 \le +j_2$ with $2j_2 + 1$ states. The combined system is written as $j_1 \otimes j_2$.

It is like the vector spaces $V_1$ and $V_2$, with dimensions $2j_2 +1$ each, are combined as $V_1 \otimes V_2$ with dimensions $(2j_1 +1)(2j_2 +1)$. The total angular momentum operator is acting on $V_1 \otimes V_2$. This operator constitutes $SU(2)$ Lie algebra.
Two quantum numbers are needed to specify an individual system and four quantum numbers to specify the combined system. 
For the whole system 
\eq{J^2 = (J_1 +J_2)^2, \qquad J_z = J_{1z} + J_{2z}.}
These operators are applied to states as
\eq{J^2\ket{j,m} &= j(j+1)\ket{j,m}\nonumber \\
	J_z\ket{j,m} &=m\ket{j,m} \nonumber\\
	J_\pm\ket{j,m} &= \sqrt{(j\mp m)(j\pm m+1)}\ket{j,m\pm 1}}
where $J_+ = J_x+iJ_y$ and $J_- = J_x -iJ_y$.
The uncoupled states $\ket{j_1j_2m_1m_2}$ are the eigenstates of operators $\left\{J_1^2,J_{1z},J^2_2, J_{2z}\right\}$ and the coupled state $\ket{j_1j_2jm}$ are the eigenstate of operators $\left\{J_1^2,J^2_2,J^2, J^2_z\right\}$.
The coupling of $j_1$ and $j_2$ is the construction of eigenfunctions of $J^2$ and $J_z$
We will write the total angular momentum basis $\ket{j_1j_2;jm}$ in terms of tensor product basis $\ket{j_1m_1}\ket{j_2m_2}$. That is done by expressing the coupled state in terms of the uncoupled state.
\eq{&\ket{(j_1 j_2)jm}= \sum_{m_1 = -j_1}^{j_1}\sum_{m_2 = -j_2}^{j_2} \ket{j_1m_1}\ket{j_2m_2}\bra{j_1m_1j_2m_2}\ket{jm}\nonumber\\
	&=\sum_{j_1j_2m_1m_2}C_{mm_1m_2}^{jj_1j_2}\ket{j_1m_1}\ket{j_2m_2}.}
where $j=\abs{j_1-j_2},...,\abs{j_1+j_2}, \ m = -j,...,j$.
The coefficients $C_{mm_1m_2}^{jj_1j_2}$
\eq{\sum_{j_1j_2m_1m_2}C_{mm_1m_2}^{jj_1j_2} = \bra{j_1m_1;j_2m_2}\ket{j_1j_2;jm} = \bra{j_1m_1;j_2m_2}\ket{jm}}
are called Clebsch-Gordon coefficients (CGC). These are non-zero only when $m = m_1 + m_2$ and $\abs{j_1-j_2} \le j \le j_1 +j_2$.

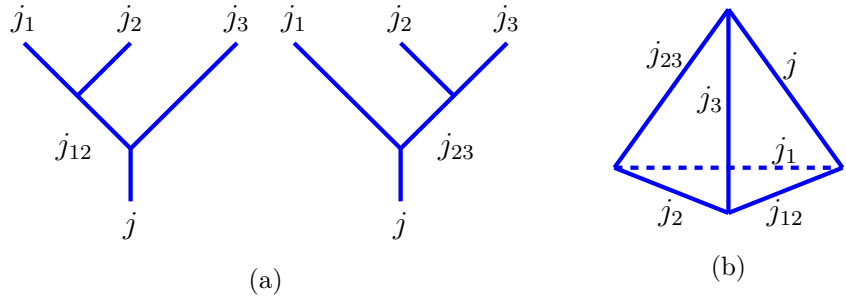
\begin{figure*}[h!]
	\centering
	\begin{subfigure}{0.5\textwidth}
		\centering
		\begin{tikzpicture}[scale=0.7]
			\draw [blue, ultra thick] (2,0) -- (2,1); 
			\draw [blue, ultra thick] (2,1) -- (4,3);
			\draw [blue, ultra thick] (2,1) -- (0,3);
			\draw [blue, ultra thick] (1,2) --(2,3);
			\node [above] at (0,3) {$j_1$};
			\node [above] at (2,3) {$j_2$};
			\node [above] at (4,3) {$j_3$};
			\node [below] at (2,0) {$j$};
			\node [below left] at (1.5,1.5) {$j_{12}$};
		\end{tikzpicture}
		\begin{tikzpicture}[scale=0.7]	
			\draw [blue, ultra thick] (10,0) -- (10,1); 
			\draw [blue, ultra thick] (10,1) -- (12,3);
			\draw [blue, ultra thick] (10,1) -- (8,3);
			\draw [blue, ultra thick] (11,2) --(10,3);		
			\node [above] at (8,3) {$j_1$};
			\node [above] at (10,3) {$j_2$};
			\node [above] at (12,3) {$j_3$};
			\node [below] at (10,0) {$j$};
			\node [below right] at (10.5,1.5) {$j_{23}$};
		\end{tikzpicture}
		\caption{}
	\end{subfigure}
	\begin{subfigure}{0.2\textwidth}
		\centering	
		\begin{tikzpicture}[scale=1.5]
			\draw[ultra thick, blue] (0,-1) -- (0,0.8);
			\draw[ultra thick, blue] (0,-1) -- (1,-0.6);
			\draw[ultra thick, blue] (0,-1) -- (-1,-0.6);
			\draw[ultra thick,dashed, blue] (-1,-0.6) -- (1,-0.6);
			\draw[ultra thick, blue] (-1,-0.6) -- (0,0.8);
			\draw[ultra thick, blue] (1,-0.6) -- (0,0.8);
			\node [left] at (0.05,0) {$j_3$};
			\node [] at (0.5,-0.45) {$j_1$};
			\node [] at (0.5,-1) {$j_{12}$};
			\node [] at (-0.5,-1) {$j_2$};
			\node [] at (0.55,0.3) {$j$};
			\node [] at (-0.55,0.4) {$j_{23}$};
		\end{tikzpicture}
		\caption{}
	\end{subfigure}
	\caption{(a) Recoupling (b) the triangle condition. }
	\label{TriCond}
\end{figure*}

The total angular momentum can have value $j = j_1 + j_2, j_1+j_2 -1,...,\abs{j_1 -j_2}$.
Any of the numbers $j_1,j_2,j$ can have values that are greater than or equal to the difference of the other two and less than or equal to the sum of the other two. This condition is called \textit{triangle condition} and is represented by $\Delta(j_1j_2j)$.
We can also write the total angular momentum basis in terms of the product basis by using the \textit{Wigner's 3j-symbols}. In that case, the coefficients are called \textit{Wigner coefficients}.
The Wigner 3j-symbol is zero unless the triangle condition is satisfied. 
The CGC are related to the 3j-symbols as
\eq{\bra{j_1m_1;j_2m_2}\ket{j_1j_2;jm} = (-1)^{-j_1+j_2 -m}\sqrt{2j+1}
	\begin{Bmatrix}
		j_1&j_2&j \\
		m_1&m_2&-m
	\end{Bmatrix}.}
On similar lines, we can couple three angular momenta
$J_1,J_2,J_3$ whose total angular momentum $J= J_1+J_2+J_3$. There are two coupling schemes as
\eq{(J_1+J_2)+J_3 = J_{12}+J_3 = J \qquad J_1+(J_2 +J_3) = J_1+J_{23} = J.} 
The total coupling may be done by 
first coupling $j_1$ and $j_2$ to $j_{12}$ and then $j_{12}$ and $j_3$ to $J$
\eq{\ket{(j_1j_2)j_{12}j_3;jm}
	= \sum_{m_{12} = -j_{12}}^{j_{12}}\sum_{m_3 = -j_3}^{j_3} \ket{(j_1j_2);j_{12}m_{12}}\ket{j_3m_3} 
	\bra{j_{12}j_3;m_{12}m_3}\ket{j_{12}j_3;jm}.}
Alternatively, we can first combine $j_2$ and $j_3$ to get $j_{23}$ and next $j_{23}$ can be combined with $j_1$ to make $J$
\eq{\ket{j_1((j_2 j_3)j_{23});jm} 
	= \sum_{m_1 = -j_1}^{j_1}\sum_{m_{23} = -j_{23}}^{j_{23}}	
	\ket{j_1m_1}
	\ket{j_2j_3;j_{23}m_{23}} \bra{j_1 j_{23};m_1m_{23}}\ket{j_1j_{23};jm}.}
These two coupling schemes are shown in Fig. \ref{TriCond} (a). The coupling scheme results in a complete orthonormal bases for the $(2j_1 + 1)(2j_2 + 1)(2j_3+1)$-dimensional space spanned by $\ket{j_1,m_1}\ket{j_2,m_2}\ket{j_3,m_3}$, $m_1 = -j_1,...,j_1, m_2 = -j_2,...,j_2$; $m_3 = -j_3,...,j_3$.

The angular momenta in two coupling schemes are related by a unitary transformation. The matrix elements of this unitary transformation are known as \textit{recoupling coefficients}. These coefficients are independent of $m$ and so we have

\eq{\ket{((j_1j_2)j_{12}j_3)jm} &= \sum_{j_{23}} \ket{(j_1(j_2j_3)j_{23})jm}\bra{(j_1(j_2j_3)j_{23})j}\ket{((j_1j_2)j_{12}j_3)j}.}
These coefficients can be written in terms of \textit{Wigner 6j-symbols},
\eq{\bra{(j_1(j_2j_3)j_{23})j}\ket{((j_1j_2)j_{12}j_3)j} = (-1)^{j_1+j_2+j_3+j} \sqrt{(2j_{12} +1)(2j_{23} +1)}\begin{Bmatrix}
		j_1 & j_2 & j_{12} \\
		j_3 & j & j_{23}
	\end{Bmatrix}.}
The 6j-symbols have a symmetry that permutation of columns or rows leaves it invariant. Similar to the 3j-symbols, 6j-symbols are not matrices. The Racah coefficients \cite{racah1942theory} are related to the recoupling coefficients as
\eq{W(j_1 j_2 j_3J; j_{12}j_{23}) = \frac{\bra{(j_1(j_2j_3)j_{23})j}\ket{((j_1j_2)j_{12}j_3)j}}{\sqrt{(2J_{12}+1)(2J_{23}+1)}}.}
Therefore, the Racah coefficients are related to the Wigner 6j-symbols by
\eq{\begin{Bmatrix}
		j_1 & j_2 & j_{12} \\
		j_3 & j & j_{23}
	\end{Bmatrix} = (-1)^{j_1+j_2+j3+j} W(j_1j_2j_3j;j_{12}j_{23})}
If $a \equiv j_1, \ b \equiv j_2, \ c \equiv j_3, \ d \equiv j, \ e \equiv j_{12}, \ f \equiv j_{23}$, we have the triangle condition as
\eq{\Delta(abc) = \sqrt{\frac{(a+b-c)!(a-b+c)!(-a+b+c)!}{(a+b+c+1)!}}.}
The right hand side is zero unless the triangle condition is satisfied. This condition is satisfied by each side of the quadrilateral in Fig. \ref{TriCond} (b). The Racah coefficient is a product of four of these factors
\eq{W(abcd;ef) = \Delta(abe)\Delta(cde)\Delta(acf)\Delta(bdf) \omega(abcd;ef),}
where $$\omega(abcd;ef) = \sum_z \frac{(-1)^{z+\beta_1}(z+1)!}{(z-\alpha_1)!(z-\alpha_2)!(z-\alpha_3)!(z-\alpha_4)!(\beta_1 - z)!(\beta_2 - z)!(\beta_3 - z)!},$$
$\alpha_1 = a+b+e, \ \alpha_2 = c+d+e, \ \alpha_3 = a+c+f \ \alpha_4 = b+d+f$
$\beta_1 = a+b+c+d, \ \beta_2 = a+d+e+f, \ \beta_3 = b+c+e+f$.
The sum over $z$ is finite over the range $max(\alpha_1,\alpha_2,\alpha_3,\alpha_4) \le z \le min(\beta_1,\beta_2,\beta_3)$
See \cite{rose1995elementary,aquilanti2009combinatorics,santos2017couplings} for detailed derivation of the above equation and coupling and recoupling of angular momenta. 

\subsubsection*{Acknowledgement}
Authors are grateful to Jack Straton and Steven Bleiler for their guidance during the preparation of this work.

	\bibliographystyle{unsrt}
	\bibliography{zRef}

\end{document}